\documentclass[letterpaper,12pt]{article}
\pdfoutput=1
\usepackage{jheppub}
\usepackage{amsmath}
\usepackage{amssymb}
\usepackage{dcolumn}
\usepackage{bm}
\usepackage{feynmp}
\usepackage{slashed}                              
\usepackage{hyperref,graphicx}           
\usepackage{soul}
\usepackage{upgreek}

\usepackage{caption}
\usepackage{subcaption}
\usepackage{color}

\numberwithin{equation}{section}
\newcommand{\be}{\begin{equation}}
\newcommand{\ee}{\end{equation}}
\newcommand{\beq}{\begin{equation}}
\newcommand{\eeq}{\end{equation}}
\newcommand{\bea}{\begin{eqnarray}}
\newcommand{\eea}{\end{eqnarray}}
\newcommand{\bear}{\begin{eqnarray}}
\newcommand{\eear}{\end{eqnarray}}

\newcommand{\ba}{\begin{array}}
\newcommand{\ea}{\end{array}}

\newcommand{\negspxhalf}{\mkern-10mu} 
\def\tev{\,{\rm TeV}}
\def\gev{\,{\rm GeV}}
\def\mev{\,{\rm MeV}}
\def\kev{\,{\rm keV}}
\def\MeV{\,{\rm MeV}}

\def\ev{\,{\rm eV}}

\def\fbi{\,{\rm fb}^{-1}}

\newcommand{\eg}{{\it e.g.}}

\def\ltap{\ \raise.3ex\hbox{$<$\kern-.75em\lower1ex\hbox{$\sim$}}\ }
\def\gtap{\ \raise.3ex\hbox{$>$\kern-.75em\lower1ex\hbox{$\sim$}}\ }
\def\lsim{\ \raise.3ex\hbox{$<$\kern-.75em\lower1ex\hbox{$\sim$}}\ }
\def\gsim{\ \raise.3ex\hbox{$>$\kern-.75em\lower1ex\hbox{$\sim$}}\ }
\newcommand{\figref}[1]{Fig.~\ref{#1}}

\newcommand{\eref}[1]{Eq.~(\ref{#1})}
\newcommand{\numN}{\ensuremath{\mathcal{N}_N}}

\begin{document}

\title{Neutrino Masses from Low Scale\\ Partial Compositeness
}

\author[a]{Zackaria Chacko} 
\author[b]{, Patrick J. Fox}
\author[b]{, Roni Harnik}
\author[a,c]{, and Zhen Liu}

\affiliation[a]{Maryland Center for Fundamental Physics, Department of Physics,\\ University of Maryland, College Park, MD 20742-4111 USA}
\affiliation[b]{Theoretical Physics Department, Fermilab, P.O. Box 500, 
Batavia, IL 60510, USA}
\affiliation[c]{School of Physics and Astronomy, University of Minnesota, Minneapolis, MN 55455, USA}

\date{June 25$^{\mathrm{th}}$, 2019}

\preprint{FC-5657, FERMILAB-PUB-20-238-T}




\abstract{ 
 We consider a class of models in which the neutrinos acquire Majorana 
masses through mixing with singlet neutrinos that emerge as composite 
states of a strongly coupled hidden sector. In this framework, the light 
neutrinos are partially composite particles that obtain their masses 
through the inverse seesaw mechanism. We focus on the scenario in which 
the strong dynamics is approximately conformal in the ultraviolet, and 
the compositeness scale lies at or below the weak scale. The small 
parameters in the Lagrangian necessary to realize the observed neutrino 
masses can naturally arise as a consequence of the scaling dimensions of 
operators in the conformal field theory. We show that this class of 
models has interesting implications for a wide variety of experiments, 
including colliders and beam dumps, searches for lepton flavor violation 
and neutrinoless double beta decay, and cosmological observations. At 
colliders and beam dumps, this scenario can give rise to striking signals 
involving multiple displaced vertices. The exchange of hidden sector 
states can lead to observable rates for flavor violating processes such 
as $\mu \rightarrow e \gamma$ and $\mu \rightarrow e$ conversion. If the 
compositeness scale lies at or below a hundred MeV, the rate for 
neutrinoless double beta decay is suppressed by form factors and may be 
reduced by an order of magnitude or more. The late decays of relic 
singlet neutrinos can give rise to spectral distortions in the cosmic 
microwave background that are large enough to be observed in future 
experiments.
 }

\maketitle
\section{Introduction\label{s.intro}}

Over the last few decades a series of experiments have conclusively 
established that neutrinos have tiny but non-vanishing masses and have 
determined their mass splittings. However, the mechanism that gives rise 
to these small masses remains a mystery. Among the many well-motivated 
proposals that have been put forward are the seesaw mechanism in its 
various different 
incarnations~\cite{Minkowski:1977sc,YanagidaSS,GellMann:1980vs,GlashowSS,Mohapatra:1979ia}, 
\cite{Konetschny:1977bn,Magg:1980ut,Schechter:1980gr,Cheng:1980qt,Lazarides:1980nt,Mohapatra:1980yp}, 
\cite{Foot:1988aq,Ma:1998dn}, and the Majoron 
model~\cite{Gelmini:1980re,Chikashige:1980ui,Georgi:1981pg}.

One attractive class of models that can naturally explain the smallness 
of the neutrino masses are those in which the neutrinos acquire mass 
through their couplings to the composite states of a strongly coupled 
sector~\cite{ArkaniHamed:1998pf,vonGersdorff:2008is,Grossman:2010iq}. 
Since the compositeness scale arises from dimensional transmutation, it 
can be parametrically lower than the Planck scale. The smallness of 
neutrino masses is explained by the fact they arise from operators of 
dimension greater than four. In this scenario the neutrinos could be 
either Dirac or Majorana particles. The composite neutrino framework has 
been linked to dark 
matter~\cite{Robinson:2012wu,Robinson:2014bma,Hundi:2011et} and to the 
origin of the baryon asymmetry of the universe~\cite{Grossman:2008xb}. 
The cosmological signals of this class of theories have been considered 
in~\cite{Okui:2004xn}.

\begin{figure}[t] 
\centering 
\includegraphics[width=0.9\textwidth]{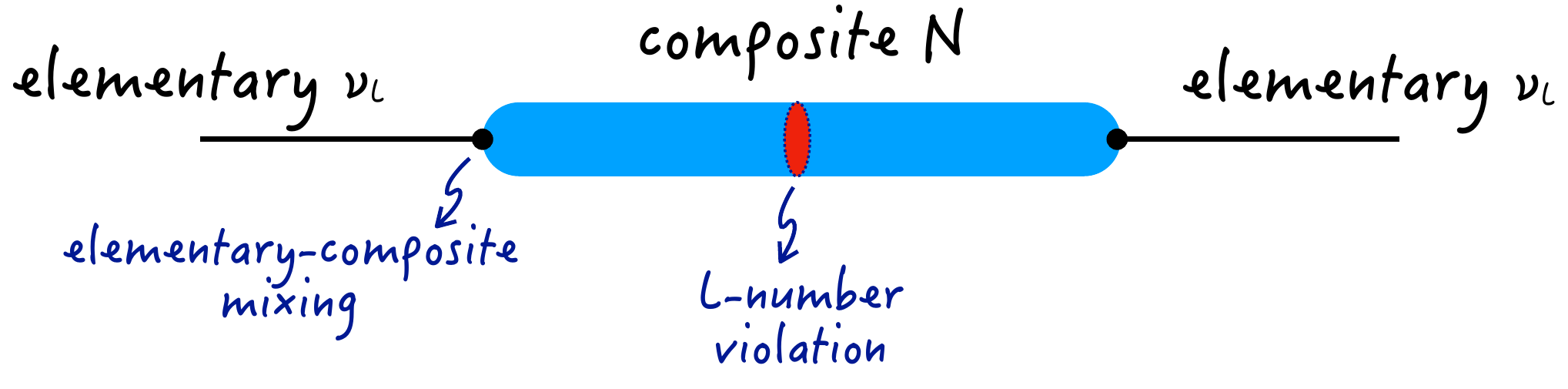} 
 \caption{A sketch of the generation of neutrino masses in our 
framework through partial compositeness. Elementary neutrinos mix with 
composite singlet neutrinos and lepton number is violated in the 
composite sector. The light neutrino mass may be explained either by 
small mixing between the elementary and composite neutrinos, or by small 
lepton number violation in the composite sector, with a continuum of 
possibilities in between.}
 \label{fig:numass}
 \end{figure}

In this paper we explore a class of models that realize this scenario 
and show that they can lead to rich phenomenological consequences. We 
consider a framework in which the neutrinos acquire Majorana masses 
through mixing with singlet neutrinos that emerge as composite states of 
a strongly coupled hidden sector, as shown schematically in 
Figure~\ref{fig:numass}. 
 From the perspective of the low energy theory, at or below the 
compositeness scale, the neutrino masses arise through the inverse 
seesaw mechanism~\cite{Mohapatra:1986aw,Mohapatra:1986bd}. The 
difference between our framework and that of a conventional inverse 
seesaw is that the singlet neutrinos are now the hadrons of a new strong 
force, and the light neutrinos are partially composite particles. Our 
approach is similar to that of composite Higgs models, (for reviews 
see~\cite{Bellazzini:2014yua,Panico:2015jxa}), in which the SM fermions 
receive their masses through partial compositeness~\cite{Kaplan:1991dc, 
Contino:2004vy}. The composite singlet neutrinos are analogous to the 
vector-like fermion resonances in composite Higgs models. However here, 
in contrast to composite Higgs models, the strongly coupled sector is 
neutral under the SM gauge groups and so we are free to consider 
compositeness scales that lie well below the electroweak scale.

In this paper we focus on the scenario in which the compositeness scale, 
which we denote by $\Lambda$, lies at or below the weak scale. This 
allows the possibility of directly producing composite states at 
colliders and other experiments. For concreteness we take the strongly 
coupled hidden sector to be a conformal field theory (CFT). The 
conformal symmetry is broken at the scale $\Lambda$. Lepton number is 
also broken by dynamics in the hidden sector. The small parameters 
required to obtain the tiny neutrino masses within the framework of a 
low scale seesaw model can be understood here as naturally arising from 
the scaling dimensions of operators in the~CFT.

The spectrum of the neutrino sector consists of the light neutrinos,
which are now partially composite and parametrically lighter than the
compositeness scale, and an infinite tower of composite neutrinos of
which the lightest states have a mass of order the compositeness scale
$\Lambda$.
 \begin{figure}[t]
\centering
\includegraphics[width=0.65\textwidth]{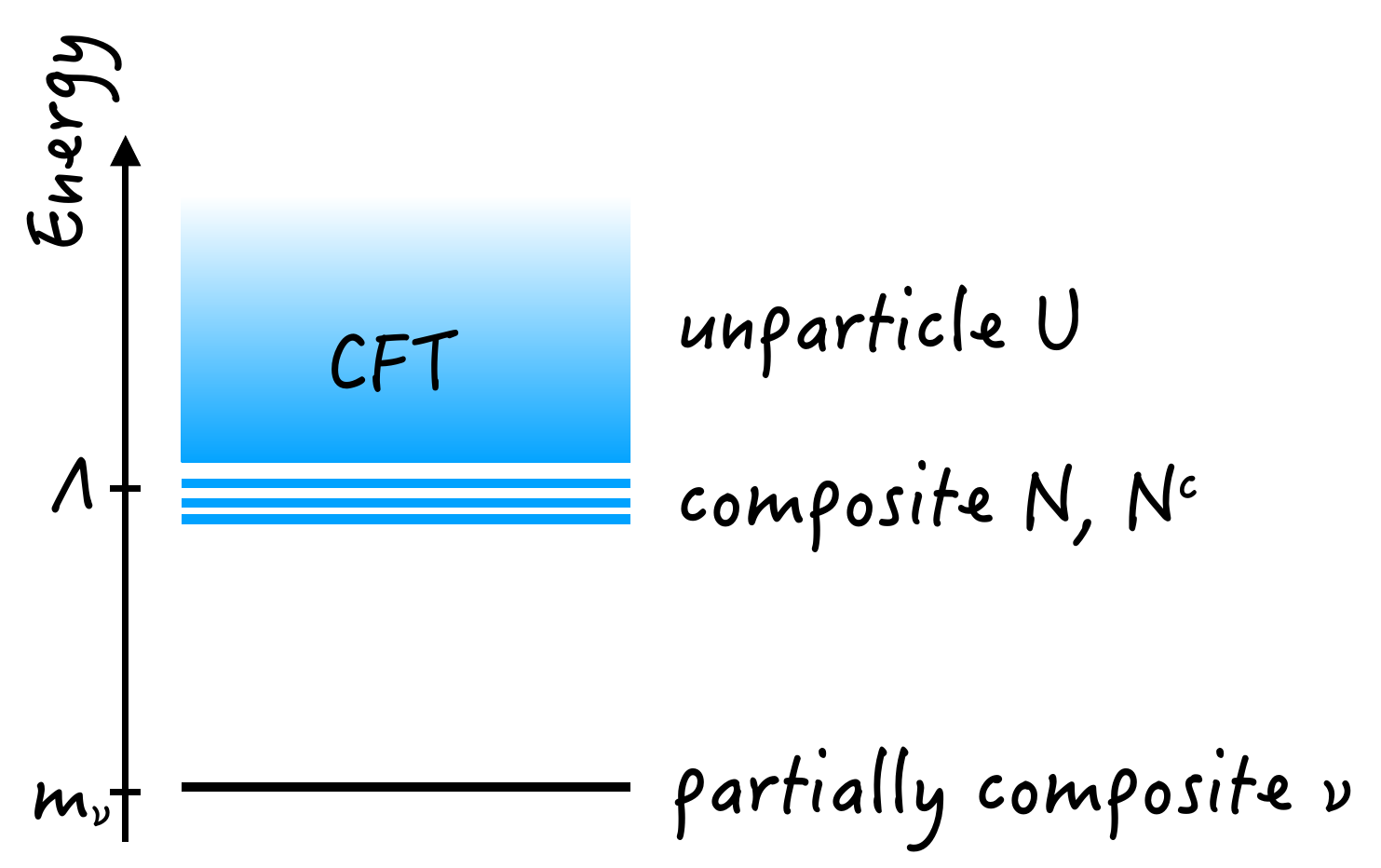}
 \caption{A sketch of the spectrum of particle masses in our framework, 
showing the light partially composite neutrinos and the composite 
resonances at the compositeness scale $\Lambda$ that blend into a 
continuum of ``unparticles" at higher scales.}
 \label{fig:spectrum}
 \end{figure}
 At scales well above $\Lambda$ the spectrum of composite states blends 
into a continuum spectrum of 
``unparticles"~\cite{Georgi:2007ek,Georgi:2007si}. The spectrum of particle
masses in this scenario is shown in Figure~\ref{fig:spectrum}. 

The AdS/CFT 
correspondence~\cite{Maldacena:1997re,Gubser:1998bc,Witten:1998qj,Klebanov:1999tb} 
relates strongly coupled CFTs to theories of gravity in higher 
dimensions. Theories of phenomenological interest in which the strong 
conformal dynamics is spontaneously broken are 
dual~\cite{ArkaniHamed:2000ds,Rattazzi:2000hs} to the two-brane 
Randall-Sundrum (RS) construction~\cite{Randall:1999ee}, originally 
proposed as a solution to the hierarchy problem of the SM. Several 
authors have explored neutrino masses within the framework of the RS 
solution to the hierarchy problem, for 
example~\cite{Grossman:1999ra,Huber:2002gp,Huber:2003sf,Gherghetta:2003hf}. 
It was shown in~\cite{Agashe:2015izu} that the AdS/CFT correspondence 
relates a certain class of these extra dimensional models to theories in 
which the neutrino masses are generated by the inverse seesaw mechanism, 
with composite states playing the role of the singlet neutrinos. These 
theories therefore share many features of our construction. However, an 
important difference is that in these theories the SM leptons are 
themselves partially or entirely composite. The compositeness scale is 
then constrained to lie above a TeV, and consequently the phenomenology 
is very different. A holographic model of neutrino masses from low scale 
compositeness that is more closely related to our construction was considered 
in~\cite{McDonald:2010jm} (see also \cite{Gripaios:2006dc}).

\begin{figure}[t]
\centering
\includegraphics[width=1.0\textwidth]{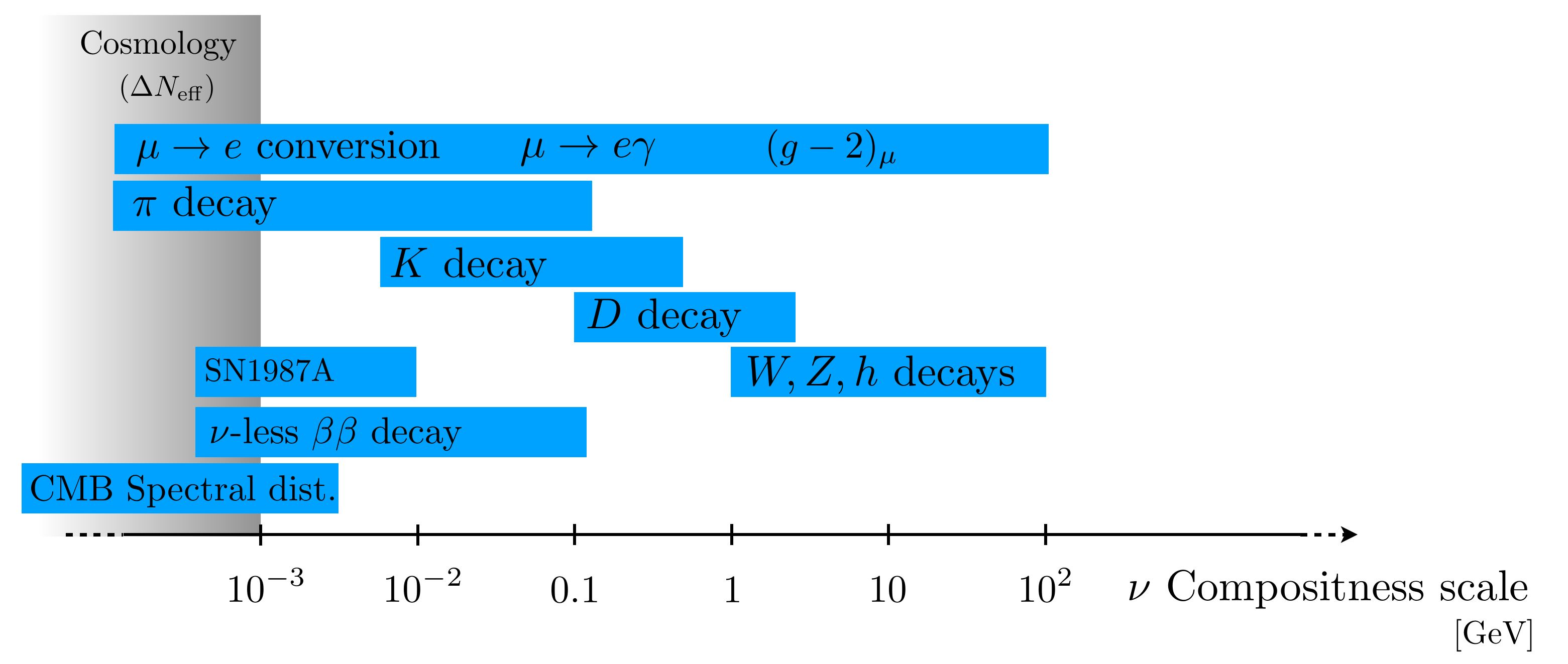}
 \caption{A sketch of the various probes of neutrino compositeness and 
the range of compositeness scales that they are potentially sensitive 
to.}
 \label{fig:range}
 \end{figure}

The class of models that we are exploring has interesting implications for a 
wide variety of current and near-future experiments, including colliders 
and beam dumps, searches for lepton flavor violation and neutrinoless 
double beta decay, as well as cosmological observations. The range of 
compositeness scales that can be probed by each of these different types 
of signals is summarized in Figure~\ref{fig:range}.
We now consider each of these different experimental probes in turn,
along with the constraints from current data.
\begin{itemize}
\item {\bf LHC signals:}
This scenario can give rise to striking signals at colliders.
A $W$ boson produced at the Large Hadron Collider (LHC) can now decay into a
hard lepton and one or more composite singlet neutrinos. The singlet
neutrinos subsequently decay into SM final states. In much of the parameter space, these decays are slow on collider timescales, giving rise to events with multiple displaced vertices. The observation of this
striking signal would allow this class of theories to be distinguished
from more conventional inverse seesaw models in which the singlet
neutrinos are elementary. We find that the HL-LHC will have excellent
sensitivity to this class of models. Codex-B, FASER-2, and MATHUSLA can
all further improve on the HL-LHC reach.
This is discussed in Section~\ref{sec:collider}.
\item {\bf Meson decays in beam dumps:}
If the singlet neutrino mass lies below a few GeV, this class of models
can also be discovered at beam dumps. Mesons produced at beam dumps can
now decay into final states that contain one or more singlet neutrinos.
The subsequent decays of these singlet neutrinos into SM final states
can lead to striking signals, particularly in the case when these decays
result in displaced vertices. We find that DUNE and SHiP are both
sensitive to this scenario.
This is discussed in further detail in Section~\ref{sec:dump}.
\item {\bf Charged lepton flavor violation:}
In general the couplings of the SM to the hidden sector
violate the flavor symmetries of the lepton sector of the SM.
Consequently, this scenario can give rise to lepton flavor violation at
loop level through the exchange of hidden sector states. The COMET and
Mu2e experiments, which are searching for the lepton flavor violating
process
$\mu \rightarrow e$ conversion,
will be able to probe this class of models.
We discuss this in detail in Section~\ref{s.loop}.
\item {\bf Neutrinoless double $\beta$ decay:}
Experiments searching for neutrinoless double beta decay are extremely
important because the observation of this process would establish that
neutrinos are Majorana particles. The characteristic momentum scale
associated with neutrinoless double beta decay is set by the nucleon
spacing in the nucleus, which is of order the pion mass. It follows that
if the scale of compositeness in the neutrino sector lies at or below a
hundred MeV, the rate for neutrinoless double beta decay is suppressed by form
factors and may be greatly reduced, with important consequences for
experiment. We discuss this in detail in Section~\ref{s.beta}.
\item {\bf Cosmology:}
In the early universe, thermal contact with the SM ensures that the
hidden sector is populated. At temperatures below the compositeness
scale the states in the hidden sector annihilate efficiently into light
neutrinos, with the result that their abundance is exponentially
suppressed. If the compositeness scale lies below an MeV, the
temperature at which the weak interactions decouple, this results in a
correction to the total energy density in neutrinos that is unacceptably
large. Values of $\Lambda$ below an MeV are therefore disfavored. The
remaining relic composite states eventually decay into final states
that, in addition to neutrinos, may also include charged particles. If
the compositeness scale lies below 50 MeV, these late decays into
charged states can give rise to spectral distortions in the cosmic
microwave background (CMB) that are potentially observable in future
experiments. The cosmological history of this class of models is
discussed in Section~\ref{s.cosmo}.
\item {\bf Supernovae:}
If the compositeness scale is lower than or of order 40 MeV, the
temperature in the core of a supernova, hidden sector states can be
produced at the core. Once produced, they get trapped inside the core and
thermalize, and so they do not contribute to energy loss. The
presence of these additional degrees of freedom would be expected to
affect the supernova dynamics. However, given the present uncertainties
in supernova dynamics we cannot robustly rule out this possibility.
This is discussed in Section~\ref{s.astro}.
\end{itemize}

In the next section we present the framework that underlies this class
of models and determine the allowed parameter space. In the subsequent
sections we discuss each of the various types of signals as listed
above. We conclude in Sec.~\ref{s.conclusion}.

 \section{An Inverse Seesaw Model from Strong Dynamics\label{s.inverse}}

In this section we describe our framework for neutrino mass generation
based on the partial compositeness of neutrinos. For concreteness, we
will take the strong dynamics to be that of a conformal field theory
(CFT) deformed by a relevant operator $\mathcal{O}_{\rm S}$. The
corresponding terms in the Lagrangian take the form,
\begin{equation}
\mathcal{L}_{\rm UV} \supset \mathcal{L}_{\rm CFT}
+ \lambda_{\rm S} \mathcal{O}_{\rm S} \;.
\label{LagCFT}
 \end{equation}
 When the deformation $\mathcal{O}_{\rm S}$ gets large, it triggers 
breaking of the CFT at a scale which we denote by $\Lambda$. We focus on 
the scenario in which there are no pions or other light composite 
states, so that $\Lambda$ corresponds to the mass scale of the lightest 
composite particles. We require that the spectrum of light composite 
states contains at least two pairs of fermions $N$ and their Dirac 
partners $N^c$ that will play the role of singlet neutrinos. We will 
return to the issue of flavor at the end of this section, but for now we 
suppress all flavor indices. The low energy Lagrangian then includes 
kinetic terms and mass terms for the composite singlet neutrinos,
 \begin{equation}
\mathcal{L}_{\rm IR} \supset
i\bar{N} \bar{\sigma}^\mu \partial_{\mu} N + i\bar{N}^c \bar{\sigma}^\mu \partial_{\mu} N^c - \left(M_N N^c N +\mathrm{h.c.}\right) ~.
\label{freeN}
 \end{equation}
 We expect that the mass parameter $M_N$ is of order $\Lambda$. 

The CFT is assumed to couple to the SM through a neutrino portal
interaction which, above the scale $\Lambda$, takes the form
\begin{equation}
\mathcal{L}_{\rm UV} \supset \frac{\hat{\lambda}}{M_{\mathrm{UV}}^{\Delta_{\mathrm{N}} - 3/2}} L H \mathcal{O}_{\rm N} + {\rm h.c.}~.
\label{intLHO}
 \end{equation} 
 In this expression the mass scale $M_{\mathrm{UV}}$ represents the 
ultraviolet cutoff of the theory, and $\hat{\lambda}$ is a dimensionless 
parameter taken to be $\mathcal{O}(1)$. Here $\mathcal{O}_{\rm N}$ is a 
primary operator of the CFT and $\Delta_{\mathrm{N}}$ is its scaling 
dimension. Unitarity places restrictions on the scaling dimensions of 
CFT primary operators which depend on their transformation properties 
under the Lorentz group. For spin 1/2 operators such as 
$\mathcal{O}_{\rm N}$, unitarity requires $\Delta_{\mathrm{N}}\ge 3/2$. 
The limiting case of $\Delta_{\mathrm{N}} = 3/2$ corresponds to 
$\mathcal{O}_{\rm N}$ being a free fermion. For $\Delta_{\mathrm{N}} \ge 
5/2$ the coupling in Eq.~(\ref{intLHO}) makes the theory ultraviolet 
sensitive, and additional counterterms involving the SM fields are 
required for consistency. We therefore focus on the range of scaling 
dimensions, $ 3/2 \le \Delta_{\mathrm{N}} < 5/2$. We use the two-point 
function to normalize the operator $\mathcal{O}_{\rm N}$, following the 
the conventions of unparticle physics for fermionic 
operators~\cite{Cacciapaglia:2007jq},
 \begin{equation}
\int d^4 x e^{ipx}
\langle 0|T\left[ \mathcal{O}_{\rm N}(x) \mathcal{O}_{\rm N}^\dagger(0)\right] |0 \rangle =
\frac{A_{\Delta_{\mathrm{N}} - 1/2}}{2 i {\rm cos} \left( \Delta_{\mathrm{N}} \pi \right) }
\frac{\sigma^\mu p_\mu}
{\left(-p^2 - i \epsilon\right)^{5/2 - \Delta_{\mathrm{N}}} }
\end{equation}
where
\begin{equation}
A_{\Delta_{\mathrm{N}} - 1/2} =
\frac{16 \pi^{5/2}}{\left(2 \pi \right)^{2 \Delta_{\mathrm{N}} - 1}}
\frac{\Gamma \left(\Delta_{\mathrm{N}}\right)}
{\Gamma \left(\Delta_{\mathrm{N}} - 3/2 \right)
\Gamma \left(2 \Delta_{\mathrm{N}} - 1 \right)}\,.
\label{eq:PhaseSpaceVol}
\end{equation}
At energies of order $\Lambda$, the interaction in \eref{intLHO}
gives rise to a term in the low-energy Lagrangian of the form,
\begin{equation}
\mathcal{L}_{\rm IR} \supset \lambda L H N + {\rm h.c.}
\label{intLHN}
 \end{equation}
 Based on our normalization of $\mathcal{O}_{\rm N}$, using the methods of
``naive dimensional analysis" (NDA)~\cite{Weinberg:1978kz,Manohar:1983md,
Georgi:1986kr} we estimate $\lambda$ to be of order,
 \begin{equation}
\lambda \sim
C_\lambda\,
\hat \lambda
\left(\frac{\Lambda}{M_{\mathrm{UV}}}\right)^{\Delta_{\mathrm{N}}-3/2}\;,
\label{lambdascaling}
 \end{equation}
where the order one multiplicative factor $C_\lambda$ is given by,
 \begin{equation}
C_\lambda = \frac{(4\pi)^{3/2-\Delta_{\mathrm{N}}}}{\Gamma(\Delta_{\mathrm{N}}-3/2)}
\sqrt{\frac{ \pi}{ (\Delta_{\mathrm{N}}-3/2)\cos (\Delta_{\mathrm{N}}\pi) }} ~.
\label{lambdascaling2}
 \end{equation}
 The terms in \eref{freeN} and \eref{intLHN} respect a global lepton 
number symmetry under which $L$ and $N^c$ carry charge $+1$ and $N$ has 
charge $-1$. We now assume that at high energies, in addition to the 
terms in \eref{LagCFT} and \eref{intLHO}, the Lagrangian contains a 
small deformation of the CFT, denoted by $\mathcal{O}_{\rm 2N^c}$, which 
explicitly violates lepton number,
 \begin{equation}
\label{Nc2deformation}
\mathcal{L}_{\rm UV} \supset \frac{\hat{\mu}^c}{M_{\mathrm{UV}}^{\Delta_{\rm 2N^c} - 4}}
\mathcal{O}_{\rm 2N^c} + {\rm h.c.}
 \end{equation} Here $\Delta_{\rm 2N^c}$ is the scaling dimension of the 
operator $\mathcal{O}_{\rm 2N^c}$, and $\hat{\mu}^c$ is a dimensionless 
parameter. The unitarity bound on the scaling dimensions of scalar 
operators restricts $\Delta_{\rm 2N^c} \geq 1$, where the limiting case 
of $\Delta_{\rm 2N^c} = 1$ corresponds to a free scalar. For the range 
of scaling dimensions $1 \leq \Delta_{\rm 2N^c} < 2$, we can normalize 
$\mathcal{O}_{\rm 2N^c}$ as per the conventions of unparticle physics, 
this time for scalar operators~\cite{Georgi:2007ek,Georgi:2007si},
 \begin{equation}
\int d^4 x e^{ipx}
\langle 0|T\left[ \mathcal{O}_{\rm 2N^c}^{\dagger}(x) \mathcal{O}_{\rm 2N^c}(0)\right] |0 \rangle =
-\frac{A_{\Delta_{\rm 2N^c}}}{2 i\, {\rm sin}
\left( \Delta_{\rm 2N^c} \pi \right) }
\frac{1}
{\left(-p^2 - i \epsilon\right)^{2 - \Delta_{\rm 2N^c}} }~,
\label{georgi-norm}
 \end{equation}
The absorptive (imaginary) part of Eq.~(\ref{georgi-norm}) simplifies
to~\cite{Kenzie-thesis},
 \begin{equation}
\left.
\int d^4 x e^{ipx}
\langle 0|T\left[ \mathcal{O}_{\rm 2N^c}^{\dagger}(x) \mathcal{O}_{\rm 2N^c}(0)\right] |0 \rangle\right|_{\mathrm{abs.}} =
 \frac{1}{2} \frac{A_{\Delta_{\rm 2N^c}}}
{\left(p^2\right)^{2 - \Delta_{\rm 2N^c}} } \theta(p^2)~.
\label{georgi-norm-Im}
 \end{equation} 
 The operator normalization in Eq.~(\ref{georgi-norm}) is only valid for 
the range of scaling dimensions $1 \leq \Delta_{\rm 2N^c} < 2$. For $ 
\Delta_{\rm 2N^c} \geq 2$, the left-hand side of Eq.~(\ref{georgi-norm}) 
diverges in the ultraviolet, and we can no longer employ this 
normalization. However, the absorptive part of Eq.~(\ref{georgi-norm}), 
Eq.~(\ref{georgi-norm-Im}), is ultraviolet safe. We therefore use 
Eq.~(\ref{georgi-norm-Im}) to normalize the operator $\mathcal{O}_{\rm 
2N^c}$ over the entire range of scaling dimensions $\Delta_{\rm 2N^c} > 
1$. Assuming this deformation carries a lepton number of $(-2)$, at 
scales of order $\Lambda$ the Lagrangian contains a lepton number 
violating term of the form,
 \begin{equation}
\mathcal{L}_{\rm IR} \supset \frac{\mu^c}{2} \left(N^c\right)^2 + {\rm h.c.}
\label{massN2}
 \end{equation} 
 The mass parameter $\mu^c$ is related to the parameters 
in the ultraviolet theory as 
 \begin{equation} 
\mu^c \sim C_\mu \hat{\mu}^c\, \Lambda \left(\frac{\Lambda}{M_{\mathrm{UV}}} 
\right)^{\Delta_{\rm 2N^c} - 4} ~, 
\label{muscaling} 
 \end{equation} 
where the multiplicative factor $C_\mu$ is given by, 
 \bea 
C_\mu &=& \frac{(4\pi)^{2-\Delta_{\rm 2N^c}}}{\Gamma(\Delta_{\rm 2N^c}-1)} 
\sqrt{\frac{1}{\Delta_{\rm 2N^c}-1}}~. 
 \eea 
 We see from Eqs.~(\ref{freeN}),(\ref{intLHN}) and (\ref{massN2}) 
that the low energy Lagrangian contains all the ingredients necessary to 
realize the inverse seesaw, 
 \begin{equation} 
 \label{L_IR} 
\mathcal{L}_{\rm IR} \supset i\bar{N} \bar{\sigma}^\mu \partial_{\mu} N 
+ i\bar{N}^c \bar{\sigma}^\mu \partial_{\mu} N^c - \left[M_N N^c N + 
\lambda L H N + \frac{\mu^c}{2} \left(N^c\right)^2 + {\rm h.c.} 
\right]~. 
 \end{equation} 
 Integrating out $N$ and $N^c$ we get a 
contribution to the masses of the light neutrinos from the inverse 
seesaw of order, 
 \begin{equation} 
{m_\nu}\mid_{{\rm inv. seesaw}} = 
\mu^c \left(\frac{\lambda v_{\rm EW}}{M_N}\right)^2 ~, 
 \end{equation} 
 with $v_{\rm EW} \approx 174\gev$.  We expect comparable but slightly 
smaller contributions to the neutrino masses from integrating out the 
higher mass singlet fermion resonances. Therefore our final expression 
for the masses of the light neutrinos takes the form, 
 \begin{equation} 
\label{eq:numass} 
 {m_\nu} \sim \mu^c \left(\frac{\lambda v_{\rm EW}}{M_N}\right)^2 
\sim \Lambda \left[C_\mu \hat{\mu}^c 
\left(\frac{\Lambda}{M_{\mathrm{UV}}}\right)^{\Delta_{2\mathrm{N^c}}-4} 
\right] 
\left[C_\lambda \hat{\lambda} \left(\frac{v_{\mathrm{EW}}}{\Lambda}\right) 
\left(\frac{\Lambda}{M_{\mathrm{UV}}}\right)^{\Delta_{\mathrm{N}}-3/2} 
\right]^2 \;. 
 \end{equation} 
 The first square bracket in \eref{eq:numass} controls the strength of 
lepton number violation in the composite sector. The second square 
bracket controls the degree to which the elementary SM neutrinos mix 
with the composite singlet neutrinos. Both of these effects must be 
present to generate Majorana neutrino masses. The lightness of the 
neutrino may be explained in this framework either by the approximate 
lepton number symmetry in the composite sector or by the smallness of 
elementary-composite mixing, with a continuum of possibilities in 
between. We see from \eref{eq:numass} that the sizes of these effects 
depend on the scaling dimensions of the operators $\mathcal{O}_{\rm N}$ 
and $\mathcal{O}_{\rm 2N^c}$. Therefore, in this framework, the scaling 
dimensions of CFT operators can provide a simple and natural explanation 
for the smallness of neutrino masses.

In obtaining \eref{eq:numass}, we have assumed that the
contributions to the neutrino mass from scales of order $\Lambda$
dominate over the contributions from higher scales. In general, in the
presence of the deformations Eqs.~(\ref{intLHO}) and
(\ref{Nc2deformation}), we expect an ultraviolet contribution to the
Weinberg operator $(LH)^2$. The size of this effect is controlled by the
CFT three-point function $\langle \mathcal{O}_{\rm N}(x)
\mathcal{O}_{\rm N} (y) \mathcal{O}_{\rm 2N^c}(z) \rangle$. From
dimensional considerations we see that the condition that there not be a
contribution to $(LH)^2$ that diverges in the ultraviolet translates
into a bound on the scaling dimensions of the operators
$\mathcal{O}_{\rm N}$ and $\mathcal{O}_{\rm 2N^c}$, $2\Delta_{\rm N}
+\Delta_{\rm 2N^c} \le 8$. We assume that this bound is satisfied, so
that the dominant contribution to the neutrino masses arises from scales
of order $\Lambda$, and is given by \eref{eq:numass}.

Another restriction on the parameter space arises from the condition
that the Majorana mass for the singlet neutrinos arising from
\eref{massN2} cannot be larger than the compositeness scale, $\mu^c
\lesssim \Lambda$. This translates into a lower bound on the coupling
$\lambda$. There is also an upper bound on $\lambda$ from the condition
that the Dirac mass for the neutrinos obtained from \eref{intLHN} be
less than the compositeness scale. From this it follows that $\lambda$
must lie in the range
\begin{equation}
\label{lambdabound}
\frac{M_N}{v_{\rm EW}} \gtrsim \lambda \gtrsim \frac{\sqrt{m_\nu M_N}}{v_{\rm EW}}\;.
\end{equation}
In the limit when $\lambda$ sits at its lower bound in
\eref{lambdabound}, the operator $\mathcal{O}_{\rm 2N^c}$ may be playing
the role of the operator $\mathcal{O}_{\rm S}$ that triggers the
breaking of conformal symmetry.

In general the deformation $\mathcal{O}_{\rm 2N^c}$ will also generate a
lepton number violating mass term for $N$ at at the scale $\Lambda$,
 \begin{equation}
\mathcal{L}_{\rm IR} \supset \frac{\mu}{2} N^2 + {\rm h.c.}
\label{othermassN2}
 \end{equation}
 The parameter $\mu$ is expected to be of order $\mu^c$. This term, 
although very similar in form to the lepton number violating mass term 
for $N^c$ shown in \eref{massN2}, does not contribute significantly to 
the neutrino masses.

The Lagrangian at the scale $\Lambda$ is also expected to contain
four-fermion interactions between the $N$'s. These take the schematic
form,
 \begin{equation}
\mathcal{L}_{\rm IR} \supset - \kappa \frac{\left({\bar{N}\sigma^\mu N}\right)^2}{\Lambda^2} + \kappa^\prime \frac{\left({N^c N}\right)^2}{\Lambda^2} + \ldots \; ,
\label{int4N}
 \end{equation}
 where we have shown two such terms. These interactions respect the 
overall lepton number symmetry. The parameters $\kappa$ and 
$\kappa^\prime$ are expected to be of order $(4 \pi)^2$. These 
nonrenormalizable terms are characteristic of the composite nature of 
the singlet neutrinos. Although not required to realize the inverse 
seesaw, they play an important role in the phenomenology of this class 
of models, particularly in the context of astrophysics and cosmology.

To generate the mass splittings necessary for neutrino oscillations 
there must be at least two pairs of $N,\, N^c$ and in general there may 
be more. Reinstating the flavor index that we have so far suppressed, 
the singlet neutrinos come in pairs $N_\alpha$ and $N^c_\alpha$ with 
$\alpha=1,\,2,\ldots ,\, \numN$, with a corresponding number of 
operators $\mathcal{O}_{\rm N}^\alpha$ in the CFT. Including this flavor 
index the couplings between the SM lepton doublets and the CFT operators 
are now given by,
 \begin{equation}
\mathcal{L}_{\rm UV} \supset \frac{\hat{\lambda}_{i\alpha}}{M_{\mathrm{UV}}^{\Delta_{\mathrm{N}} - 3/2}} L^i H \mathcal{O}_{\rm N}^\alpha + {\rm h.c.}~.
\label{eq:intLHOwithflavour}
 \end{equation}
 The flavor index on the lepton doublets runs over the SM generations, 
$i=1,2,3$. We work in the basis in which the charged lepton masses are 
flavor diagonal.

Keeping track of flavor indices, the low energy Lagrangian in 
Eq.~(\ref{L_IR}) now takes the form,
 \begin{equation}
\label{eq:L_IRwithflavour}
\mathcal{L}_{\rm IR} \supset
i\bar{N}_\alpha \bar{\sigma}^\mu \partial_{\mu} N_\alpha + i\bar{N}^c_\alpha \bar{\sigma}^\mu \partial_{\mu} N^c_\alpha -
\left[\left(M_N\right)_{\alpha\beta} N^c_\alpha N_\beta + \lambda_{i\alpha} L^i H N^\alpha + \frac{\left(\mu^c\right)_{\alpha\beta}}{2} N^c_\alpha N^c_\beta + {\rm h.c.} \right]
 \end{equation}
 The $N_\alpha, N^c_\alpha$ may be rotated to make $M_N$ diagonal. In 
this basis, the SM neutrinos will, in general, couple to all the 
$N_\alpha$. We therefore expect that there will be both flavor 
preserving and flavor violating processes involving the heavy neutrinos. 
The mixing between the neutrinos is conventionally parametrized in terms 
of a mixing matrix $U$. In the limit of small mixing between the active 
and the composite singlet neutrinos, the corresponding elements of the 
mixing matrix are related to the parameters in \eref{eq:L_IRwithflavour} 
as,
 \be
U_{N_\alpha\ell_i} = \frac{\lambda_{i\alpha}\,v_{\mathrm{EW}}}{M_{N_\alpha}}~.
\label{eq:Ulambdarelation}
 \ee
 In our study of the phenomenology of this class of models, we will make 
certain simplifying assumptions. We assume that there is a discrete 
symmetry in the CFT that relates the different $\mathcal{O}_{\rm 
N}^\alpha$ operators so that they all have the same scaling dimension, 
$\Delta_{\rm N}$. We further assume that this symmetry is preserved in 
the breaking of conformal invariance by the operator $\mathcal{O}_{\rm 
S}$, so that all the composite singlet neutrinos are approximately 
degenerate. For concreteness, we assume that the $N_\alpha$ and 
$N_\alpha^c$ are the lightest states in the hidden sector, so that their 
only kinematically allowed decay modes are to SM particles. In order to 
relate measurements performed at energies above the compositeness scale 
to the parameters in the low energy theory, we will take 
Eqs.~(\ref{lambdascaling}), (\ref{muscaling}) and (\ref{eq:numass}) to 
be exact equalities rather than estimates. Then, although the bounds and 
projections we obtain are only accurate up to $\mathcal{O}(1)$ factors, 
this should suffice to give a sense of the current constraints on this 
class of models and the expected reach of future experiments.

In Table~\ref{tab:benchmarks}, we present a few benchmark sets of model 
parameters that lead to neutrino masses in the right range. We will 
refer to the benchmarks as I-V in the order that they appear in 
the table. All the benchmark models have compositeness scales that lie 
well below the electroweak scale. The benchmarks fall into two broad 
categories that provide different explanations for the smallness of 
neutrino masses. In the first set of benchmarks, the parameters 
$\hat{\lambda}$ and $\hat{\mu}^c$ are order one at $M_{\rm UV}$. 
The smallness of neutrino masses is explained by the scaling dimensions 
of the operators $\mathcal{O}_{\rm N}$ and $\mathcal{O}_{\rm 2N^c}$, 
which control the extent to which the SM neutrinos mix with their 
composite counterparts and the extent of overall lepton number violation 
respectively. In the second set of benchmarks, the smallness of neutrino 
masses is explained, in part, by approximate global symmetries of the 
theory in the ultraviolet. There are two symmetries that can play a role 
here. In the limit that $\hat{\mu^c}$ vanishes, the theory has an exact 
lepton number symmetry. In the limit that $\hat{\lambda}$ vanishes, the 
SM decouples from the hidden sector and the lepton number symmetry of 
the SM is restored. Hence approximate symmetries can explain why either 
$\hat{\mu^c}$ or $\hat{\lambda}$ is small and thereby provide an 
explanation for the smallness of neutrino masses.

 \begin{table}[t]
   \centering
   \resizebox{\columnwidth}{!}{
   \begin{tabular}{@{} c|cccccc|ccc@{}} 
   \hline\hline
    & $\hat{\lambda}$ & $\hat{\mu}^c$  & $\Delta_{\mathrm N}$ & $\Delta_{\rm 2N^c}$ & $\Lambda\,[\MeV]$ & $M_{\mathrm{UV}}\, [\tev]$ & $\lambda$ & $\mu^c\,[\MeV]$ & $m_\nu\, [\ev]$ \\
   \hline
   I & $1$ & $1$ & $1.85$ & $3.9$ & $400$ & $M_{\rm Pl}$ & $8\times 10^{-8}$ & $80$ & $0.1$ \\
  II & $1$ & $2$ & $1.8$ & $4.05$ & $40$ & $M_{\rm Pl}$ & $4\times 10^{-7}$ & $10^{-2}$ & $0.045$ \\ \hline
  III &  $5\times 10^{-5}$ & $1$ & $1.9$ & $3.8$ & $40$ & $2\times 10^3$ & $2\times 10^{-8}$ & $7$ & $0.05$ \\
  IV & $1$ & $4\times10^{-14}$ & $2.4$ & $2.25$ & $400$ & $2\times 10^3$ & $3\times 10^{-7}$ & $5$ & $0.08$ \\ 
 V & $10^{-2}$ & $5\times10^{-3}$ & $2.25$ & $3.4$ & $400$ & $2\times 10^3$ & $3\times 10^{-8}$ & $300$ & $0.05$ \\
   \hline\hline
   \end{tabular}
   }
   \caption{Some representative choices of parameters that lead to 
neutrino masses of parametrically the right size. Shown are the 
dimensionless couplings $\hat{\lambda}$ and $\hat{\mu}^c$, the scaling 
dimensions $\Delta_{\mathrm N}$ and $\Delta_{\rm 2N^c}$, the 
compositeness scale $\Lambda$, and the cutoff scale $M_{\mathrm{UV}}$.  
Also shown are some derived parameters, in particular the coupling 
constant $\lambda$, the Majorana mass term for the composite singlet 
neutrinos $\mu^c$, and the resulting mass scale of the light neutrinos. 
For each set of parameters all the constraints outlined in the paper are 
satisfied. In the upper set of benchmarks, the smallness of the neutrino 
masses is fully explained by the running of parameters from the Planck 
scale cutoff down to the infrared. In the lower set of benchmarks, the 
ultraviolet cutoff is lowered to 2~PeV, and the light masses of the 
neutrinos are explained, at least in part, by symmetries.
}
 \label{tab:benchmarks}
 \end{table}

In the first set of benchmarks the cutoff $M_\mathrm{UV}$ is taken to 
very high, of order the Planck scale, $M_{\rm Pl} = 1.2 \times 10^{19}$ 
GeV. The dimensionless couplings $\hat \lambda$ and $\hat \mu$ are of 
order one, with the small parameters necessary to realize the scale 
seesaw mechanism at low scales arising from renormalization group 
evolution from the Planck scale down to the compositeness scale 
$\Lambda$. In this scenario, even small modifications to the scaling 
dimensions of the operators $\mathcal{O}_{\rm N}$ and $\mathcal{O}_{\rm 
2N^c}$ can lead to large changes in the values of the infrared 
parameters. It is instructive to track the size of overall lepton number 
violation, which is set by $\hat \mu^c$ in the ultraviolet and by 
$\mu^c/\Lambda$ at the compositeness scale, and the size of SM lepton 
number violation, which is set by $\hat{\lambda}$ in the ultraviolet and 
by $\lambda v_{\rm EW}/M_N$ at the compositeness scale. Within this 
class of models, both these symmetries are violated by order one at the 
Planck scale, in line with the expectation that quantum gravity violates 
all global symmetries. The tight experimental constraints on mixing 
between SM neutrinos and singlet neutrinos imply that SM lepton number 
must necessarily be a symmetry to very good accuracy at low energies. 
This tends to naturally occur in our framework since the interaction $LH 
\mathcal{O}_{\rm N}$ that generates this mixing is always irrelevant. 
For high values of $M_{\rm UV}$, even if $\hat{\lambda}$ is of order one 
in the ultraviolet, the mixing can easily be sufficiently suppressed so 
as to result in suitably small masses for the light neutrinos. This 
remains true even if $\mathcal{O}_{\rm 2N^c}$ is a (slightly) relevant 
operator, so that overall lepton number is violated by order one in the 
infrared, as illustrated in Benchmark I in the table. However, if 
$\mathcal{O}_{\rm 2N^c}$ is instead a (slightly) irrelevant operator, 
overall lepton number can also emerge as an accidental symmetry at the 
compositeness scale, in which case it will also play a role in the 
suppression of neutrino masses. This is illustrated in Benchmark II in 
the table.

In the second set of benchmarks we take $M_\mathrm{UV}$ to be low, of 
order the flavor scale, $2 \times 10^{3}$ TeV. There is now less room 
for renormalization group evolution to generate the small parameters 
necessary to accommodate a low scale seesaw model. Therefore, in this 
scenario, symmetries are also expected to play a role in suppressing the 
neutrino masses. Neutrino masses in the right range can be obtained if 
$\hat{\lambda}$ is very small at $M_\mathrm{UV}$, so that SM lepton 
number is a symmetry to very good accuracy in the ultraviolet. This can 
happen even if overall lepton number is violated by order one in the 
infrared, as illustrated in Benchmark III in Table~\ref{tab:benchmarks}. 
Alternatively, a realistic spectrum of neutrino masses can be obtained 
if overall lepton number is a symmetry to very good accuracy at 
$M_\mathrm{UV}$, even if SM lepton number is violated by order one at 
that scale. This is illustrated in Benchmark IV. It is also possible 
that both $\hat{\mu}^c$ and $\hat{\lambda}$ are small, so that both 
overall lepton number and SM lepton number are only mildly violated at 
the cutoff scale, as in Benchmark~V.

It is important to note that the symmetries at $M_\mathrm{UV}$ in the 
low scale models could naturally emerge from dynamical effects. For 
example, one could easily imagine a different CFT above $M_\mathrm{UV}$ 
that serves as an ultraviolet completion for the low scale models, and 
for which the scaling dimensions are such that these symmetries emerge 
at low energies. This CFT could be broken at $M_\mathrm{UV}$ resulting 
in low scale theories with the features we present in the table, in 
which SM lepton number, hidden sector lepton number, or both, are 
approximate symmetries.

In summary, our framework naturally allows for the generation of the 
small couplings required to realize the inverse seesaw mechanism at low 
scales, and a wide range of low energy parameters are possible. In the 
following sections we will connect the parameter space to specific 
experimental observations.

\section{Colliders}
\label{sec:collider}

In this section we discuss the phenomenology of this class of models at 
colliders. We focus on the case in which the composite singlet neutrinos 
have masses at or below the weak scale. We begin with a concrete and 
detailed calculation of the decay rate of $W$ and $Z$ gauge bosons into 
composite singlet neutrinos. We then translate the existing constraints 
from searches for elementary heavy neutral leptons (HNLs) to determine 
the range of parameter space that is currently allowed by the data. We 
also determine the reach of similar searches in the future. Finally, we 
sketch the novel collider signatures of this class of models, which 
include $W$ decays that give rise to multiple displaced vertices, and 
discuss the prospects of the HL-LHC for these signals.

\subsection{$W$ and $Z$ decays}

The hidden sector interacts with the SM through the neutrino portal 
interaction, \eref{intLHO}. Hidden sector states can therefore be 
produced in decays of the $W^\pm$ and $Z$ bosons through diagrams 
involving an off-shell neutrino. After being produced, the hidden sector 
hadronizes into final states consisting of one or more composite singlet 
neutrinos. In much of parameter space the composite singlet neutrinos 
are unstable on detector timescales and so the search is for their decay 
products.

\begin{figure}[t]
  \centering
    \includegraphics[width=0.3\textwidth]{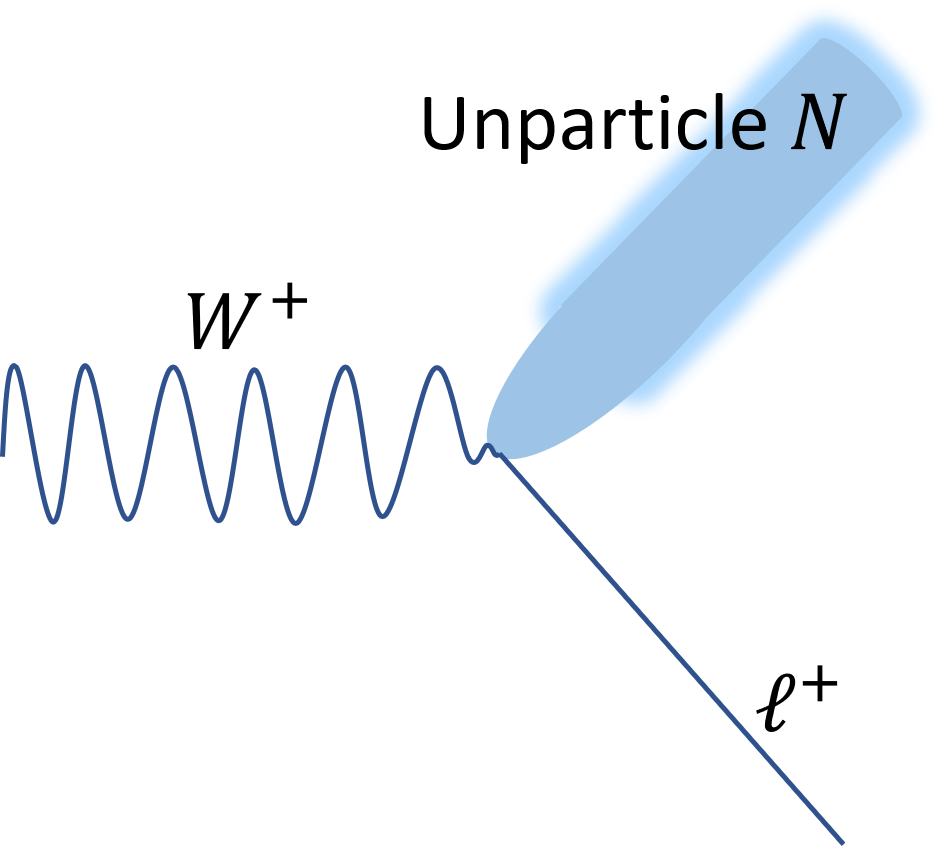}
    \caption{Unparticle production from W boson decays.}
    \label{fig:Feyn}
  \end{figure}

We first consider the inclusive decay of a $W$ boson into a single 
flavor of lepton and unparticles, $W^\pm\rightarrow \ell^\pm + 
\mathcal{U}$, as shown in \figref{fig:Feyn}. The leading order matrix 
element contains a single insertion of the interaction \eref{intLHO},
 \begin{equation}
i\mathcal{M} = 
\frac{ig}{2 \sqrt{2}} 
\bar{u}(p_N) 
\left(\frac{-i \hat{\lambda} v_{\rm EW}}{M_{\mathrm{UV}}^{\Delta_{\mathrm N} - 3/2}} \right)
\frac{i \slashed{p}_N}{p_N^2} 
\gamma^\mu \left( 1 - \gamma^5 \right) 
v(p_\ell) 
\epsilon_\mu(p_W)~.
\label{eq:Wdecaymatrixelementsq}
 \end{equation}
 Here $g$ is the $SU(2)_{\rm L}$ gauge coupling while $p_W$ represents 
the four-momentum of the decaying $W$ boson, $p_\ell$ the four-momentum 
of the charged lepton, and $p_N = p_W - p_\ell$ the momentum carried 
away by the unparticles. In the rest frame of the decaying $W$ the spin 
averaged matrix element, after summing over the spins of the final 
state particles, is given by,
 \begin{equation}
\frac{1}{3} \sum |\mathcal{M}|^2 =  
\frac{g^2}{3} \frac{\hat{\lambda}^2 v_{\rm EW}^2}{M_{\mathrm{UV}}^{2\Delta_{\mathrm N} - 3}}
\frac{E_\ell}{m_W} \frac{3 m_W - 2 E_\ell}{m_W - 2 E_\ell}~.
\label{eq:WdecayM2}
 \end{equation}
 Here $E_\ell$ represents the energy of final state charged lepton. The 
partial decay width is determined by integrating over the phase space 
densities of the charged lepton and the unparticles,
 \be
d\Gamma = \frac{|\mathcal{M}|^2}{2m_W}(2\pi)^4\delta^{(4)}(p_W-p_\ell-p_N) d \Phi_\ell d \Phi_{\mathcal{O}}~. 
 \ee
 The phase spaces of the charged lepton and the unparticles are given by
 \be 
d \Phi_\ell = \frac{1}{\left(2 \pi \right)^3} \frac{1}{2 E_\ell} d^3 p_\ell~, \quad
d \Phi_{\mathcal{O}} = \frac{A_{\Delta_{\mathrm N} - 1/2}}{\left(2 \pi \right)^4} 
\theta(p_N^0) \theta(p_N^2-\mu_{IR}^2) 
\left(p_N^2-\mu_{IR}^2 \right)^{\Delta_{\mathrm N} - 5/2} d^4 p_N \;.
 \ee
 The expression for $A_n$, the phase space volume factor, was given in 
\eref{eq:PhaseSpaceVol}.  In the expression for the unparticle phase 
space we have introduced an infrared cutoff, $\mu_{IR}$, to account for 
the fact that conformal invariance is broken at the compositeness scale. 
Performing the integration over the angular variables we obtain an 
expression for the energy distribution of the final state lepton,
 \beq
\frac{d \Gamma}{d E_\ell}=\frac {g^2} {96\pi^2} \frac {|\hat \lambda v_{\rm EW}|^2} {m_W^2} 
\!\left(\!\frac {m_W}{M_{\mathrm{UV}}}\!\right)^{\negspxhalf 2 \Delta_{\mathrm N} - 3}
\!\!\!\!\!\!\!\!\!\!
A_{\Delta_{\mathrm N}-1/2}\left(\frac {\!3m_W-2E_\ell} {m_W-2E_\ell}\right)\!\!\left(\frac {2 E_\ell} {m_W}\!\right)^{\!\!\!2}
\!\!\!
\left (\!1-\frac {2E_\ell} {m_W}-\frac {\mu_{IR}^2} {m_W^2}\!\right)^{\!\!\!\Delta_{\mathrm N}-5/2}
\label{eq:Wdiff}
 \eeq
 For $\Delta_{\mathrm N} < 5/2$ the integral over $E_\ell$ for the total width is 
infrared divergent, with the divergence arising from the neighborhood of 
$E_\ell = m_W/2$. Physically, this divergence is regulated by the fact 
that the conformal symmetry is broken and so the final state particles 
in the unparticle sector, the composite singlet neutrinos, are not 
massless but have masses of order the compositeness scale $\Lambda$. 
Therefore the point $E_\ell = m_W/2$ can never actually be reached. By 
choosing a nonzero value of $\mu_{IR}^2$ we regulate this divergence. In 
practice, there is an additional cutoff arising from the requirement 
that the decay products of $N$ be sufficiently energetic to be observed 
in the collider environment.  The partial width is then given by,
 \bea
\Gamma(W\rightarrow \ell\,\mathcal{U}) &&= m_W\frac {|\hat \lambda|^2} {48\pi^2} \left(\frac {m_W} {M_{\mathrm{UV}}} \right)^{2\Delta_{\mathrm N}-3} A_{\Delta_{\mathrm N}-1/2}
f(\Delta_{\mathrm N},\frac {\mu_{IR}^2} {m_W^2}).
\label{eq:Wwidth}
 \eea
 The function $f(\Delta_{\rm N},{\mu_{IR}^2}/{m_W^2})$ captures the 
dependence on the scaling dimension $\Delta_{\rm N}$ and $\mu_{IR}$. Its 
detailed form is given in \eref{eq:fnew} of the appendix. After 
rewriting the $W$-partial width as a function of the infrared 
parameters, it can be expressed in terms of the conventional HNL mixing angle using 
Eqs.~(\ref{lambdascaling}) and (\ref{eq:Ulambdarelation}),
 \bea
\Gamma(W\rightarrow \ell\,\mathcal{U}) 
&&= \sum_{\alpha=1}^{\numN} m_W \frac {g^2 |U_{N{_\alpha}\ell}|^2} {96\pi^2 C_\lambda^2} \left(\frac {M_{N}} {m_W}\right)^2 \left(\frac {m_W} {\Lambda} \right)^{2\Delta_{\mathrm N}-3} \! \! A_{\Delta_{\mathrm N}-1/2}
f(\Delta_{\mathrm N},\frac {\mu_{IR}^2} {m_W^2}).  
\label{eq:Wwidth}
 \eea
 Since the final result is sensitive to the value of $\mu^2_{IR}$, this 
expression must be considered as just an estimate. As a consistency 
check we consider the limit that $\Delta_{\mathrm N}\rightarrow 3/2$,
 \beq
\lim_{\Delta_{\mathrm N}\rightarrow3/2}A_{\Delta_{\mathrm N}-1/2}f(\Delta_{\mathrm N},\mu_{IR})=\frac{2 \pi  \left(\mu_{IR}^6-3 \mu_{IR}^2 m_W^4 +2m_W^6\right)}{\mu_{IR}^2 m_W^4 }.
 \eeq
 In this limit, $C_\lambda \rightarrow 1$. Then, setting $\mu_{IR} = 
m_N$, \eref{eq:Wwidth} reproduces the standard result for an elementary 
HNL, as expected. Going forward, we shall set set $\mu_{IR} 
= m_N$ in this section when presenting results.

\begin{figure}[t]
  \centering
  \includegraphics[width=0.6\textwidth]{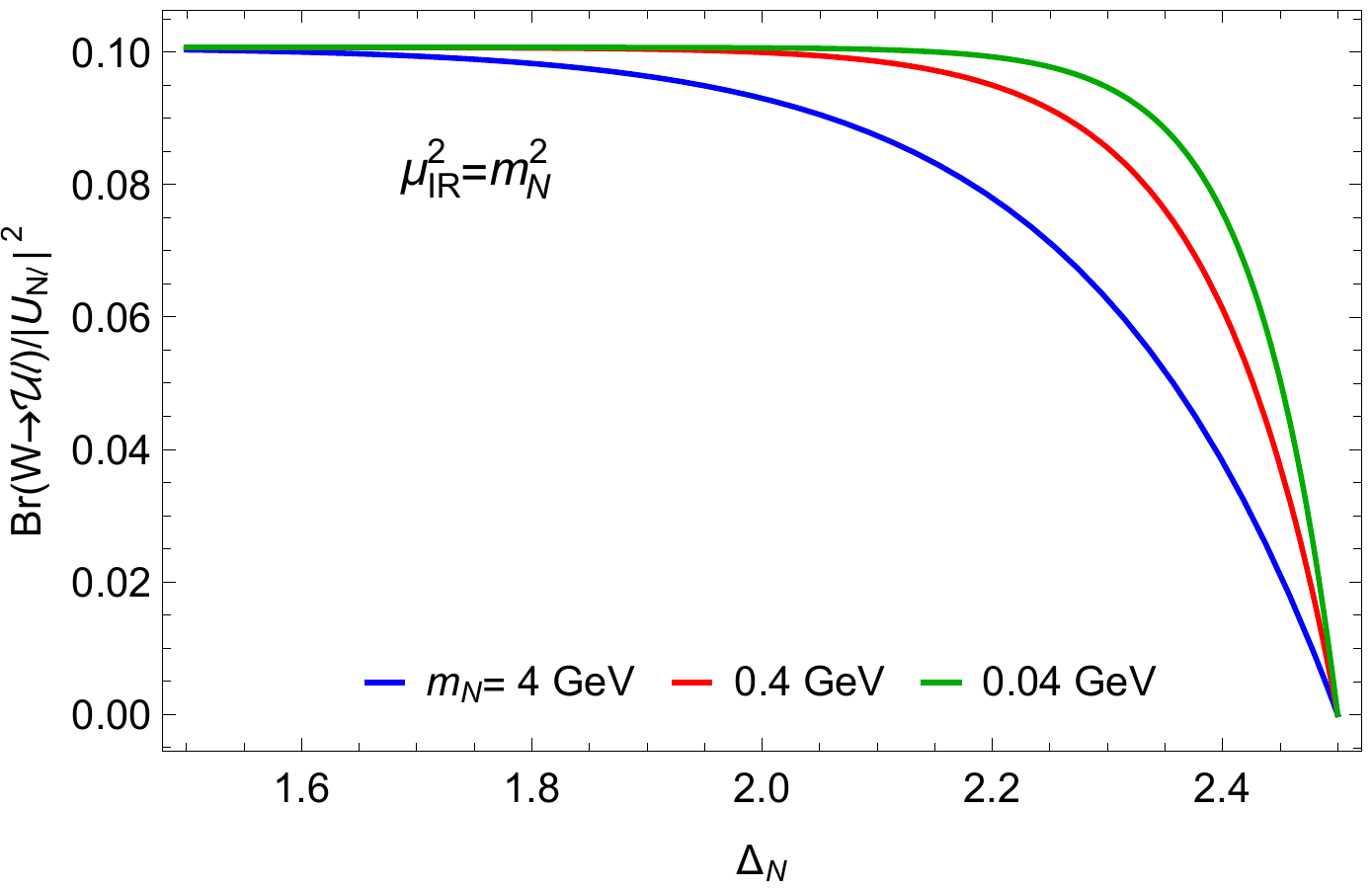}
  \caption{The branching fraction of $W\to \ell\, \mathcal{U}$, normalized by the IR mixing angle $\sum_\alpha |U_{N_\alpha\ell}|^2$, for benchmark choices of $m_N$, as a function of the scaling dimension $\Delta_{\mathrm N}$. 
  }
  \label{fig:Wwidth2}
  \end{figure}

\begin{figure}[t]
  \centering
  \includegraphics[width=0.49\textwidth]{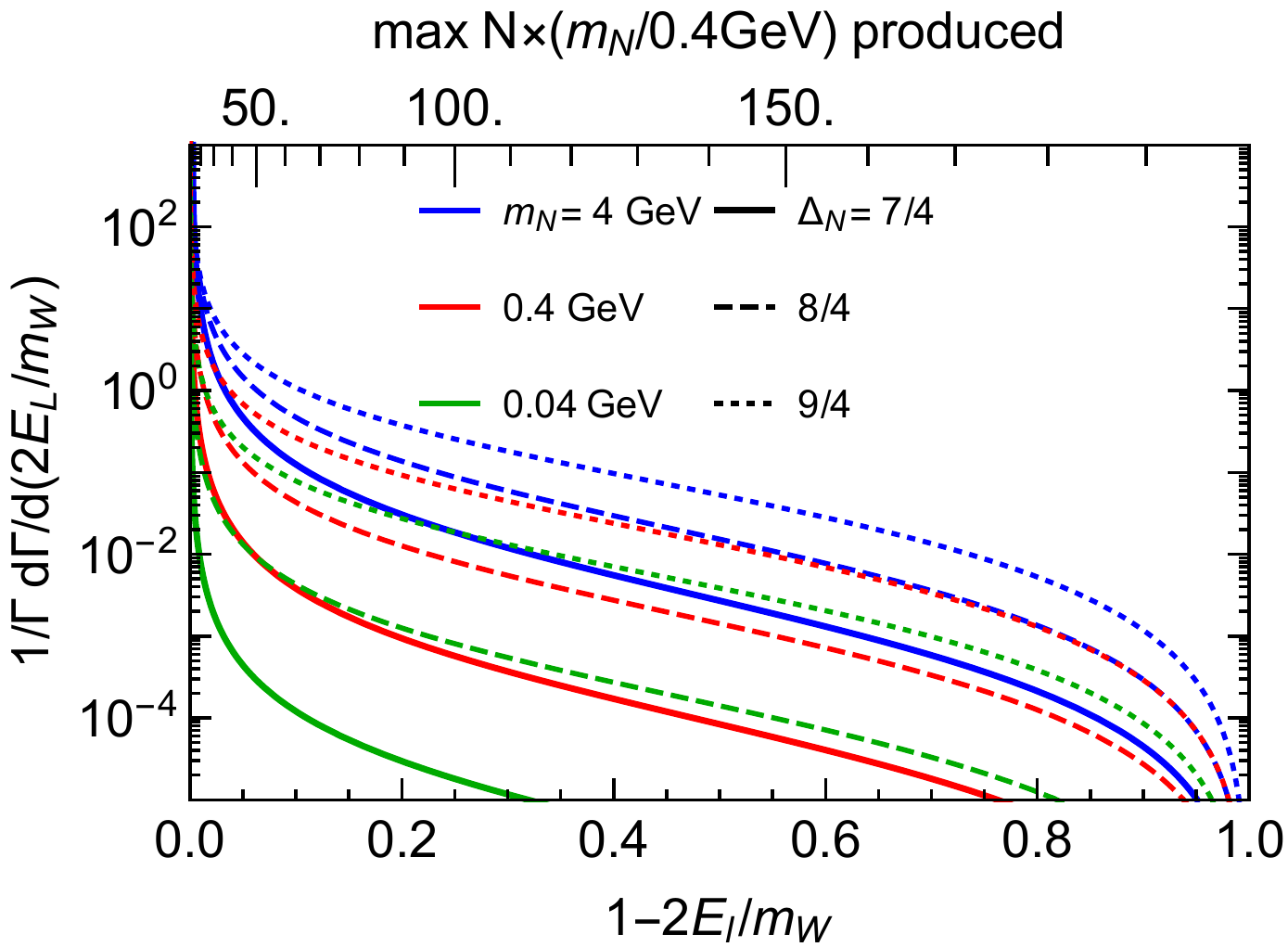}
  \includegraphics[width=0.49\textwidth]{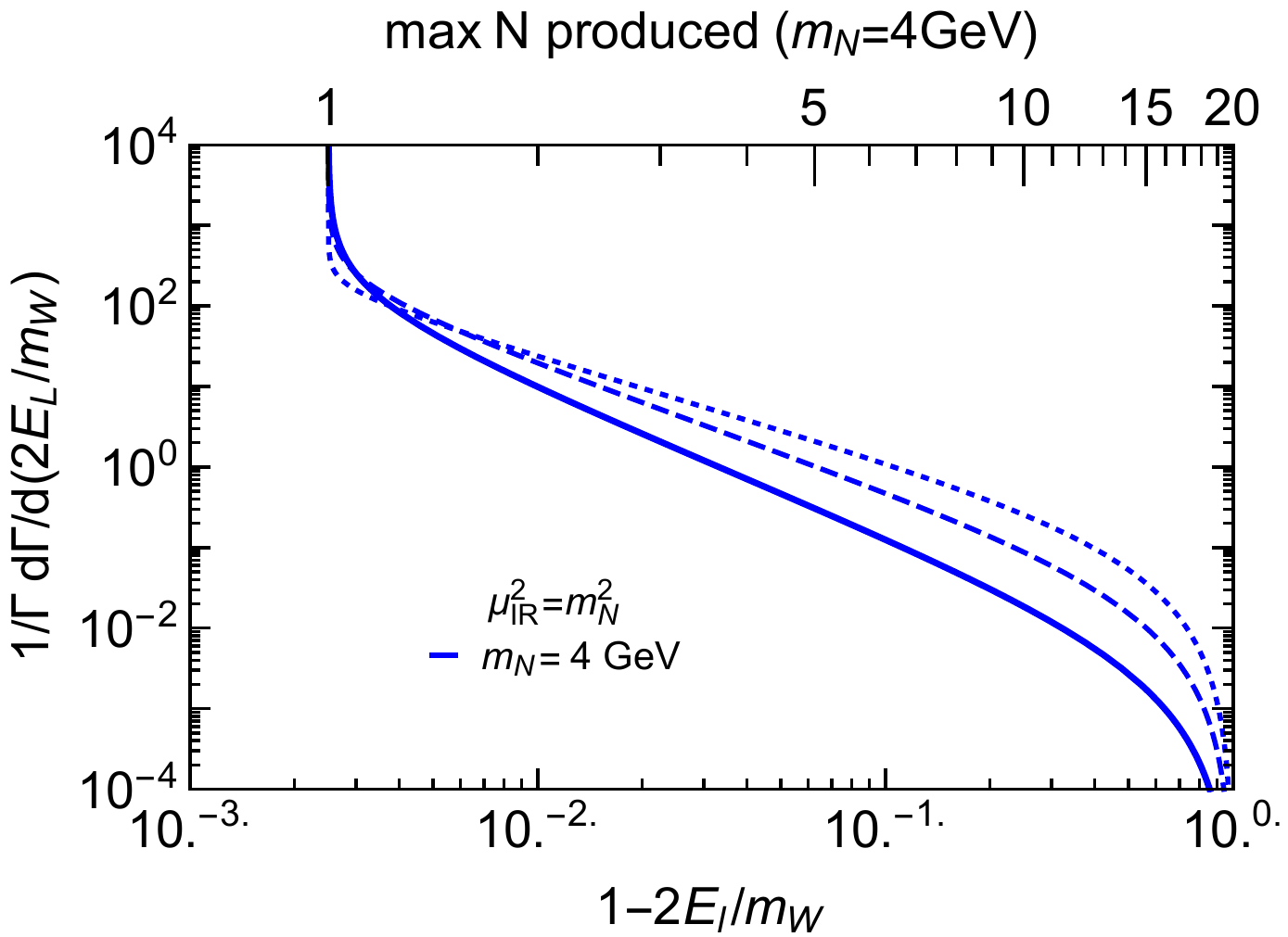}
  \includegraphics[width=0.49\textwidth]{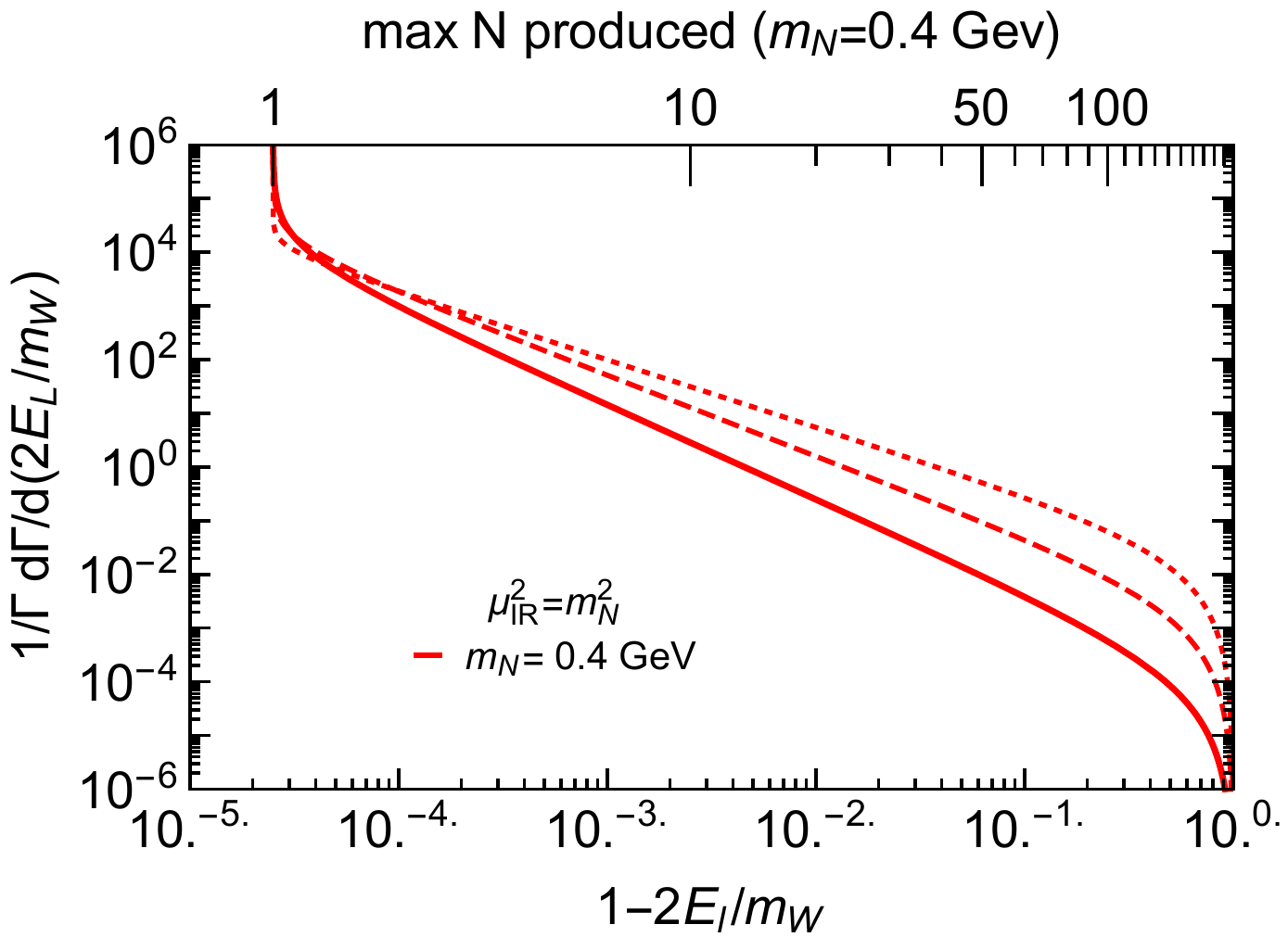}
  \includegraphics[width=0.49\textwidth]{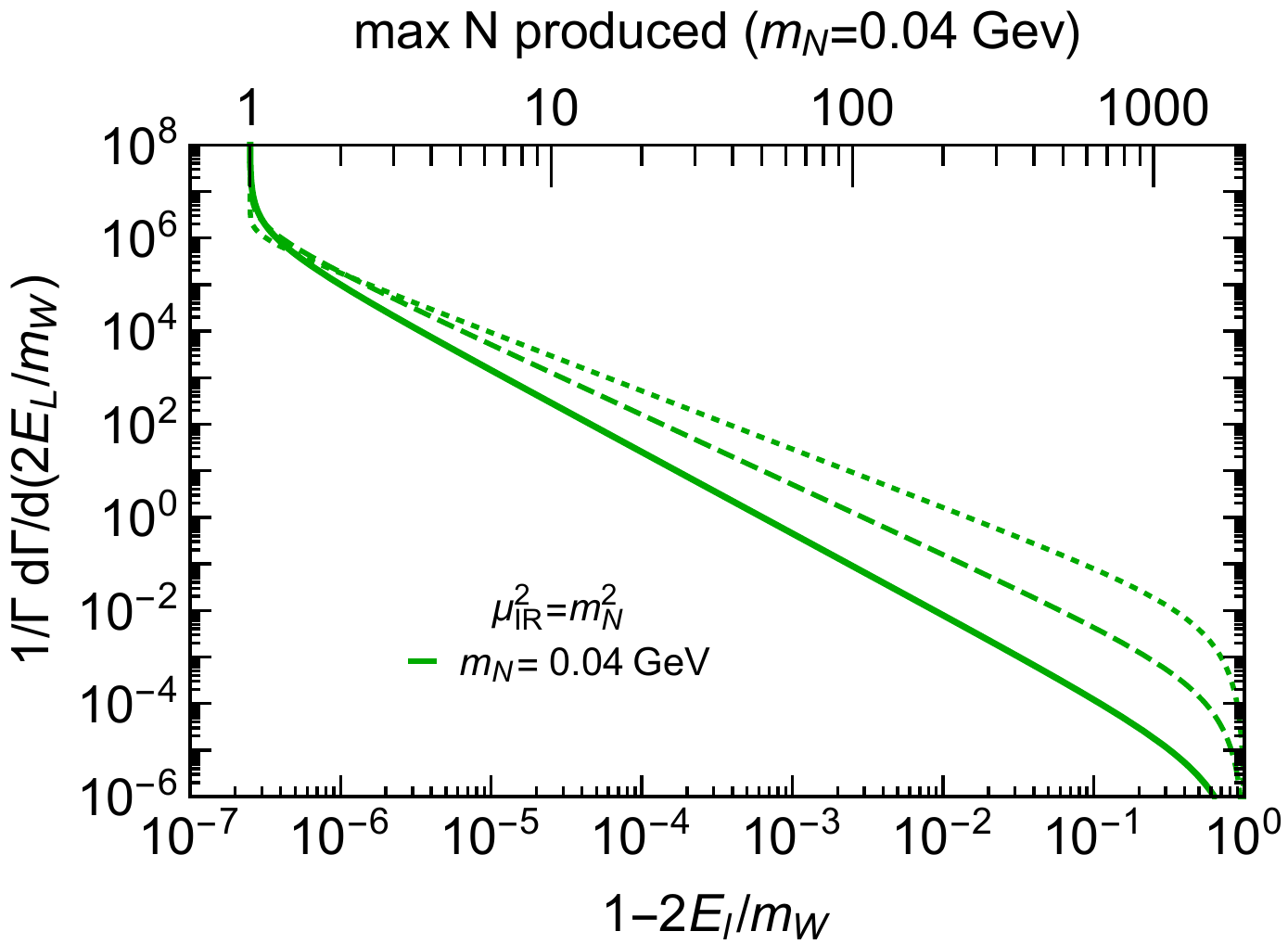}
  \caption{The differential width of $W\to \ell\, \mathcal{U}$ as a function of the charged lepton energy $E_\ell$ for various benchmark choices of $m_N$ and scaling dimension $\Delta_{\mathrm N}$. 
The solid, dashed, and dotted lines represent benchmark scaling dimension of $\Delta_{\mathrm N}$ of 7/4, 2, and 9/4, respectively. 
The green, red, and blue curves corresponds to singlet neutrino masses of 0.04, 0.4, and 4~GeV, respectively. 
  On the upper axis of the figure we also show the  maximum number of singlet neutrinos allowed by kinematics. 
 The upper left panel shows the overall distribution in log-linear scale, while the rest of panels zoom in to near the maximum $E_\ell$ value and show the distribution in log-log scale.
   }
  \label{fig:Wwidth1}
  \end{figure}

In Fig.~\ref{fig:Wwidth2} we show the branching fraction of $W\to \ell\, 
\mathcal{U}$, normalized by the infrared mixing angle $\sum 
|U_{N_\alpha\ell}|^2$, as a function of the scaling dimension $\Delta_{\mathrm N}$ 
for various benchmark choices of $M_N$. This branching fraction is 
sensitive to the choice of the infrared cutoff. The dependence on 
scaling dimensions, as a function of $\Delta_{\mathrm N}$, is dominated by two 
effects, the increase from $1/C_{\lambda}$ and the decrease with 
multi-body phase space of $A_{\Delta_{\mathrm N}-1/2}\times f(\Delta_{\mathrm N}, \mu_{\rm 
IR}^2/m_W^2)$. The multi-body phase space dominates the behavior and 
drives the decrease of the branching fractions into unparticle states.

So far we have discussed the inclusive rate for $W$ bosons to decay to 
unparticles.  However, below the scale at which conformal symmetry is 
broken the unparticles must ``hadronize" into singlet neutrinos so the 
ultimate fate of the decay is $W^\pm \rightarrow \ell^\pm + n N$, with 
$n$ an odd integer. The maximum number of final state singlet neutrinos 
$N$ that is kinematically allowed, for a given charged lepton energy 
$E_\ell$ in the $W$-boson center of mass frame, is given by,
 \beq
n_{\rm max}=\frac {m_W} {m_N} \sqrt{1-\frac {2 E_\ell} {m_W}}.
\label{eq:nmax}
 \eeq 
 In \figref{fig:Wwidth1} we show the differential width of the $W$ to 
final states that include a charged lepton and an arbitrary number of 
singlet neutrinos as a function of the charged lepton energy. The form
of these distributions is independent of the parameter $\hat 
\lambda$. The results are presented for three benchmark values of the 
scaling dimension, $\Delta_{\mathrm N} = 7/4, 8/4$ and $9/4$, and for three 
benchmark masses of the composite singlet neutrinos, $M_N = 0.04, 0.4, 
$ and $4$ GeV. We omitted the case of $\Delta_{\mathrm N}=3/2$ since it is simply a 
delta function. In the top left panel, we show the overall distribution 
as a function of $E_\ell$, which is plotted on a linear scale, for the 
benchmark values of the masses and $\Delta_{\mathrm N}$. In the rest of the 
panels, we show the same distribution for each of the benchmark values 
of $M_N$ separately, but now with $E_\ell$ plotted on a logarithmic 
scale, zooming into the threshold region. From this figure we can see 
that the charged lepton energy favors the kinematic threshold of $E_\ell 
\simeq m_W/2$, and is separated from it only by the non-vanishing mass 
of the HNL. Note that the range of the horizontal axis 
extends to different values of $E_\ell$ in these three panels, 
reflecting the different benchmark composite singlet neutrino masses. As a result 
of this behavior and the unitarity of the area under the differential 
curve, the heavier the composite singlet neutrinos, or equivalently the 
larger the infrared cut off $\mu_{IR}$, the greater the fraction of 
events with charged lepton energies at lower values. Due to the shape of 
this distribution, the unparticle decays of $W$ bosons strongly favor 
low numbers of singlet neutrinos in the final state.

The rate for $Z$ boson decay into hidden sector states can be calculated 
in the same manner as for $W$ decay.  The square of the resulting matrix 
element takes a form very similar to that for $W$ decay, 
\eref{eq:WdecayM2},
 \begin{equation}
  \frac{1}{3} \sum |\mathcal{M}|^2 =  
  \frac{(g^2+g^{\prime 2})}{6} \frac{\hat{\lambda}^2 v_{\rm EW}^2}{M_{\mathrm{UV}}^{2\Delta_{\mathrm N} - 3}}
  \frac{E_\ell}{m_Z} \frac{3 m_Z - 2 E_\ell}{m_Z - 2 E_\ell}~.
  \label{eq:ZdecayM2}
 \end{equation}
 The partial width of the $Z$ boson is then given by,
 \bea
\Gamma(Z\!\rightarrow \bar\nu_\ell\, \mathcal{U})\!
= \! \sum_{\alpha=1}^{\numN} \! m_Z \frac {(g^2+g^{\prime 2}) |U_{{N_\alpha}\ell}|^2} {192\pi^2 C_\lambda^2} \! \left(\frac {M_{N}} {m_Z}\right)^{\!\!2}\!\! \left(\!\frac {m_Z} {\Lambda} \!\right)^{\!\!2\Delta_{\mathrm N}-3}\!\!\!\! A_{\Delta_{\mathrm N}-1/2}
f(\Delta_{\mathrm N},\frac {\mu_{IR}^2} {m_Z^2})~.
\label{eq:Zwidth}
 \eea

\subsection{Current Constraints and Future Reach from HNL Searches}

\begin{figure}[t]
  \centering
    \includegraphics[width=0.99\textwidth]{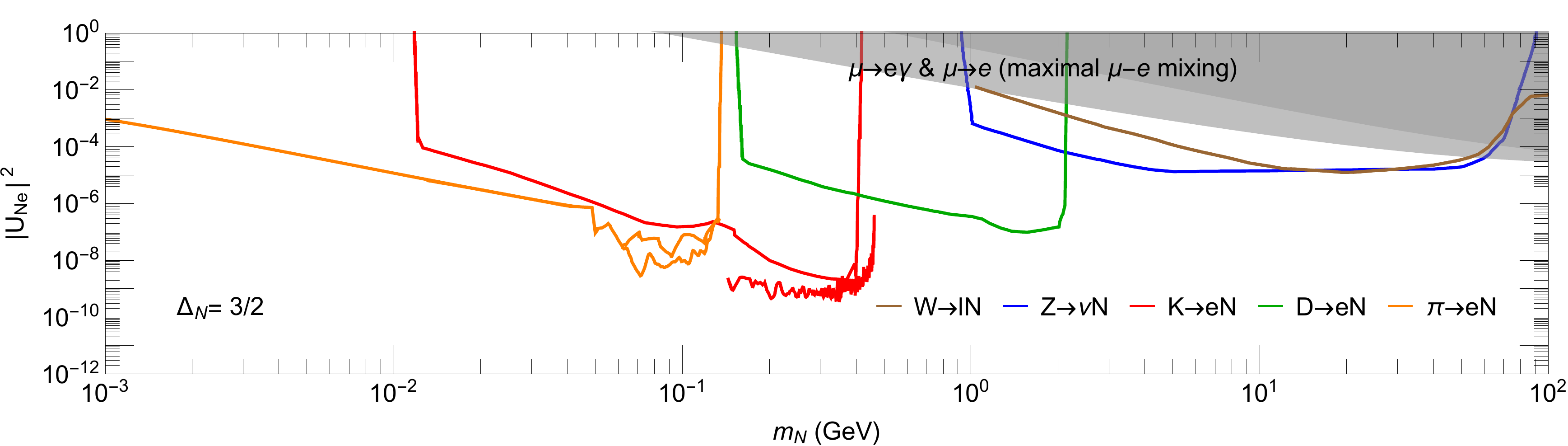}
    \includegraphics[width=0.99\textwidth]{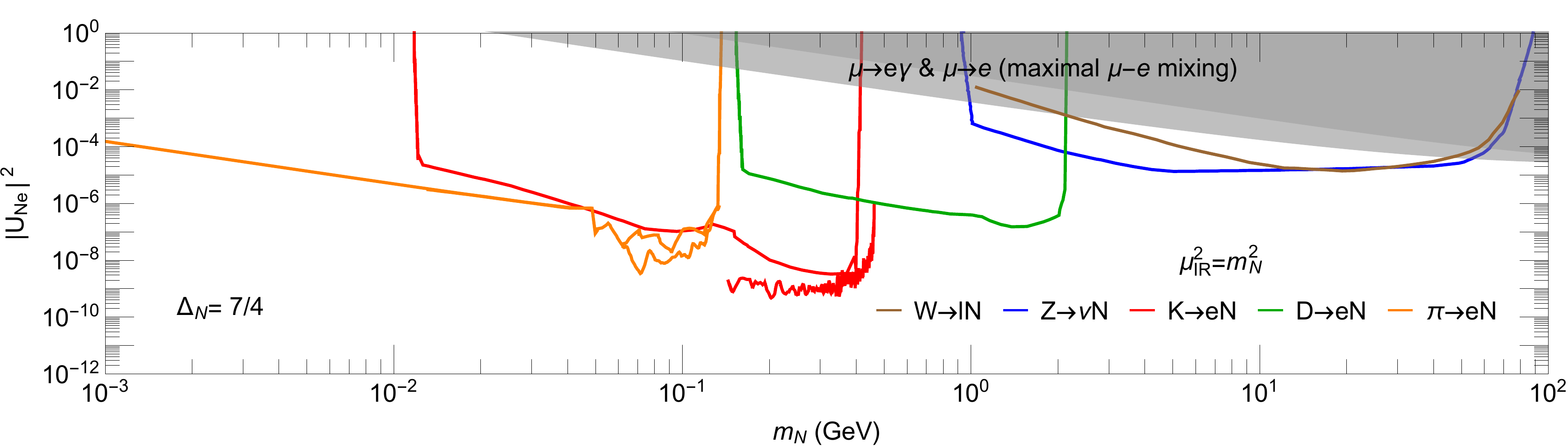}
    \includegraphics[width=0.99\textwidth]{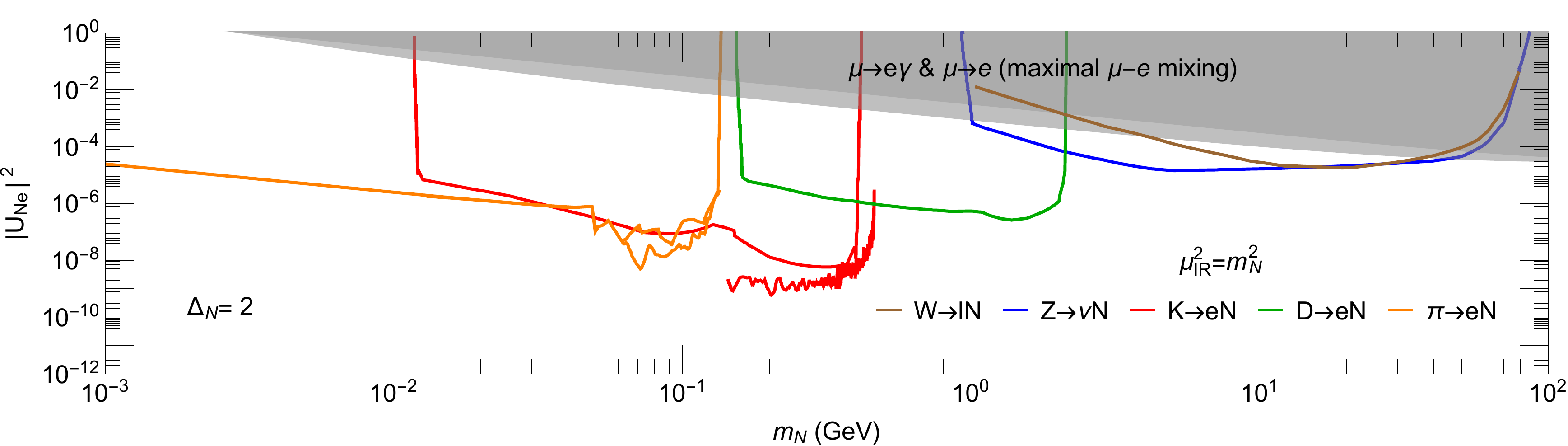}
    \includegraphics[width=0.99\textwidth]{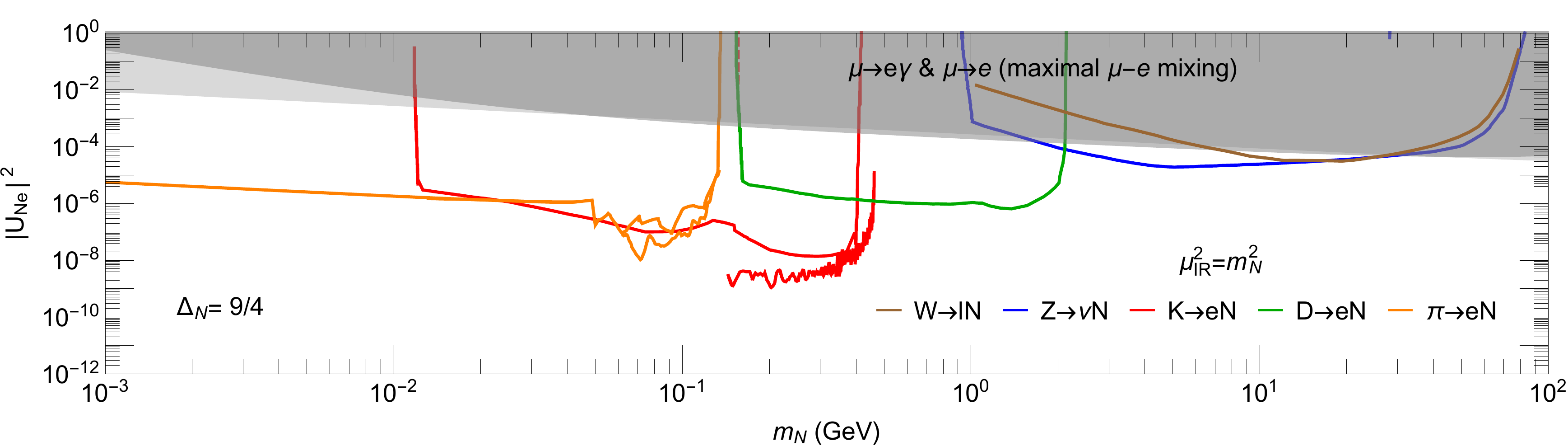}
    \caption{The existing limits on the IR mixing angle as a function of composite neutrino mass ($m_N\sim \Lambda$, the compositeness scale) for final states involving electrons.  These bounds are a reinterpretation of constraints on HNLs, as discussed in the text.  
     The lower and upper shaded gray regions correspond to the flavor dependent constraints from the $\mu\rightarrow e$ conversion and $\mu\rightarrow e\gamma$ measurements in the maximal electron-muon flavor mixing scenario, which is discussed in the next section.
}
    \label{fig:oldlimits}
\end{figure}
\begin{figure}[t]
  \centering
    \includegraphics[width=0.99\textwidth]{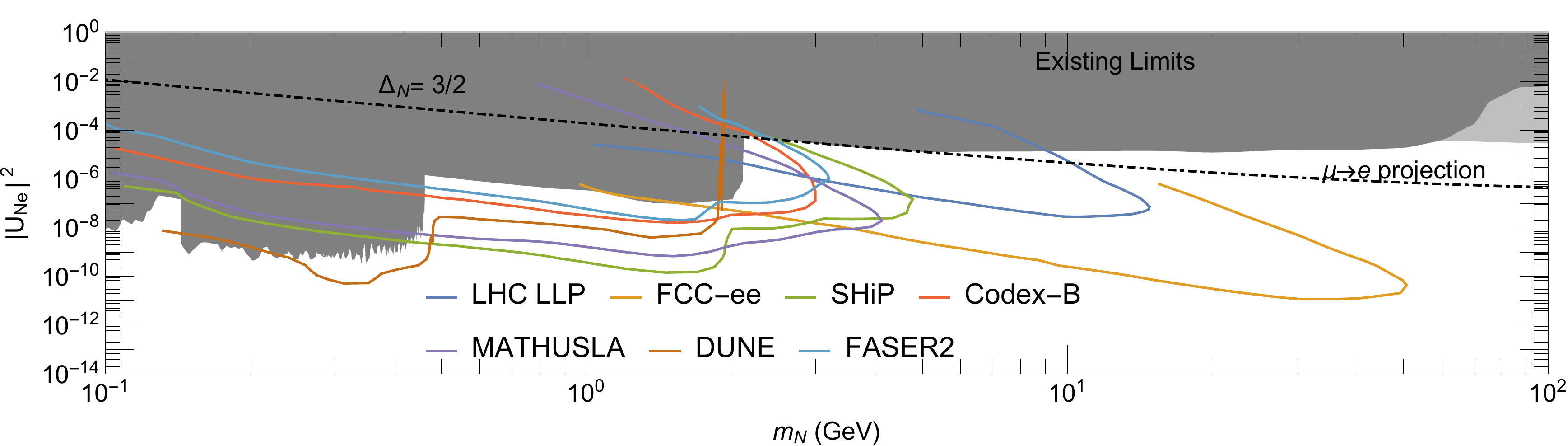}
    \includegraphics[width=0.99\textwidth]{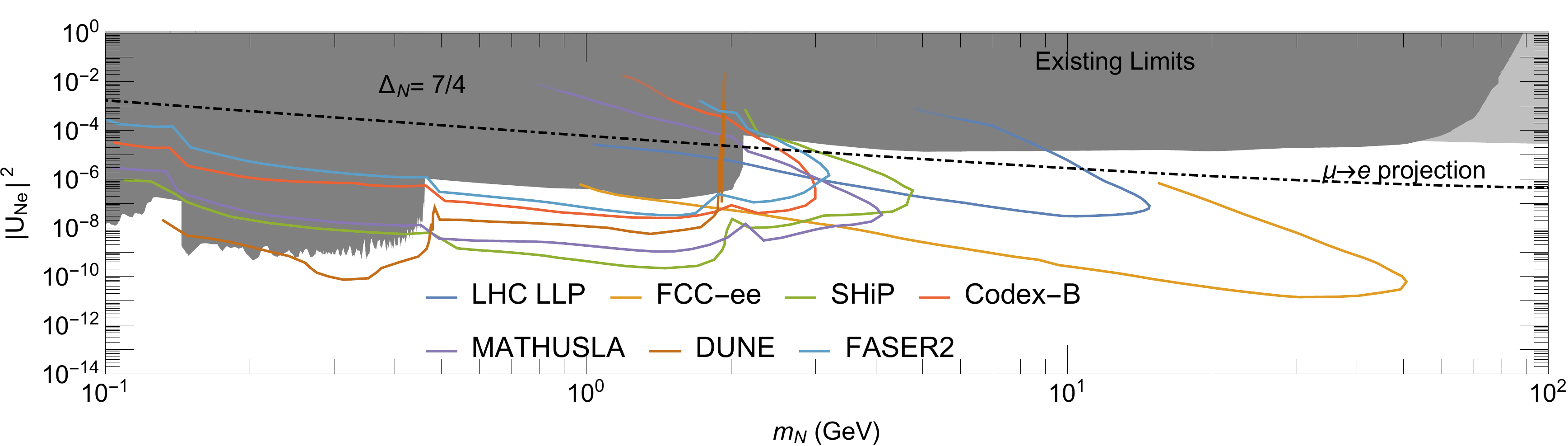}
    \includegraphics[width=0.99\textwidth]{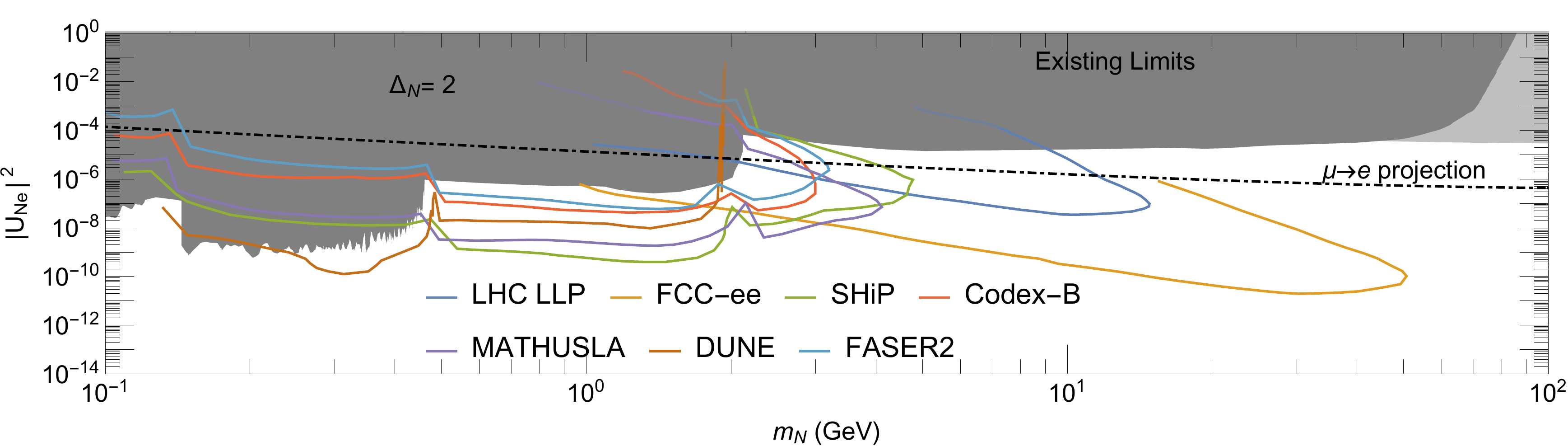}
    \includegraphics[width=0.99\textwidth]{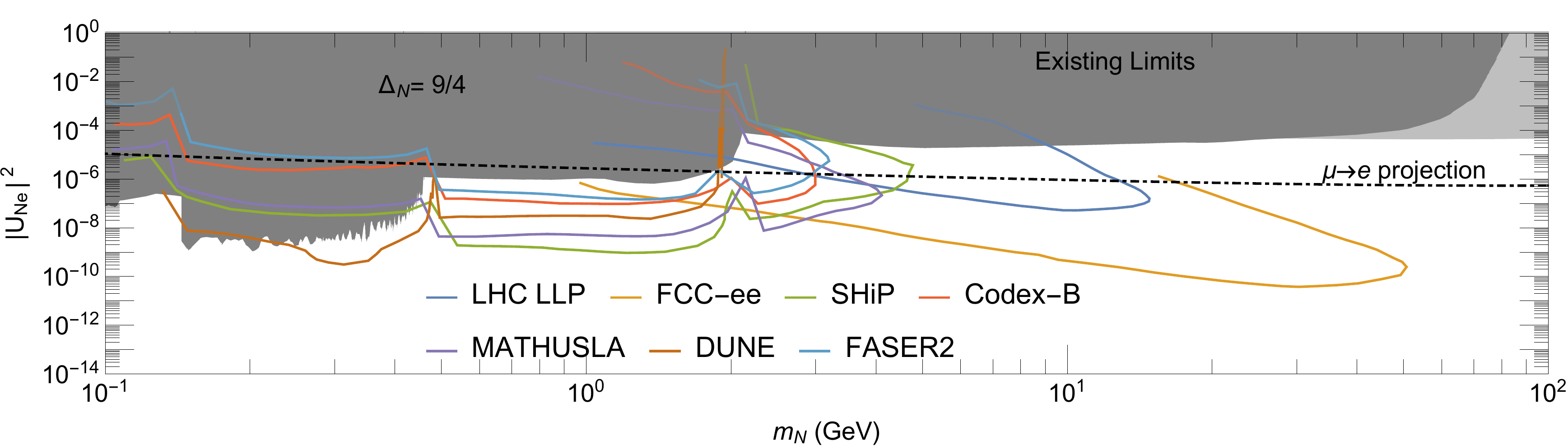}
    \caption{
    Future projections for constraints on the composite neutrino parameter space, in the singlet neutrino mass $m_N$-mixing angle squared $|U_{Ne}|^2$ plane, for different benchmark choices of the scaling dimensions $\Delta_{\mathrm N}$. The current limits detailed in \figref{fig:oldlimits} are shaded in gray. Details of the future projections can be found in the text. The flavor dependent current limit and future projections from the $\mu$-$e$ conversion experiments are in light gray shade and dot-dashed line, respectively.
}
    \label{fig:futureprojections}
\end{figure}


In this subsection we determine the current bounds on this class of 
models by recasting existing searches for the decay of $W$ and $Z$ 
bosons into elementary HNLs that mix with the SM neutrinos. 
We also discuss the expected reach of similar searches in the future. 
Our results are obtained under the assumption that all decays into the 
hidden sector result in the production of just a single composite 
singlet neutrino, which allows us to compare the rates for unparticle 
production to the corresponding rates for production of elementary 
HNLs~\cite{Atre:2009rg}. Although the resulting limits will 
be stronger than the actual bounds, the differences are expected to be 
modest since the results of the previous subsection indicate that decays 
into states with very few composite singlet neutrino are heavily 
favored.

In our analysis, we consider only the bounds coming from final states 
that include electrons, which tend to be the strongest in 
flavor-democratic models. Furthermore, for simplicity we assume that all 
the $N_\alpha$ are degenerate, $M_{N_\alpha}=M_N$, and present the bound 
in terms of the effective mixing angle squared,
 \be
|U_{Ne}|^2\equiv \sum_{\alpha=1}^{\numN}  |U_{{N_\alpha}e}|^2~.
 \label{effective_mixing_angle}
 \ee

 The current constraints on this class of models are displayed in 
\figref{fig:oldlimits}. The constraints are expressed as limits on the 
mixing angle $U_{Ne}$ as a function of the mass $m_N$ of the composite 
singlet neutrino. The bounds are shown for four different choices of the 
scaling dimension $\Delta_{\mathrm N}$.  
The limits from $W$ and $Z$ boson decay shown in the 
figure are based on the following searches:
 \begin{itemize}
 \item {\bf LHC}: Above a few GeV the 
constraints come from the current LHC searches for promptly decaying 
HNLs produced in $W$ boson 
decays~\cite{Sirunyan:2018mtv,Aad:2019kiz}, shown as the brown curves.
 \item {\bf LEP}: LEP dominates the current constraints in the 2-40~GeV 
regime. The bounds on HNL production from rare $Z$ 
boson decays searched for by the DELPHI experiment at 
LEP~\cite{Abreu:1996pa} are shown in blue. The LEP search covers a broad 
range of signature spaces, including HNLs giving rise 
to prompt decays, displaced vertices and energy deposition in the 
calorimeters. In our analysis, we consider only the bounds from prompt 
decays, since the constraints from long-lived particle searches at LEP 
are weaker than those from a combination of prompt searches at LEP 
and D-meson decays at beam dumps.
 \end{itemize}

The future reach of HNL searches for this class of models 
is shown in \figref{fig:futureprojections}. In the figure the current 
limits in the composite neutrino mass versus mixing angle plane are 
shaded in gray, with the flavor-dependent constraints in a lighter 
shade. The projected future limits from $W$ and $Z$ bosons decays shown 
in the figure are based on the following proposed searches:
 \begin{itemize} 
 \item {\bf HL-LHC}: The projected HL-LHC reach is shown in 
Fig.~\ref{fig:futureprojections} as the blue curves, using a 
conservative analysis of long-lived particles produced in $W$-boson 
decays~\cite{Liu:2019ayx}.  The analysis employs a dilepton trigger with 
an additional displaced lepton as the signal, based on the calibration 
data for the displaced dilepton analysis carried out by 
CMS~\cite{CMS-PAS-EXO-16-022}. New search ideas at the HL-LHC will be 
able to further improve the coverage. For a more optimistic projection 
without background, see, \eg, Ref.~\cite{Drewes:2019fou,Chun:2019nwi}.
 \item {\bf FCC-ee}: Projections for FCC-ee are shown in orange in 
Fig.~\ref{fig:futureprojections}, based on a long-lived particle 
analysis from its $Z$-pole run~\cite{Beacham:2019nyx}.
 \end{itemize}

\clearpage
\subsection{Novel Signatures}

This class of theories can give rise to exotic signals that are not 
present in conventional models with an elementary singlet neutrino. If 
the $W$ boson mass is hierarchically larger than the compositeness 
scale, $m_W \gg \Lambda$, multiple singlet neutrinos can be produced in 
the same decay, $W^{\pm} \rightarrow \ell^{\pm} + n \, N$, where $n$ is 
an odd number. These composite singlet neutrinos are not stable, but 
decay to SM particles through the weak interactions. At colliders, 
decays to final states that contain charged leptons, $N \rightarrow \nu 
+ \ell^{+} + \ell^{-}$, are especially promising. In certain regions of 
parameter space the singlet neutrinos are long-lived, leading to 
striking displaced vertex signatures that have not been constrained in 
previous experiments~\cite{Abreu:1996pa, 
Liu:2019ayx,Cottin:2018nms,Aad:2019kiz, Dev:2019anc, Drewes:2019fou}. 
Therefore, in this class of models, the decays of $W$ bosons can give 
rise to highly characteristic signals involving multiple displaced 
vertices. Since the LHC is expected to produce of order $10^{12}$ $W$ 
bosons over its lifetime, it is expected to have excellent reach for 
this class of models in much of parameter space.

 The lifetime of the composite singlet neutrinos is given approximately 
given by,
 \bea
c\tau_\alpha&\approx&\left(\sum_\ell \kappa_\ell \frac {G_F^2 m_N^5} {192\pi^3} |U_{N_\alpha \ell}|^2\right)^{-1} \label{eq:lifetime} \nonumber \\
 &\sim& 60 \times \left(\frac {\gev} {M_N}\right)^5 \left(\frac {0.01{\rm eV}} {m_\nu}\right) \left(\frac {\mu^c} {\kev} \right){\rm meter}.
 \eea
 The parameter $\kappa_i$ counts, for a specific lepton flavor $i$, the 
number of kinematically allowed decay channels mediated by the off-shell 
$W$, as well as the number of appropriately weighted decay channels 
mediated by the off-shell $Z$.\footnote{In calculating $\kappa_i$, it is 
necessary to take into account interference between the $W$ and 
$Z$-mediated channels when considering final states involving charged 
leptons.} For a composite singlet neutrino mass of $M_N=1$~GeV, 
$\kappa_e \approx \kappa_\mu \approx 8.25$. The $Z$-mediated process also 
contributes to $\kappa_\tau\approx 2.8$.  The mixing angle scales as 
$|U_{N_\alpha \ell}|^2 \sim m_\nu/\mu^c$ and corresponds to a benchmark 
value of $10^{-5}$ for the above choices of $m_\nu$ and $\mu^c$. In our 
analysis we use the results presented in Ref.~\cite{Bondarenko:2018ptm} 
to model the details of threshold effects, finite fermion mass 
corrections, etc., on the lifetime of the composite singlet neutrinos.

\begin{figure}[t]
\centering
\includegraphics[width=0.45\textwidth]{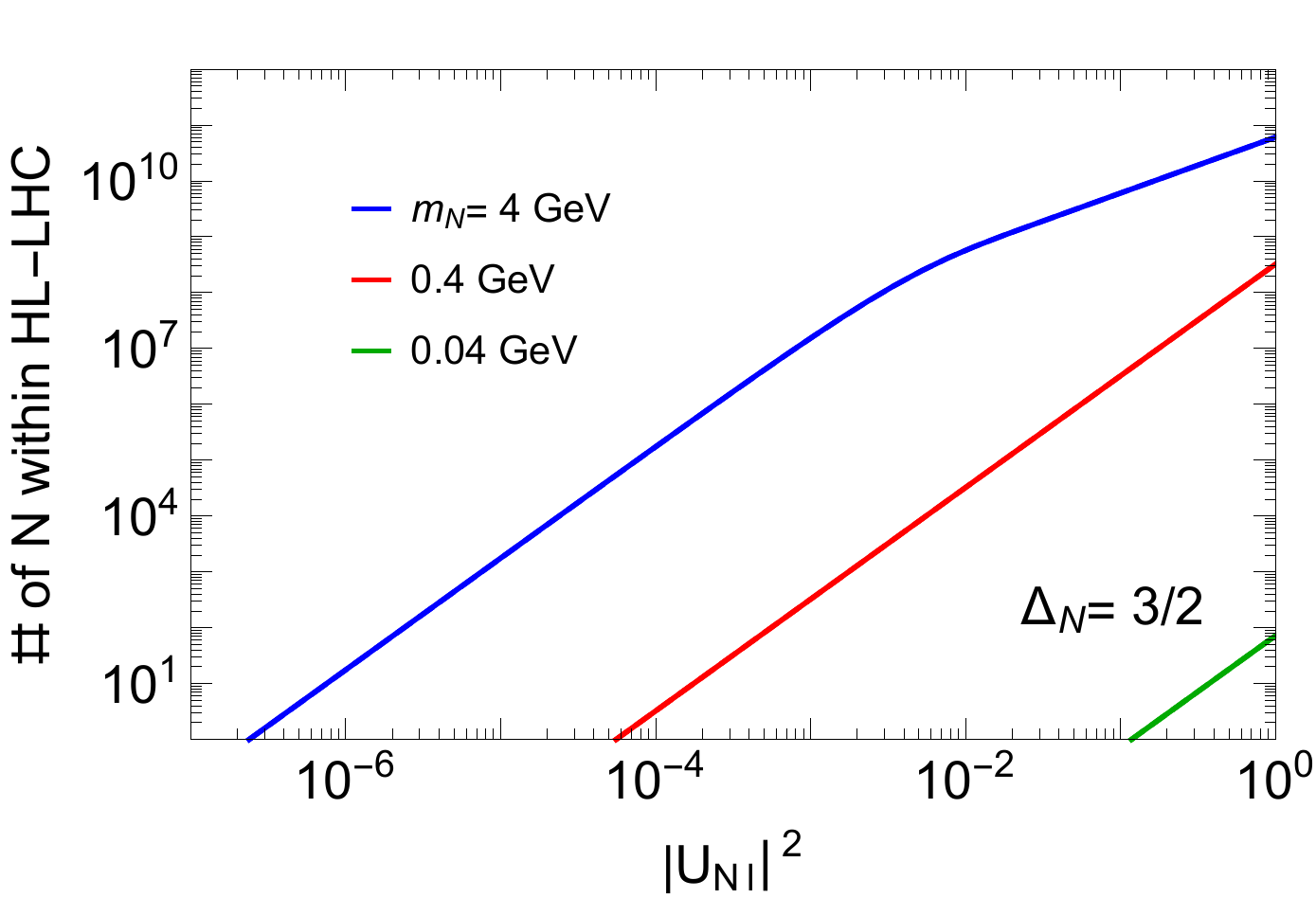}
\includegraphics[width=0.45\textwidth]{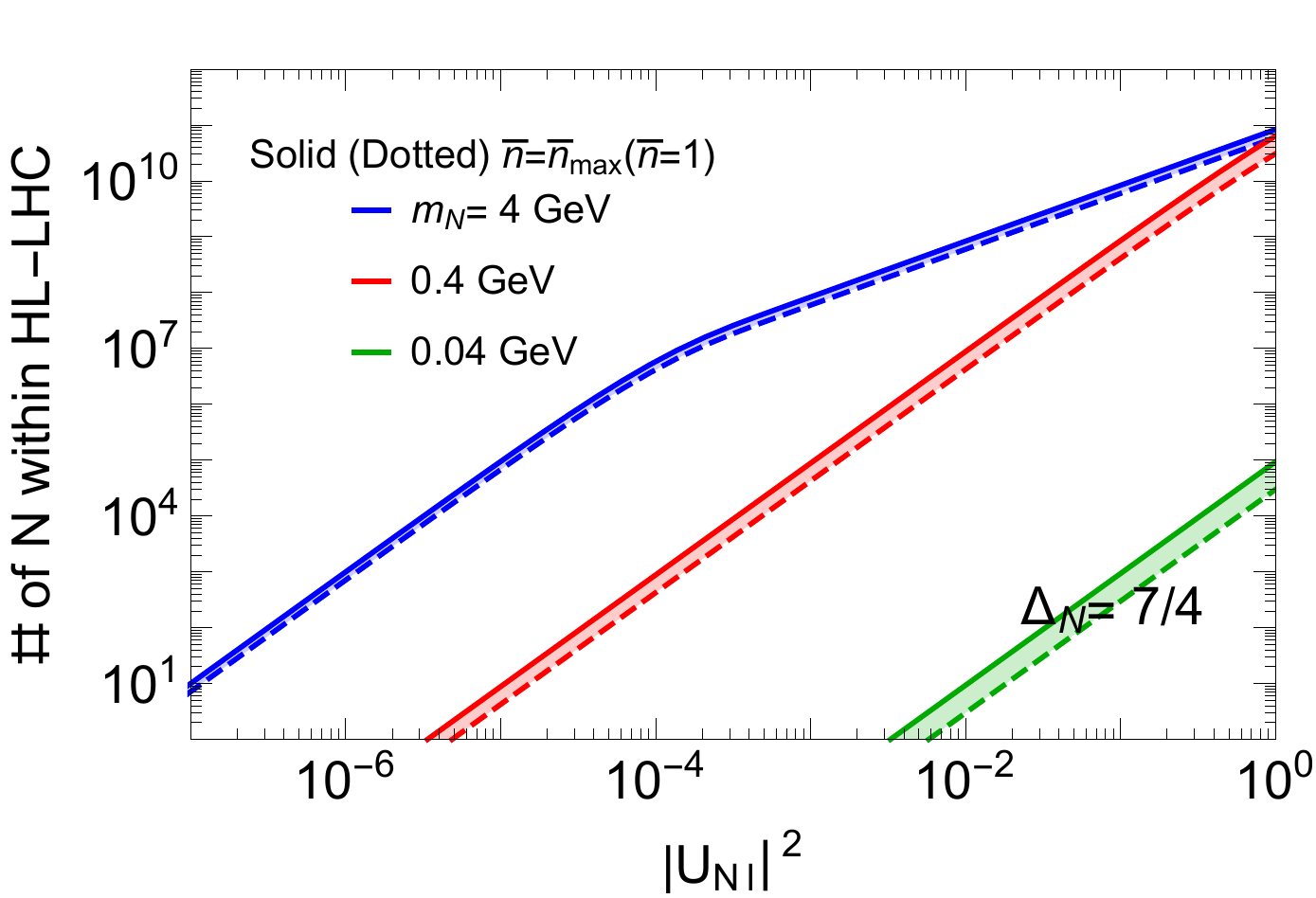}
\includegraphics[width=0.45\textwidth]{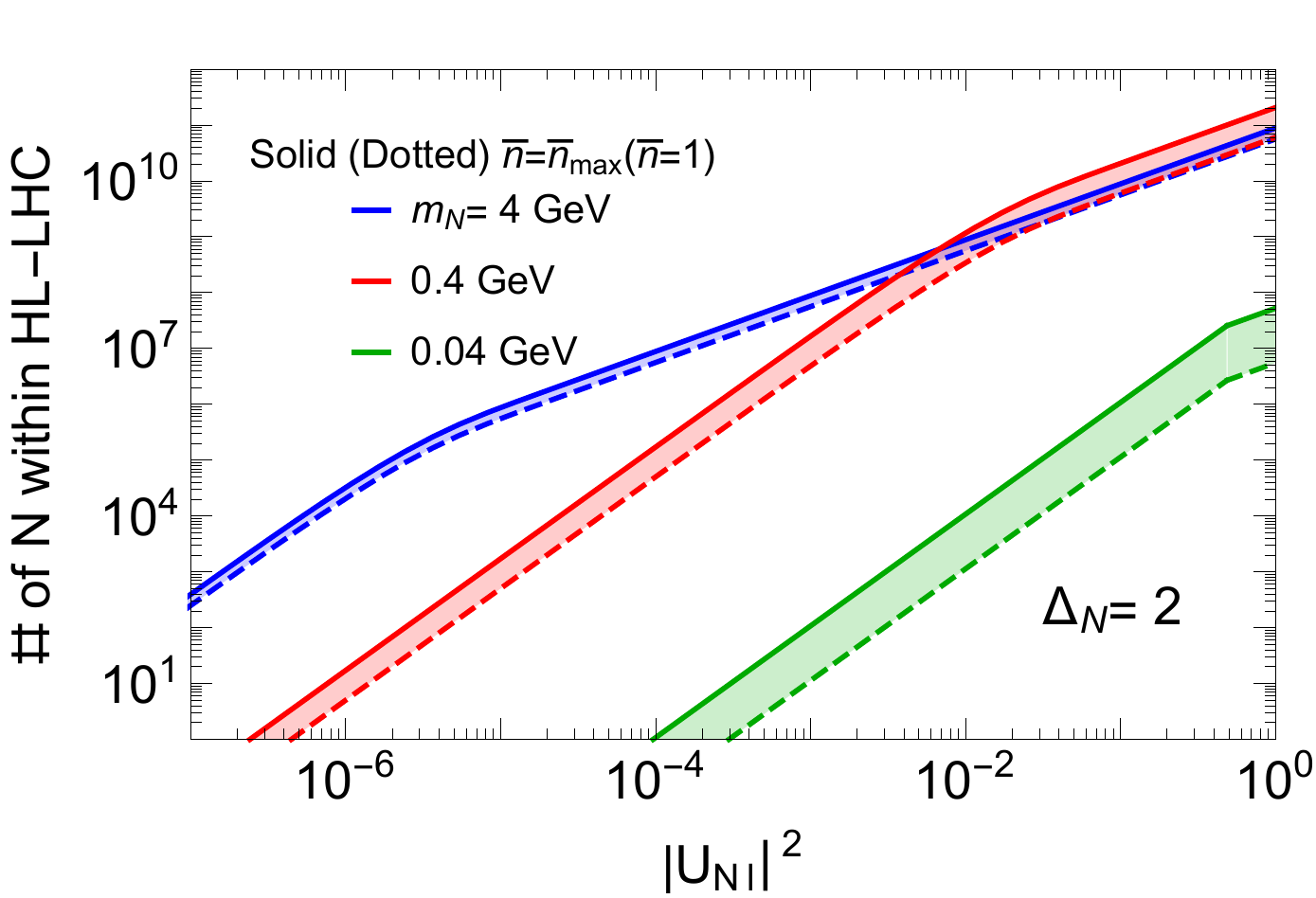}
\includegraphics[width=0.45\textwidth]{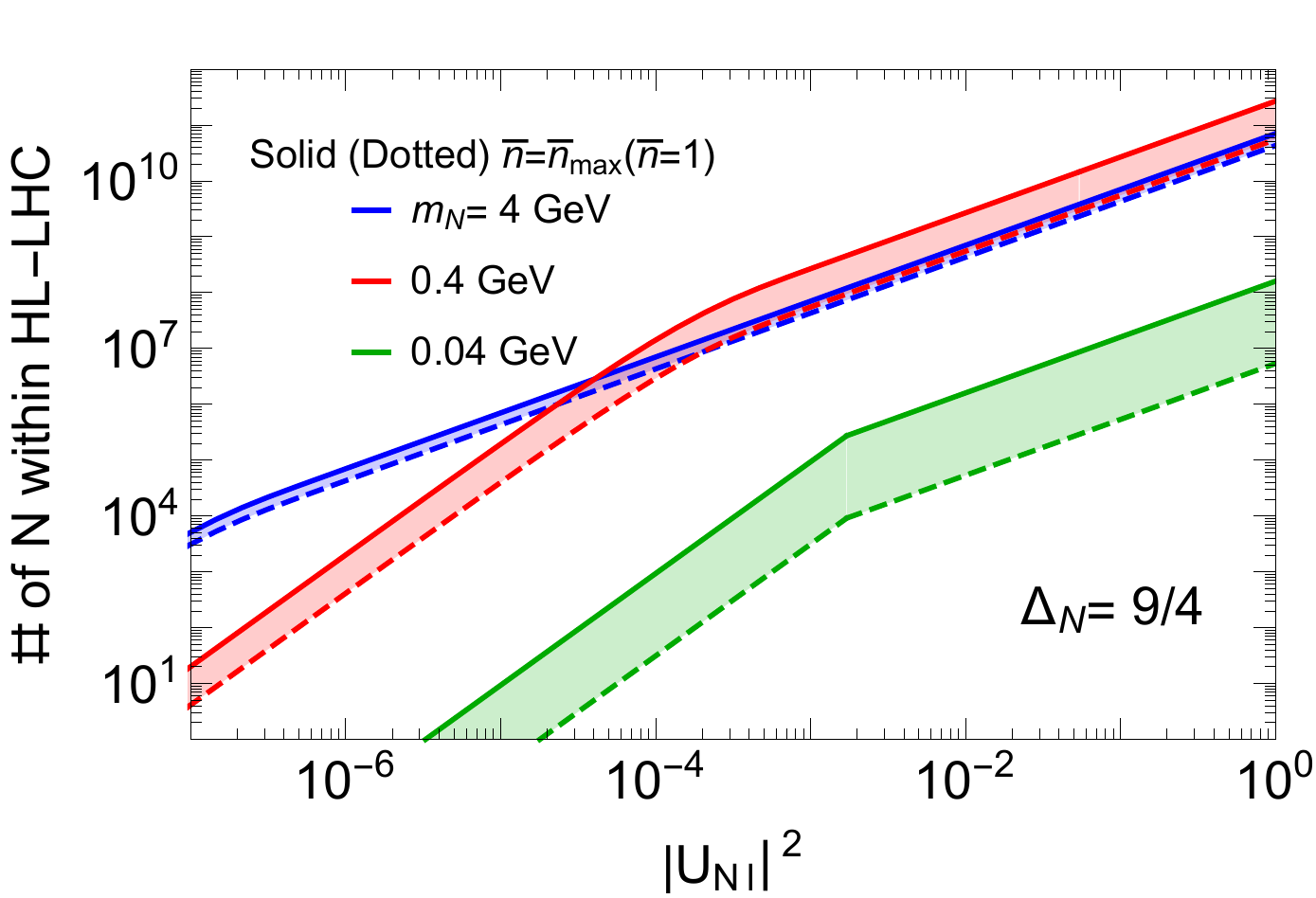}
\caption{The expected number of composite neutrinos decays at the HL-LHC for various benchmark choices of the $m_N$ and scaling dimensions $\Delta_{\mathrm N}$. The solid or dashed lines show how the results vary by assuming each unparticle produced results in $n_{\rm max}$ or one singlet neutrino, respectively.
}
\label{fig:NrateHLLHC}
\end{figure}

Given the large proper lifetime of the composite singlet neutrinos and 
their energy distribution as indicated from \figref{fig:Wwidth1}, we can 
place an approximate upper limit on the total number of composite 
singlet neutrino decay events within a detector radius $R_{\rm LHC}$ at 
the HL-LHC as,
 \beq
\langle n_N \rangle \lesssim n_W\times \frac {\Gamma_W^{\ell N}(M_N,\Delta_{\mathrm N},\mu_{IR})} {\Gamma_W^{tot}}\times {\overline n_{\rm max} (m_N,\Delta_{\mathrm N})}\times \left(1-\exp\left[-\frac {R_{\rm LHC}} {\gamma(\bar n_{max})\beta(\bar n_{max}) c\tau}\right]\right).
 \eeq
 Here $\bar n_{\rm max}$ represents the maximum number of composite 
singlet neutrinos that can be produced in the decay of a $W$ boson to 
the hidden sector after averaging over the kinematics of individual 
events. In \figref{fig:NrateHLLHC} we show the total number of composite 
neutrino decays within a detector radius of $R_{\rm LHC}=10$ meters at 
the 13 TeV LHC with an integrated luminosity of $3000~\fbi$. This figure 
should be considered together with \figref{fig:oldlimits} to determine 
the regions of parameter space that are not already excluded. The 
results are presented for three different benchmark values of the composite 
singlet neutrino mass, 0.04, 0.4, and 4 GeV, shown in green, red and 
blue respectively. To explore the dependence of the result on the number 
of composite singlet neutrinos produced per $W$ decay, we consider two 
limiting cases. In the first limiting case, this number is taken to be 
the maximally kinematically allowed value $n_{max}$ defined in 
\eref{eq:nmax}, averaged over the differential distribution in 
\eref{eq:Wdiff}. In the second of the two limiting cases, the number of 
composite singlet neutrinos produced per $W$ decay is simply taken to be 
$1$. The kinks in the curves correspond to the transition between prompt 
signals (larger mixing angles) and collider metastable signals (smaller 
mixing angles). We see that in general, if we require about 100 singlet 
neutrinos decays within the LHC for discovery, the HL-LHC can probe 
values of the mixing angle $|U_{N\ell}|^2$ in the range from $10^{-7}$ 
to $10^{-5}$. This shows that the HL-LHC can indeed probe unexplored 
regions of the parameter space. Note that the phenomenology of 
unparticle neutrino is closely related to that of dark 
showers~\cite{Strassler:2006im,Han:2007ae,Schwaller:2015gea,Pierce:2017taw,Alimena:2019zri,Yuan:2020eeu}, 
where the shower particles are often hidden pions or hidden photons. 
Instead, our model predicts an HNL shower, and in particular, in 
lepton(charged and neutral)-rich final states. Here, in addition, the 
unparticle calculation allows for an approximate prediction of the 
accompanying charged lepton differential cross section from $W$ decay, 
which could be an crucial consistency check for the underlying theory 
and dynamics.

\section{Beam Dumps}
\label{sec:dump}

For compositeness scales below a few GeV, the rare decays of pions,
kaons and B-mesons constitute a powerful probe of this class of models.
In this section we discuss the current bounds on this scenario from the
rare decays of mesons. In addition, we explore the reach of future beam
dump experiments and the proposed LHC-based experiments FASER2, Codex-B
and MATHUSLA.

 For the meson decay $\mathfrak{m}\rightarrow \ell\, \mathcal{U}$, the
square of the spin averaged matrix element is given by,
 \be
|\overline{\mathcal{M}}|^2 = 4\,G_F^2 |V_{qq'}|^2 f_\mathfrak{m}^2\left(\frac{\hat{\lambda}v_{\mathrm{EW}}}{M_{\mathrm{UV}}^{\Delta_{\mathrm N}-3/2}}\right)^2 m_\mathfrak{m}E_\ell~,
 \ee
 where $f_\mathfrak{m}$ is the meson decay constant. The corresponding
partial width is then,
\beq
\Gamma(\mathfrak{m}\rightarrow \ell\, \mathcal{U})
=\sum_{\alpha=1}^{\numN} \frac {m_\mathfrak{m}G_F^2  f_{\mathfrak{m}}^2m_N^2 |V_{qq^\prime}|^2 |U_{N_\alpha\ell}|^2} {32\pi^2 C_\lambda} \left(\frac {m_\mathfrak{m}} {M_{\rm UV}}\right)^{2\Delta_{\mathrm N}-3} A_{\Delta_{\mathrm N}-1/2}\times g\left(\Delta_{\mathrm N},\frac {\mu_{IR}^2} {m_\mathfrak{m}^2}\right),
\eeq
 where $g(\Delta_{\mathrm N},\frac {\mu_{IR}^2} {m_W^2})$ captures the dependence 
on the scaling dimension $\Delta_{\mathrm N}$ and the infrared cutoff $\mu_{IR}$. 
The details may be found in the appendix in \eref{eq:gnew}. As before, 
we can express the partial width in terms of the conventional sterile 
neutrino mixing angle $U_{N\ell}$ using 
Eq.~(\ref{effective_mixing_angle}).

As with the calculation of $W$ and $Z$ boson decays into composite 
singlet neutrinos detailed in the previous section, we translate the 
existing constraints on elementary HNLs into constraints on the 
parameter space of our model. As before, our limits are obtained under 
the assumption that all decays to the hidden sector result in just a 
single composite singlet neutrino. This allows the bound to be obtained 
from a straightforward comparison of the rate for unparticle production 
to the rate for production of elementary sterile 
neutrinos~\cite{Atre:2009rg}.

The current bounds on the effective mixing angle squared 
$|U_{Ne}|^2\equiv \sum_{\alpha=1}^{\numN}
|U_{{N_\alpha}e}|^2$ as a function of composite neutrino mass $m_N$ 
from rare meson decays are shown in \figref{fig:oldlimits}. The results 
are presented for four different choices of $\Delta_{\mathrm N}$, the scaling 
dimension. With the exception of the pion decay bound, the charged meson 
searches look for the decay products of elementary HNLs 
into final states involving charged leptons or charged pions. For 
concreteness, in our analysis we have only considered constraints 
arising from final states involving electrons. Since the heavy neutrino 
provides the chirality flip required for scalar meson decays, and the 
electron is the lightest charged lepton, these constraints tend to be 
the strongest in flavor-democratic models. The limits we present are 
based on the following searches for elementary HNLs.
 \begin{itemize}
 \item {\bf TRIUMF}: The strongest constraints below the pion mass 
threshold are from searches at TRIUMF for soft positrons from positively 
charged pion decays and the overall 
pion exclusive decay rate measurement~\cite{Britton:1992xv}. These are shown as the orange curves 
in \figref{fig:oldlimits}. We also include the bounds from the recent 
PIENU~\cite{Aguilar-Arevalo:2017vlf} experiment at TRIUMF shown in the 
same color which sets a stronger constraint in regions where the limits 
are provided.
 \item {\bf PS191}: Searches for charged kaon decays into a charged 
lepton plus a HNL performed at CERN's PS191 
experiment~\cite{Vannucci:679670} lead the constraints in the 
130-400~MeV regime. These are shown in red in \figref{fig:oldlimits}.
 \item {\bf CHARM}: Searches for HNLs from D-meson 
decays were conducted at the CHARM experiment~\cite{Bergsma:1985is}.
The resulting bounds are 
shown in green in \figref{fig:oldlimits}.
 \item {\bf NA62}: The NA62~\cite{NA62:2020mcv} experiment searched for 
long-lived HNL decays from a high intensity beam dump with 
$1.74\times10^{18}$ Proton-on-Target (PoT). It is sensitive to HNLs produced 
in kaon decays, leading to the limits shown as the lower red curves in 
Fig.~\ref{fig:oldlimits}.\footnote{T2K ND280~\cite{Abe:2019kgx} also 
searched for HNLs from kaon decays. The resulting 
limits are weaker than those from NA62 and we do not explicitly include 
them here.}
 \end{itemize}

In \figref{fig:futureprojections} we show the projected reach of future 
experiments. In this plot, the existing limits on the parameter space 
are shaded in gray. Included in the figure are the projected 
sensitivities to composite singlet neutrinos produced from meson decays 
in several different future experiments, including SHiP, Codex-B, 
MATHUSLA, DUNE and FASER-2. Most of the projections are based on 
Ref.~\cite{Beacham:2019nyx}.
 \begin{itemize}
 \item {\bf SHiP}: The SHiP experiment~\cite{Alekhin:2015byh}, which has 
a higher beam energy than other beam dump proposals, can probe 
long-lived HNL decays. It is sensitive to HNL 
masses up to 5 GeV from the decays of B-mesons. The sensitivity is shown 
as the green curves in Fig.~\ref{fig:futureprojections}.
 \item {\bf DUNE}: The DUNE near detector can search for the decays of 
long-lived particles produced from the beam dump used to produce 
neutrinos~\cite{Adams:2013qkq}. The highly intense beam with 
$3\times10^{22}$ PoT can copiously produce 
$D$-mesons, kaons and pions, all of which can act as sources for singlet 
neutrinos. DUNE has greater sensitivity reach below the kaon mass 
threshold than other proposals, as can be seen from 
Fig.~\ref{fig:futureprojections}.
 \item {\bf FASER2, Codex-B, MATHUSLA}: The projected sensitivities of 
these LHC satellite experimental proposals are shown in 
Fig.~\ref{fig:futureprojections} as the blue, orange, and purple curves 
respectively. These limits, which are taken from 
Ref.~\cite{Beacham:2019nyx}, arise from the decays of mesons into 
long-lived HNLs.
 \end{itemize}

One complication in translating the limits on elementary singlet 
neutrinos into bounds on our model arises from the fact that, in 
general, each of these future experiments presents their results after 
combining all the different meson production modes. It then becomes a 
challenge to scale the results appropriately for each of the different 
contributions. In our analysis we have assumed that production through 
decays of the lightest available meson dominates, which is true for most 
of the parameter regions. For instance, in the DUNE projections, in the 
case of singlet neutrino masses above the kaon mass we assume the limits 
are entirely from $D$ meson decays, while for singlet neutrino masses 
between the pion and kaon masses, we take the limits as arising purely 
from kaon decays (neglecting the contribution from $D$ mesons).

\section{Muon Magnetic Moment and Lepton Flavor Violation}\label{s.loop}
\begin{figure}[t] 
   \centering
   \includegraphics[width=0.7\textwidth]{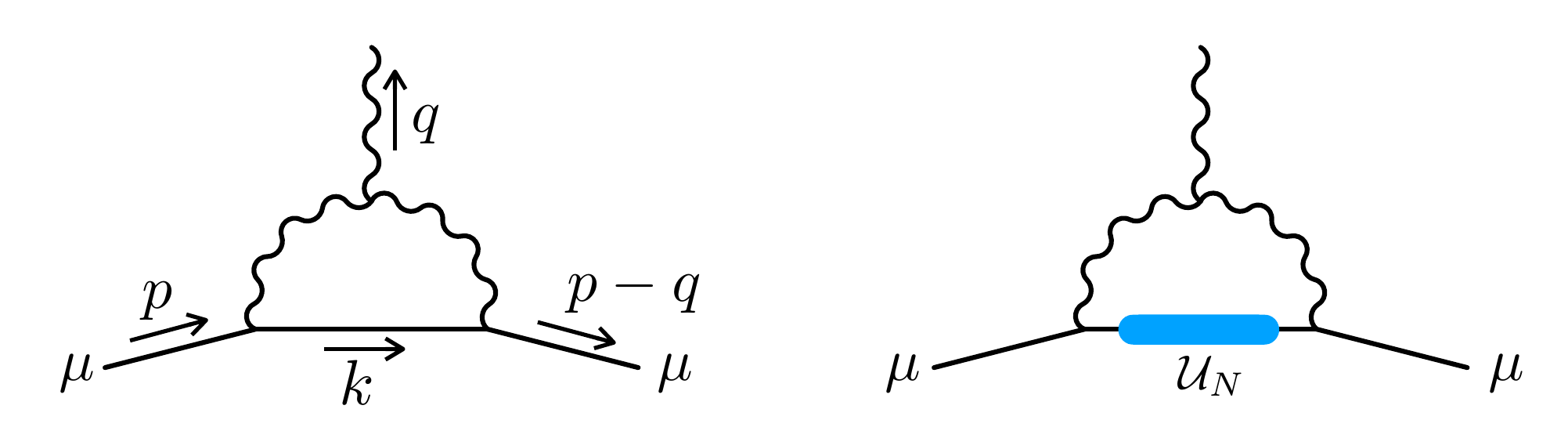} 
   \caption{Diagrams contributing to $(g-2)_\mu$ with an elementary 
neutrino in the loop (left), and the correction to this from unparticles 
(right). Our momentum labels are shown for 
the elementary case. Similar diagrams with an outgoing electron on the 
external leg will contribute to~$\mu \rightarrow e\gamma$.}
   \label{fig:diagram}
\end{figure}

Through its couplings to the SM neutrinos the hidden sector contributes 
at loop level to the anomalous magnetic moment of the muon and to lepton 
flavor violating processes. In this section we determine the size of 
these effects and explore the prospects for future experiments to detect 
them.

The measured value of the anomalous magnetic moment of the muon differs 
at approximately $\sim4\sigma$ from the SM 
prediction~\cite{Blum:2013xva,Lindner:2016bgg}. This long standing 
anomaly can be explained by the presence of new light states and will 
soon be tested by Fermilab's E989~\cite{Grange:2015fou}.  The lepton 
flavor violating processes $\mu\to e \gamma$ and $\mu-e$ conversion will 
also be probed with increased precision at the upcoming COMET and Mu2e 
experiments \cite{Adamov:2018vin,Bartoszek:2014mya}.  The calculation of 
composite neutrino contributions to these processes is quite similar. We 
focus first on the calculation of $(g-2)_\mu$ and on lepton flavor violation in the next subsection.

\subsection{Muon Magnetic Moment} 

\begin{figure}[t] 
  \centering
  \includegraphics[width=0.6\textwidth]{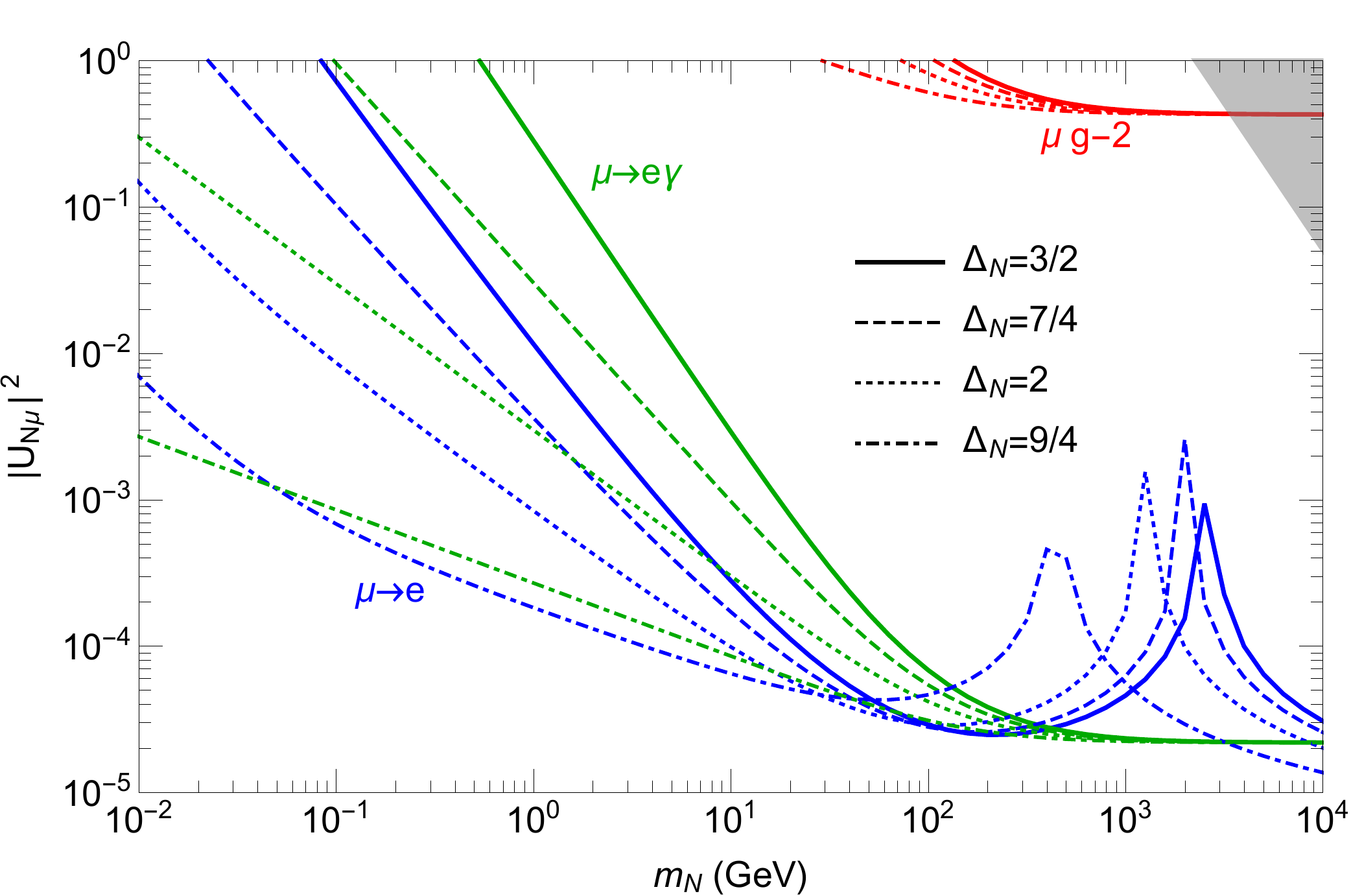} 
  \caption{Constraints on neutrino mixing as a function of the singlet neutrino mass scale from muon g-2 (red lines), $\mu\to e\gamma$ (green lines), and $\mu$-$e$ conversion (blue lines) experimental measurements, for benchmark choices of $\Delta_{\mathrm N}$ of 3/2 (solid lines), 7/4 (dashed lines), 2 (dotted lines), 9/4 (dot-dashed lines).
  In the gray region the boundary condition in \eref{lambdabound} is violated.
  }
  \label{fig:loopinduced}
\end{figure}

 A neutral fermion of mass $m_f$ that couples to the muon and the $W$ 
boson gives rise to a one loop contribution to $(g-2)_\mu$ through the 
diagram shown on the left of \figref{fig:diagram}.  The resulting one loop amplitude 
can be written in the form,
 \begin{equation}
 \bar{u}(p-q,m_\mu) \Gamma^\alpha(p-q,p)u(p,m_\mu)
 \end{equation}
 where $q$ represents the momentum of the outgoing photon.
 In unitary gauge the expression for $\Gamma^\alpha(p-q,p)$ is proportional to, 
 \begin{equation}
\int \frac{d^4 k}{16\pi^2} \frac{\gamma^\nu P_L \slashed{k}\gamma^\mu P_L}{(k^2-m_f^2)} 
D_{F \; \mu \lambda} (k -p) 
D_{F \; \delta \nu} (k-p+q) V^{\lambda \delta \alpha}(k-p, k-p+q, q)
 \label{eq:loop}
 \end{equation}
 where $D_{F \; \mu \lambda} (k -p)$ and $D_{F \; \delta \nu} (k-p+q)$ 
represent the $W$ boson propagators and $V^{\lambda \delta \alpha}(k-p, 
k-p+q, q)$ denotes the contribution to the amplitude from the triple 
gauge boson vertex. We can decompose $\Gamma^\alpha(p-q,p)$ as,
 \begin{equation}
  \Gamma^\alpha(p-q,p) = \gamma^\alpha F_1(q^2) 
+ \frac{i \sigma^{\alpha \beta} q_{\beta}}{2 m_\mu} F_2(q^2) \;
 \end{equation}
 where $F_1(q^2)$ and $F_2(q^2)$ are form factors.
 In this expression the anomalous magnetic moment arises from the 
contribution proportional to $ \sigma^{\alpha\beta} q_\beta$.  This 
effect has been calculated \cite{Moore:1984eg},
 \be
a_\mu = \frac{G_F m_\mu^2}{8\sqrt2 \pi^2} \int_0^1 dx \frac{2x\left(2r_f^2 +x^2 \left(r_f^2+r_\mu^2+2\right)-x\left(3r_f^2+r_\mu^2-2\right)\right)}{r_\mu^2 x^2 +r_f^2 -x(r_f^2+r_\mu^2-1)}
 \ee
 where $r_{f,\mu} \equiv m_{f,\mu}/M_W$. We denote the integral over $x$ as 
$I(r_f,r_\mu)$.  Expanding in the small parameter $r_\mu$ leads to a 
tractable integral with
 \be 
 I(r_f,r_\mu) =
\frac{10-43r_f^2+78r_f^4-49r_f^6+4r_f^8+18r_f^6\log 
r_f^2}{3\left(r_f^2-1\right)^4} + \mathcal{O}(r_\mu^2)~.
 \ee
 We now turn our attention to the case with unparticles in the loop. If 
we work in the insertion approximation, the contribution to the muon 
magnetic moment arises from a diagram of the form shown on the right of 
\figref{fig:diagram} that contains two insertions of the 
neutrino-composite neutrino mixing operator given in \eref{intLHO}. The 
single fermion propagator in \eqref{eq:loop} is now replaced by two 
massless neutrino propagators and an unparticle propagator. For the 
range of scaling dimensions of interest, $3/2 \le \Delta_{\mathrm{N}}< 
5/2$, we can express the unparticle propagator as an integral over a 
spectral function,
 \be
\Delta(p) = \frac{A_{\Delta_{\mathrm{N}}-1/2}}{2\pi} \int_{\mu_{IR}^2}^\infty dM^2 \left(M^2-\mu^2_{IR}\right)^{\Delta_{\mathrm{N}}-5/2} \frac{i\,\slashed{p}}{p^2-M^2 +i\epsilon}~,
 \ee
 with $A_n$ as given in \eref{eq:PhaseSpaceVol}.  Since the conformal 
symmetry is broken at low scales it is necessary to cut off this 
integral. We have chosen to do this in a simple way following 
\cite{Fox:2007sy}, cutting off contributions to the spectral function in 
the infrared from scales below $\mu_{IR}^2 \sim \Lambda^2$. Then the 
contribution of the unparticle sector to the muon magnetic moment can be 
obtained by making the replacement,
 \be
\frac{i \slashed{k}}{k^2-m_f^2}\longrightarrow 
\left(\frac{\lambda v}{C_\lambda \Lambda^{\Delta_{\mathrm{N}}-3/2}}\right)^2
\frac{A_{\Delta_{\mathrm{N}}-1/2}}{2\pi} \int_{\mu^2_{IR}}^\infty dM^2 \left(M^2-\mu^2_{IR}\right)^{\Delta_{\mathrm{N}}-5/2} \frac{-i \slashed{k} \slashed{k} \slashed{k}}{k^2 (k^2-M^2)k^2 }~,
 \ee
 in Eq.~(\ref{eq:loop}). The insertion of the mixing angles and the 
additional propagators makes the loop integral more complicated. 
However, taking advantage of the identities $\slashed{k}\slashed{k}=k^2$ 
and
 \be
\frac{1}{k^2 \left(k^2-M^2\right)} = \frac{1}{M^2}\left(\frac{1}{k^2-M^2} - \frac{1}{k^2}\right)~,
 \ee
 the unparticle diagram can be evaluated as an integral over the 
difference of two simpler diagrams,
 \be
a_\mu^{\mathcal{U}} = \frac{G_Fm_\mu^2}{16\sqrt2 \pi^2}
\left(\frac{\lambda v}{C_\lambda \Lambda^{\Delta_{\mathrm{N}}-3/2}}\right)^2
\frac{A_{\Delta_{\mathrm{N}}-1/2}}{2\pi} \int_{\mu_{IR}^2}^\infty \frac{dM^2}{M^2} \left(M^2-\mu^2_{IR}\right)^{\Delta_{\mathrm{N}}-5/2} \Delta I~,
 \ee
 where $\Delta I \equiv I(M/M_W,m_\mu/M_W)- I(0,m_\mu/M_W)$.
 Again, expanding in the small parameter $m_\mu/M_W$ leads to a 
tractable result
 \bea
a_\mu^{\mathcal{U}} =&& -\sum_{\alpha=1}^{\numN} \frac{G_Fm_\mu^2}{16\sqrt2 \pi^3} 
\left(\frac{|U_{N_\alpha\mu}|^2}{C_\lambda^2}\right) \left(\frac{M_W}{M_{N}}\right)^{2\Delta_{\mathrm{N}}-5}
 \!\! \!\! \!A_{\Delta_{\mathrm{N}}-1/2} \nonumber \\
&&\int_{z_{IR}}^\infty dz\, (z^2-z_{IR}^2)^{\Delta_{\mathrm{N}}-5/2}\left[\frac{2z^7+3z^5-6z^3+z-6z^5 \log(z^2)}{(z^2-1)^4}\right]~,
 \eea
 with $z_{IR}=\mu_{IR}/M_W$ 
and $\Lambda\sim m_N$.  Unfortunately, this is of the wrong sign to 
explain the discrepancy in the muon anomalous magnetic 
moment~\cite{Jegerlehner:2017lbd}. It therefore places a constraint on 
the parameter space of the model. The observed excess in the muon 
anomalous magnetic moment is $\Delta a_{\mu}=(31.3 \pm 7.7) \times 
10^{-10}$. Using the CLs method and requiring that the composite 
neutrino model hypothesis not be excluded at the 95\% confidence level 
as compared to the SM-only hypothesis, we are able to place a bound on 
the hidden sector contribution, $|a_\mu^{\mathcal{U}}|\ltap5\times 
10^{-10}$. This translates into limits on the mixing angle 
$|U_{N\,\mu}|$ shown as the red lines in \figref{fig:loopinduced}.

\subsection{Lepton Flavor Violation}

In general, the couplings of the hidden sector to the SM neutrinos are 
not expected to be flavor aligned and can therefore give rise to flavor 
violating processes involving the charged leptons. The lepton flavor 
violating muon decay $\mu\to e \gamma$ and the $\mu-e$ conversion 
process will be probed with greatly increased precision at the upcoming 
COMET and Mu2e experiments \cite{Adamov:2018vin,Bartoszek:2014mya}. In 
this subsection we determine the hidden sector contributions to these 
processes.

The process $\mu\to e \gamma$ arises from a diagrams of the same form as 
in \figref{fig:diagram}. The resulting amplitude can be evaluated in the 
exactly same way, resulting in a contribution to the branching ratio
given by,
 \bea
BR^{\mathcal{U}}(\mu\rightarrow e \gamma) =&&\left(\frac{3\,\alpha}{128\,\pi^3\, C_\lambda^4}\right)
\frac{\left|\sum_{\alpha=1}^{\numN}U^*_{N_\alpha\mu} U_{N_\alpha e}\right|^2}{M_{N}^{4\Delta_{\mathrm{N}}-6}}\nonumber\\
&&\times
\left(A_{\Delta_{\mathrm{N}}-\frac{1}{2}} \int_{\mu_{IR}^2}^\infty \frac{dM^2}{M^2} \left(M^2-\mu^2_{IR}\right)^{\Delta_{\mathrm{N}}-5/2} \Delta I\right)^2~.
 \eea
 The current limit on this branching ratio from the MEG experiment 
\cite{TheMEG:2016wtm} places a constraint on the matrix product 
$\left|\sum_{\alpha=1}^{\numN}U^*_{N_\alpha\mu} U_{N_\alpha 
e}\right|^2$. This is shown in~\figref{fig:oldlimits} as a bound on 
$|U_{N\ e}|$ alone, under the simplifying assumption 
$\left|\sum_{\alpha=1}^{\numN}U^*_{N_\alpha\mu} U_{N_\alpha e}\right|^2 
=|U_{N\,e}|^2$.

\begin{figure}[t] 
  \centering
  \includegraphics[width=\textwidth]{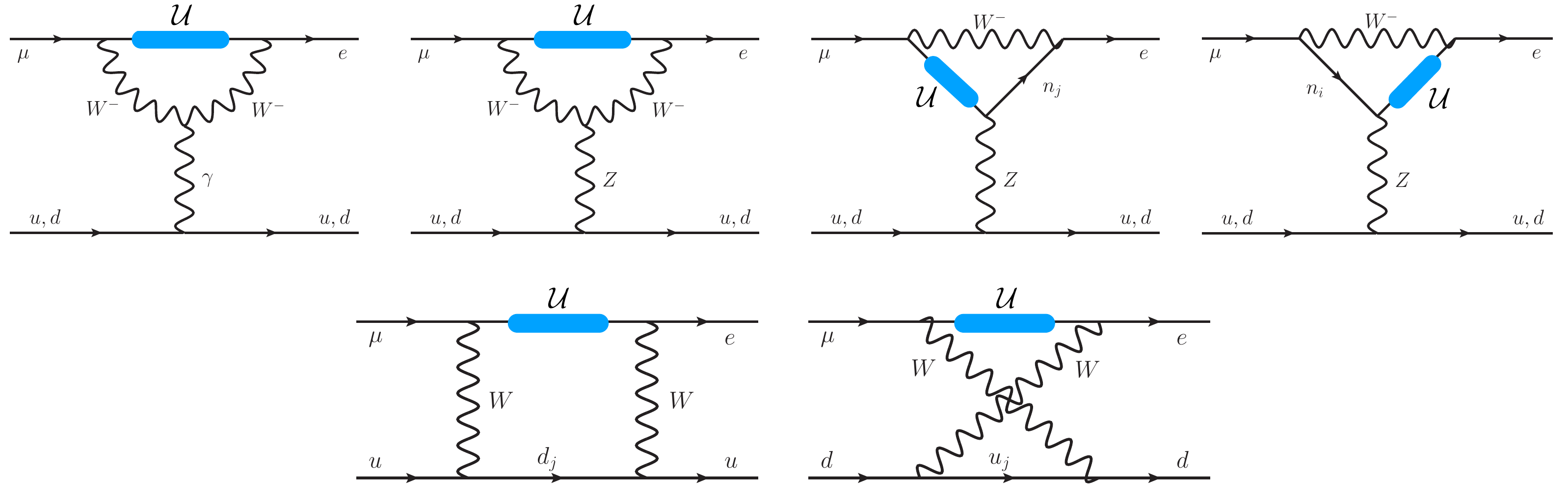} 
  \caption{Loop diagrams contributing to $\mu\to e$ transitions with unparticle neutrino particles in the loop. 
  }
  \label{fig:mu2e-diagrams}
 \end{figure}

There are several one loop diagrams that that contribute to $\mu-e$ 
conversion. They are shown in~\figref{fig:mu2e-diagrams}. They include 
$W$-box, penguin and and $Z$-penguin diagrams.
 As before, we work in the insertion approximation. In this limit the 
amplitudes of the individual diagrams can be related to the 
corresponding amplitudes for the case of elementary HNLs propagating in the loop.  This rate, $R_{\mu\rightarrow e}$, 
has been carefully calculated in \cite{Alonso:2012ji} and is given by,
 \be
R_{\mu\rightarrow e} = \frac{2G_F^2\alpha_W^2}{\pi^2 \Gamma_{capt}}\left(\frac{V^{(p)}}{Z}\right)^2
\left|(A+Z)F_u + (2A-Z)F_d \right|^2~,
 \ee
 for a nucleus of atomic number $Z$, mass number $A$ and muon capture 
rate $\Gamma_{capt}$. The details of the nuclear form factor are encoded 
in $V^{(p)}$.  The functions $F_{u,d}$, given in \cite{Alonso:2012ji}, 
are form factors determined by the one loop diagrams and are a function 
of the mass of the HNL.  There is a value of the neutrino
mass for which the neutron and proton contributions approximately 
cancel, $F_u/F_d= (Z-2A)/(A+Z)\approx -1$, and the rate for muon conversion 
is suppressed, resulting in a drop in sensitivity for that element at 
that mass. As before, for the case of unparticles running in the loop 
we make the replacement,
 \be
F_{u,d}\rightarrow \sum_{\alpha=1}^{\numN} U_{N_\alpha\mu}^*U_{N_\alpha e}\left(\frac{M_{N}}{\sqrt2 M_W C_\lambda}\right)^2\left(\frac{M_W}{\Lambda}\right)^{2\Delta_{\mathrm{N}}-3}\frac{A_{\Delta_{\mathrm{N}}-\frac{1}{2}}}{2\pi}
\int_{x_{IR}}^\infty \!\!\! dx \frac{(x-x_{IR})}{x}^{\Delta_{\mathrm{N}}-5/2}\!\! \! \!F_{u,d}(x)~,
\ee 
 with $x_{IR}=(\mu_{IR}/M_W)^2$. Cancellations between the neutron and 
proton contributions to the rate can still occur in the case of 
composite neutrinos, but at masses that are shifted relative to the case 
of elementary singlet neutrinos.

Currently, the strongest bound on $R_{\mu\rightarrow e}$ comes from the 
SINDRUM II experiment \cite{Bertl:2006up}, which is based on 
measurements on gold. This limit stands at $R_{\mu\rightarrow 
e}^{Au}<7\times 10^{-13}\, (90\%\,\mathrm{CL})$ but is expected to be 
considerably improved by COMET~\cite{Adamov:2018vin} and 
Mu2e~\cite{Bartoszek:2014mya,Bernstein:2019fyh} to the level of 
$R_{\mu\rightarrow e}^{Al}\ltap10^{-16}$.  In \figref{fig:loopinduced} 
we show the current constraints from $\mu\to e\gamma$ and $\mu$-$e$ 
conversion in green and blue lines, respectively, over a broad range of 
composite singlet neutrino masses $M_N$. These bounds are prsented as 
limits on $|U_{N\ \mu}|$, under the simplifying assumption 
$\left|\sum_{\alpha=1}^{\numN}U^*_{N_\alpha\mu} U_{N_\alpha e}\right|^2 
=|U_{N\,\mu}|^2$. We can see from the figure the complementarity between 
these two types of lepton flavor violating experiments.

In Figures~\ref{fig:oldlimits} and~\ref{fig:futureprojections} we 
present the current and future constraints on this class of models, with 
the limits from the lepton flavor violating processes $\mu\to e\gamma$ 
and $\mu$-$e$ conversion included. The bounds are presented as limits on 
$U_{N\,e}$, under the assumption 
$\left|\sum_{\alpha=1}^{\numN}U^*_{N_\alpha\mu} U_{N_\alpha e}\right|^2 
=|U_{N\, e}|^2$. We see that in the limit of large mixing, $\mu-e$ 
conversion currently gives the strongest bound on this class of models 
for $M_N\gtap 70\ \mathrm{GeV}$. In the future, assuming large mixing, 
Mu2e will provide greater sensitivity than the current limits for 
compositeness scales above a few GeV and will have the greatest reach of 
all future experiments above 50 GeV. Future collider and beam dump 
experiments are expected to have greater sensitivity for $M_N$ below 50 
GeV, showing the complementarity of these different types of searches. 
In the event of a discovery, multiple measurements from different 
experiments may help uncover the nature of the underlying theory.


\section{Neutrinoless Double Beta Decay \label{s.beta}}

 In this section we consider the implications of this class of models 
for neutrinoless double beta decay. At scales below the compositeness 
scale $\Lambda$, the spectrum of light states in the neutrino sector 
consists of just the SM neutrinos and antineutrinos, which now possess 
small Majorana masses from the nonrenormalizable Weinberg operator. The 
characteristic momentum scale in neutrinoless double beta decay is of 
order 100 MeV, which corresponds to the spacing between nucleons in the 
nucleus. Since the Weinberg operator is the lowest dimension term that 
can be added to the SM which violates lepton number, it follows that if 
the compositeness scale $\Lambda$ lies well above 100 MeV, the rate for 
neutrinoless double beta decay is determined by the mass matrix of the 
light neutrinos in the low energy effective theory, just as in 
conventional high scale seesaw models. However, if $\Lambda$ lies below 
a 100 MeV, the situation is more complicated, and the effects of the 
strong dynamics must be taken into account when calculating the rate for 
neutrinoless double beta decay.

Before determining the rate for neutrinoless double beta decay in this 
scenario, it is useful to first consider this process in the context of 
inverse seesaw models. This has been studied in, for example, 
\cite{LopezPavon:2012zg,Bolton:2019pcu}. In this framework the rate for 
neutrinoless double beta decay depends on the lepton number violating 
contribution to the off-shell neutrino propagator. As can be seen from 
the Lagrangian for the inverse seesaw, \eref{L_IR}, this depends on a 
matrix element of the form,
 \begin{equation}
  \label{matrixelement1a}
  \mathcal{M}_N = \langle \nu(p) | T\left[
  \lambda L(x) H(x) N(x) \;
  \frac{\mu^c}{2} \left(N^c(y)\right)^2 \;
  \lambda L(z) H(z) N(z) 
  \right] | \nu(p) \rangle \;,
 \end{equation}
 where we are treating the lepton number violating parameter $\mu^c$ in 
the insertion approximation. Setting the Higgs to its VEV, we see that 
for $M_N \gg p$ this matrix element scales as
 \begin{equation}
\label{melement1a}
\mathcal{M}_N\sim \frac{\lambda^2 v^2_{\rm EW} \mu^c}{M_N^2} = m_{\nu} \; . 
 \end{equation}
 Therefore, for $M_N$ much greater than 100 MeV, the rate for 
neutrinoless double beta decay in inverse seesaw models is set by the 
masses of the light neutrinos, exactly as expected from effective field 
theory considerations.

If a term of the form shown in \eref{othermassN2} is added to the 
inverse seesaw model, there is an additional contribution to 
neutrinoless double beta decay proportional to the lepton number 
violating parameter $\mu$. The corresponding matrix element takes the 
form,
 \begin{equation}
  \label{matrixelement1b}
\widetilde{\mathcal{M}}_N = \langle \nu(p) | T\left[
  \lambda L(x) H(x) N(x) \;
  \frac{\mu}{2} {N^\dagger}^2(y) \;
  \lambda L(z) H(z) N(z)
  \right] | \nu(p) \rangle \;,
 \end{equation}
 where, as before, we are treating $\mu$ in the insertion approximation. 
Setting the Higgs to its vacuum expectation value (VEV) we see that for 
$M_N \gg p$ this matrix element scales as
 \begin{equation}
\label{melement1b}
\widetilde{\mathcal{M}}_N \sim \frac{\lambda^2 v^2_{\rm EW} \mu}{M_N^2} \frac{p^2}{M_N^2} =
\frac {\mu} {\mu^c} \frac{p^2}{M_N^2} m_{\nu}.
 \end{equation}
 For $\mu \sim \mu^c$ and $M_N \gg p$, we see that this contribution to 
neutrinoless double beta decay is suppressed compared to that in 
\eref{melement1a}. This is exactly as expected from effective field 
theory considerations, since the contribution in \eref{melement1b} 
must necessarily arise from an operator of higher dimension than the 
Weinberg operator.

We now turn our attention to the opposite limit, $M_N \ll p$. The matrix 
element in \eref{matrixelement1a} now scales as,
 \begin{equation}
 \label{melement2a}
\mathcal{M}_N\sim \frac{\lambda^2 v^2_{\rm EW} M_N^2 \mu^c}{p^4} = 
m_{\nu} \left(\frac{M_N}{p} \right)^4 \;.
 \end{equation}
 Comparing this to the matrix element in \eref{melement1a}, it follows 
that in inverse seesaw models, if the mass of the singlet neutrinos 
lies below 100 MeV, the rate for neutrinoless double beta decay is 
suppressed. If a term of the form shown in \eref{matrixelement1b} is 
also present, the matrix element in \eref{matrixelement1b}, in the 
$m_N\ll p$ limit, provides an additional contribution that scales as,
 \begin{equation}
 \label{melement2b}
  \widetilde{\mathcal{M}}_N \sim \frac{\lambda^2 v^2_{\rm EW} \mu}{p^2} = 
  m_{\nu} \frac {\mu} {\mu^c} \left(\frac{M_N}{p} \right)^2 \;.
 \end{equation}
 For $\mu \sim \mu^c$ and $M_N \ll p$, we see that this contribution to 
neutrinoless double beta decay, although also suppressed, is larger than 
that in \eref{melement2a}. This example serves to illustrate that if the 
scale at which the neutrino masses are generated lies at or below 100 
MeV, the rate of neutrinoless double beta decay need not be controlled 
by the Weinberg operator.

We now consider the scenario in which the singlet neutrinos are 
composite. Both the Majorana masses of the light neutrinos and the rate 
for neutrinoless double beta decay arise from the lepton number 
violating contributions to the neutrino propagator from the composite 
sector. These are controlled by the matrix element,
 \begin{equation}
\label{matrixelement2}
\mathcal{M}_{\mathcal{U}} = \langle \nu(p) | T \left[
\hat{\lambda} L(x) H(x) \mathcal{O}_N(x) \;
{\hat{\mu}^c} \mathcal{O}_{2N^c}(y) \;
{\hat{\lambda}} L(z) H(z) \mathcal{O}_N(z)
\right] | \nu(p) \rangle \;.
 \end{equation}
 If the compositeness scale $\Lambda$ lies well above 100 MeV, both the 
neutrino masses and neutrinoless double beta decay depend on the value 
of this matrix element evaluated at momenta $p \ll \Lambda$.  At 
energies below the compositeness scale $\Lambda$, the matrix element in 
\eref{matrixelement2} will generate nonrenormalizable lepton number 
violating interactions among the SM fields. Since the Weinberg operator 
is the unique lowest dimension lepton number violating operator, 
effective field theory dictates that the rate for neutrinoless double 
beta decay will be determined by the form of the mass matrix for the 
light neutrinos.

If the scale $\Lambda$ lies below 100 MeV, the situation is more 
complicated. While the neutrino masses still depend on the value of the 
matrix element in \eref{matrixelement2} at scales $p \ll \Lambda$, 
the rate for neutrinoless double beta decay is now controlled by scales 
$p \gg \Lambda$. We can estimate the rate for neutrinoless double beta 
decay by evaluating the matrix element in \eref{matrixelement2} 
using the general expression for the fermion-fermion-scalar 3-point 
function in a CFT \cite{Todorovetal,Fradkin:1996is},
 \begin{equation}
\langle \mathcal{O}_{\rm N}(x_1) \mathcal{O}_{\rm N}(x_2) \mathcal{O}_{\rm 2N^c}(x_3) \rangle
= {\bf{\{ I + II \} }}
 \end{equation}
 where
 \begin{equation}
{\bf{ I }} =
C_1 
\frac{ 
\left(\slashed{x}_1 - \slashed{x}_2 \right) 
}
{
\left(x_{13}^2\right)^{\Delta_{\rm 2N^c}/2}
\left(x_{23}^2\right)^{\Delta_{\rm 2N^c}/2}
\left(x_{12}^2\right)^{\left(2\Delta_{\rm N} - \Delta_{\rm 2N^c} + 1 \right)/2}
}
 \end{equation}
and 
 \begin{equation}
{\bf{ II }} =
C_2
\frac{
\left(\slashed{x}_1 - \slashed{x}_3 \right) \left(\slashed{x}_3 - \slashed{x}_2 \right)
}
{
\left(x_{13}^2\right)^{\left(\Delta_{\rm 2N^c} + 1 \right)/2}
\left(x_{23}^2\right)^{\left(\Delta_{\rm 2N^c} + 1 \right)/2}
\left(x_{12}^2\right)^{\left(2\Delta_{\rm N} - \Delta_{\rm 2N^c} \right)/2}
}
 \end{equation}
 Here $C_1$ and $C_2$ are undetermined coefficients that depend on the 
details of the CFT and $\slashed{x} \equiv \bar{\sigma}^\mu x_\mu$. 
Based on this, we see that the corrections to the neutrino propagator 
from \eref{matrixelement2} are either ultraviolet divergent or 
infrared divergent, depending on the scaling dimensions of 
$\mathcal{O}_{\rm N}$ and $\mathcal{O}_{\rm 2N^c}$. However, if the 
contributions to the neutrino mass are to be finite in the ultraviolet, 
we require $2\Delta_{\rm N}+\Delta_{\rm 2N^c} \le 8$. Then, using the fact that 
the neutrino mass is generated at scales of order $\Lambda$ to estimate 
the prefactor, we find that the matrix element in 
\eref{matrixelement2} scales as
 \begin{equation}
\mathcal{M}_{\mathcal{U}} \sim m_{\nu} 
\left( \frac{\Lambda^2}{p^2} \right)^{4 - \Delta_{\rm N} - \Delta_{\rm 2N^c}/2}
 \end{equation} 
 Since $2\Delta_{\rm N} +\Delta_{\rm 2N^c} \le 8$, we see that the 
contribution to neutrinoless double beta decay from scales above 
$\Lambda$ are suppressed.

It follows from this discussion that for $\Lambda \gg 100\,\mev$, the rate for neutrinoless double beta decay is determined by the 
neutrino mass matrix, just as in conventional high scale seesaw models. 
However, for $\Lambda \lsim 100\,\mev$, the rate for 
neutrinoless double beta decay is suppressed by form factors, and may be
below the reach of next generation experiments.  

\section{Cosmology\label{s.cosmo}}

In this section we explore the cosmological history of this class of
models. We determine the limits that cosmology places on the parameter
space of these theories and show that this scenario can lead to
observable signals in the CMB.

In the early universe the hidden sector is in thermal contact with the
SM, which populates the states in the hidden sector.
We find that for $\Lambda \lesssim 1$ GeV, the couplings between the SM
and the hidden sector are large enough that the two sectors are
necessarily in thermal equilibrium at the compositeness scale. At
temperatures below the compositeness scale, the composite singlet
neutrinos annihilate away efficiently into light neutrinos, leaving
behind only an exponentially suppressed relic population. These relic
singlet neutrinos eventually decay into final states that may include
charged leptons. For compositeness scales below about 50 MeV, we find
that these decays may occur late enough that they happen after the
photons in the SM thermal bath have gone out of chemical equilibrium. This
results in distortions to the CMB that may be large enough to be
observed in future experiments.

Cosmological observations can be used to place constraints on the 
compositeness scale $\Lambda$. Precision measurements of the CMB have 
established that the neutrinos are free streaming at temperatures of 
order an 
eV~\cite{Archidiacono:2013dua,Forastieri:2015paa,Blinov:2020hmc}. For 
values of $\Lambda$ less than 100 eV, the partially composite character 
of the neutrinos leads to neutrino self-interactions that are too large 
to admit free streaming. Therefore these observations disfavor 
compositeness scales $\Lambda$ below 100 eV. In addition, the CMB also 
places tight limits on the total energy density in radiation during the 
epoch of acoustic oscillations~\cite{Aghanim:2018eyx} that can be 
translated into a lower bound on the compositeness scale. These bounds 
are usually expressed as limits on the effective number of neutrinos, 
$N_{\rm eff}$. We find that the CMB bounds on $N_{\rm eff}$ disfavor 
scenarios in which the hidden sector is in equilibrium with the 
neutrinos at temperatures below an MeV, the scale at which the neutrinos 
decouple from the rest of the SM. Since the neutrinos are in equilibrium 
with the hidden sector at temperatures of order the compositeness scale 
for any $\Lambda \lesssim 1$ GeV, values of $\Lambda$ below an MeV are 
disfavored.

\subsection{Cosmological History}

In the early universe there are several processes that can serve to
populate the hidden sector. These include $\nu\, \nu \leftrightarrow
\mathcal{U}$, $\nu\, \mathcal{U} \leftrightarrow \mathcal{U}$ and $\nu\,
\mathcal{U} \leftrightarrow \nu\, \mathcal{U}$. We shall focus on the
process $\nu \nu \rightarrow \mathcal{U}$, which does not require an
initial population of unparticles.

We can estimate the inclusive rate for $\nu \nu \rightarrow \mathcal{U}$ 
by noting that, at momentum scales of order the compositeness scale 
$\Lambda$, the cross section for this process is expected to be of 
roughly the same size as the cross section for $\nu \nu \rightarrow N 
N$. This allows us to estimate the cross section for the process $\nu 
\nu \rightarrow \mathcal{U}$ at temperatures $T$ of order $\Lambda$ as
 \begin{equation} 
\sigma_{\nu \nu \rightarrow \mathcal{U}} \sim 
\sigma_{\nu \nu \rightarrow N N} \sim 
\frac{1}{4 \pi}\left(\frac{\kappa}{\Lambda^2}\right)^2 
\left(\frac{\lambda v_{\rm EW}}{M_N}\right)^4 T^2~. 
\label{sigma(nu,nu-to-U)} 
 \end{equation} 
 Here the parameter $\kappa\sim 16\pi^2$. Since the initial state only 
involves SM neutrinos, which are ultrarelativistic, the rate for this 
process can be estimated as, 
 \begin{equation} 
n_\nu \langle \sigma_{\nu \nu \rightarrow 
\mathcal{U}} v\rangle \sim \frac{\kappa^2}{4 \pi} 
\left(\frac{\lambda v_{\rm EW}}{M_N}\right)^4 \frac{T^5}{\Lambda^4}~. 
 \label{rate(nu,nu-to-U)} 
 \end{equation} 
 This process will ensure that the hidden sector is in equilibrium with 
the SM neutrinos at temperatures of order the compositeness scale 
provided the rate is faster than the Hubble expansion rate, $H \sim T^2/ 
M_{\rm Pl}$, when $T \sim \Lambda$. Setting $\lambda$ to its lower bound 
in \eref{lambdabound}, taking $m_\nu \sim 0.01$ eV and recalling that 
$M_N \sim \Lambda$, we see that this process will necessarily bring the 
two sectors into equilibrium provided $\Lambda \lesssim 1$ GeV. For 
values of $\Lambda \gtap 1$ GeV this process by itself may not be 
sufficient to bring the two sectors in equilibrium below the confinement 
scale. Nevertheless, it will partially populate the CFT states. However, 
for these higher values of $\Lambda$ the composite singlet neutrinos 
decay prior to neutrino decoupling, as can be seen from 
Eq.~(\ref{eq:lifetime}), and therefore do not leave any observable 
cosmological effects.

We expect that the hidden sector transitions to the hadronic phase at a 
temperature $T$ of order the compositeness scale $\Lambda$. Once the 
temperature falls below the compositeness scale, the hadrons in the 
hidden sector begin to go out of the bath. The decay of the composite 
singlet neutrinos is relatively slow, and so the reduction in their 
number density occurs primarily through annihilation to SM particles. 
When the strongly coupled sector is in the hadronic phase, the dominant 
interactions between the SM and hidden sector arise from the mixing 
between the SM and composite neutrinos in \eref{intLHN}, when combined 
with the self-interactions of the $N$, \eref{int4N}. As emphasized 
earlier, the large self-interactions between the singlet neutrinos are 
characteristic of this framework and distinguish it from conventional 
inverse seesaw models. These give rise to not just the conventional 
annihilation process $N N\rightarrow \nu \nu$, but also 
semi-annihilation $N N\rightarrow N \nu$, and co-annihilation $\nu 
N\rightarrow \nu \nu$. Since the mixing between the SM and composite 
neutrinos is small, $\mathcal{O}\left(\lambda 
v_{\mathrm{EW}}/M_N\right)$, while the dimension-6 self-couplings, which 
are of order $\mathcal{O}\left(16\pi^2/ \Lambda^2\right)$, are sizable 
at temperatures $T$ of order $\Lambda$, the cross sections for these 
processes diminish with the number of external $\nu$ legs. The 
conventional annihilation process begins from the same initial state as 
semi-annihilation, but involves an additional $\nu$ in the final state. 
It is therefore mixing-suppressed and subdominant relative to 
semi-annihilation. The co-annihilation process, although suppressed by 
even more powers of $\nu-N$ mixing, can take advantage of the background 
bath of relativistic SM neutrinos. However, for $\Lambda \gtrsim 1$ MeV and mixings allowed by existing constraints (see Figure~\ref{fig:oldlimits}), 
this is not sufficient to overcome the suppression from mixing.

We therefore focus on the depletion of composite singlet neutrinos 
through the semi-annihilation process, $N N \rightarrow N \nu$. We can 
obtain an estimate of the relic density of composite singlet neutrinos, 
which we denote by $n_F$, by equating the annihilation rate at freeze 
out to the Hubble expansion rate, 
 \begin{equation} 
n_F \langle \sigma_{NN \rightarrow \nu N} v \rangle|_{T_F} = H|_{T_F} \;. 
 \end{equation} 
 Here $T_F$ denotes the temperature at freeze out. We 
estimate the annihilation rate as, 
 \begin{equation} 
\langle \sigma_{NN \rightarrow N \nu} v \rangle|_{T_F} 
\sim \frac{\kappa^2}{4 \pi} \frac{\lambda^2 v_{\rm EW}^2}{\Lambda^4} 
\gtap \frac{\kappa^2}{4 \pi} \frac{m_\nu}{M_N^3}~. 
 \end{equation} 
 To obtain the inequality we have set $\lambda$ to its lower bound in 
Eq.~(\ref{lambdabound}) and used the fact that $M_N \sim \Lambda$. The 
non-relativistic number density of $N$ at freeze out is of order 
 \begin{equation} 
 n_F \sim \left(M_N T_F\right)^{3/2} e^{-M_N/T_F}~. 
 \end{equation} 
 From this we find that $T_F \lesssim M_N/15$, for all $M_N$ below 100 
GeV. This shows that the number density of relic composite singlet 
neutrinos is exponentially suppressed at temperatures below the 
compositeness scale.  However, as we shall see later, these may still 
give rise to observable effects.

\subsection{Cosmological Bounds on the Compositeness Scale}

Cosmological observations can be used to place constraints on the 
compositeness scale $\Lambda$. There are two effects that we need to 
consider.
 \begin{itemize}
 \item
 The sizable self-interactions that arise from the partial compositeness 
of neutrinos can prevent the neutrinos from free streaming during the 
CMB era. This is disfavored by CMB measurements, leading to a lower 
bound on $\Lambda$.
 \item
 The CMB also places bounds on the total energy density in radiation 
during the epoch of acoustic oscillations. These can be translated into
a lower bound on the compositeness scale $\Lambda$.
 \end{itemize}
 We now consider these effects in turn.

\subsubsection*{Limits on Neutrino Free Streaming}

Precision measurements of the CMB have established that the neutrinos 
are free streaming during the epoch of acoustic 
oscillations~\cite{Archidiacono:2013dua,Forastieri:2015paa,Blinov:2020hmc}. 
This can be used to place limits on the size of neutrino 
self-interactions. In particular, the mean free path of the neutrinos 
must be greater than the size of the universe at temperatures of order 
an eV. In our framework, the cross section for elastic neutrino-neutrino 
scattering at temperatures below the compositeness scale $\Lambda$ is of 
order,
 \begin{equation}
\sigma_{\nu \nu \rightarrow \nu \nu} \sim
\frac{\kappa^2}{4 \pi} \left(\frac{\lambda v_{\rm EW}}{m_N}\right)^8
\frac{T^2}{\Lambda^4}
 \end{equation}
Then the inverse of the mean free path for neutrino scattering is of order,
\begin{equation}
n_\nu \sigma_{\nu \nu \rightarrow \nu \nu} \sim
\frac{\kappa^2}{4 \pi} \left(\frac{\lambda v_{\rm EW}}{m_N}\right)^8
\frac{T^5}{\Lambda^4}
\gtrsim \frac{\kappa^2}{4 \pi} \left(\frac{m_\nu}{m_N}\right)^4
\frac{T^5}{\Lambda^4} \; ,
\end{equation}
 where the inequality is obtained by setting $\lambda$ to its lower 
bound in Eq.~(\ref{lambdabound}). To satisfy the constraint on free 
streaming this must be less than the Hubble scale, $H = g_* T^2/M_{\rm 
Pl}$, at temperatures of order an eV. This translates into a lower bound 
on the compositeness scale in this scenario, $\Lambda \gtrsim 100$ eV.

\subsubsection*{Limits on the Energy Density in Radiation}

The predictions of Big Bang Nucleosynthesis (BBN) in the SM are in 
excellent agreement with data~\cite{Cyburt:2015mya}. This can be used 
to place a strong limit on any additional energy density in radiation 
from physics beyond the SM at temperatures of order an MeV, the time at 
which the neutrinos decouple from the rest of the SM. This bound implies 
that the hidden sector cannot be in equilibrium with the SM at 
temperatures of order an MeV. This constraint is automatically satisfied 
if the compositeness scale lies above an MeV, so that the hadrons of the 
strongly coupled sector exit the bath prior to BBN. In principle, the 
bound can also be satisfied if the compositeness scale lies below an 
MeV, provided that the hidden sector is initially at a lower temperature 
than the SM and comes into equilibrium with the neutrinos only after 
BBN.

There are also strong limits on the total energy density in radiation 
during the epoch of acoustic oscillations based on precision 
measurements of the CMB~\cite{Aghanim:2018eyx}. The current upper bound 
stands at $N_{\rm eff} \lesssim 3.5$. As we now explain, this constraint 
disfavors compositeness scales $\Lambda$ less than an MeV, even in the 
case that the hidden sector comes into equilibrium with the neutrinos 
only at temperatures below an MeV. The SM neutrinos decouple from the 
rest of the SM bath at temperatures of order an MeV. If the 
compositeness scale lies below an MeV, then the only SM particles that 
the composite sector will be in equilibrium with at temperatures of 
order $\Lambda$ are the neutrinos. When the strongly coupled sector 
comes into equilibrium with the neutrinos, the comoving energy density is 
conserved, while the comoving entropy density increases 
\cite{Chacko:2003dt,Chacko:2004cz,Berlin:2017ftj}. However, when the 
composite states go out of the bath, the comoving entropy density is 
conserved, while the comoving energy density increases. Therefore, if 
the compositeness scale lies below an MeV, we expect an increase in the 
total comoving energy density in the light neutrinos at the time of CMB 
over the SM prediction. The size of this effect depends on the number of 
degrees of freedom in the hidden sector. Unless this number is very 
small (less than the equivalent of 2 Weyl fermion degrees of freedom), 
this class of theories is disfavored by the current CMB bound on $N_{\rm 
eff}$. We therefore focus our attention on the regime $\Lambda \gtrsim 
1$ MeV.

\subsection{Spectral Distortions of the CMB From Late Decays}

Through their mixing with the SM neutrinos, the relic composite singlet
neutrinos eventually decay through the weak interactions into final
states composed of SM particles. The decay width of the singlet
neutrinos can be estimated from~\eref{eq:lifetime},
 \begin{equation}
\Gamma_N \sim
\frac{G_F^2}{192 \pi^3}
\left(\frac{\lambda v_{\rm EW}}{M_N}\right)^2 M_N^5
\gtrsim
\frac{G_F^2 M_N^4}{192 \pi^3} m_\nu ~,
\end{equation}
 where the inequality is obtained by setting $\lambda$ to its lower 
bound in \eref{lambdabound}. For $M_N \lesssim 1$ GeV the composite 
singlet neutrinos decay when the temperature of the SM bath is below an 
MeV. This could potentially affect the successful predictions of BBN by dissociating nuclei. However, our 
calculation of the relic abundance of the composite singlet neutrinos 
shows that their number density is too small to have an observable 
effect on BBN.

For $M_N \lesssim 50$ MeV and $\lambda$ set to its minimum value, the 
relic singlet neutrinos decay at even later times, when the temperature 
of the bath is at or below a keV. One of the primary decay channels is 
$N \rightarrow e^+ e^- \nu$. The resulting injection of electromagnetic 
energy into the bath at these late times gives rise to spectral 
distortions in the CMB. Although photon-number conserving processes are 
still active at these low temperatures, photon-number changing processes 
such as double Compton scattering are comparatively slow. Any injection 
of electromagnetic energy at this time will manifest itself in a 
$\upmu$-distortion of the CMB spectrum.\footnote{Due to the letter $\mu$ has been taken as a Lagrangian parameter in this model, here we use $\upmu$ to represent the $\mu$-distortion of the CMB spectrum.} For an electromagnetic decay 
fraction $\mathfrak{f}$, the shift in $\upmu$ is approximately given by 
\cite{Hu:1993gc,Chluba:2011hw}, 
 \be 
\upmu \approx 8\times 10^2 
\left(\frac{\tau_N}{1\mathrm{sec}}\right)^{1/2} 
\left(\frac{m_N}{\mathrm{GeV}}\right) 
\left[\frac{n_N(\tau_N)}{n_\gamma(\tau_N)}\right] 
g\left(\frac{\tau_{\mathrm{dC}}}{\tau_N}\right) 
\mathfrak{f}~, \ee 
 with 
 \be 
g(x) \approx \begin{cases} 
e^{-x^{5/4}} &  x\ltap 1
 \\ 
2.1 x^{5/9} e^{-2x^{5/9}} & x\gtap 1 
 \end{cases} ~. 
 \ee 
 Here the double Compton time scale is given by 
$\tau_{\mathrm{dC}}\approx 7\times 10^6\,\mathrm{sec}$. Since $N 
\rightarrow e^+ e^- \nu$ is one of the primary decay modes, 
$\mathfrak{f}\sim 1$. For $\Lambda \lesssim 50$ MeV, these effects can 
be as large as $\upmu\sim 10^{-8}$. Although consistent with current 
observations, a spectral distortion of this size is large enough to be 
seen at next generation CMB experiments such as 
PIXIE~\cite{Kogut:2011xw}. For a given value of $M_N$, increasing 
$\lambda$ above its lower bound in \eref{lambdabound} will hasten the 
decay of the composite singlet neutrinos and reduce the 
$\upmu$-distortion. Therefore, in contrast to the other signals we have 
studied, spectral distortions are more sensitive to lower values of the 
mixing angle.

\section{Astrophysics\label{s.astro}}

We now turn our attention to the limits on the compositeness scale
$\Lambda$ from astrophysical considerations. We expect that the effects
of neutrino compositeness are only important in astrophysical scenarios
where either the temperature or the density are comparable to $\Lambda$.
Since the cosmological bounds already constrain $\Lambda \gtrsim 1$ MeV,
which is many orders of magnitude higher than the temperature in a
typical star, we expect that neutrino compositeness will not have a
significant impact on stellar dynamics. However, the temperature $T_{\rm
SN}$ at the core of a supernova is of order 40 MeV. Then, if the
compositeness scale is low enough, composite singlet neutrinos can be
produced in the core, and could potentially affect the supernova
dynamics. In particular, the emission of composite singlet neutrinos now
offers a new mechanism for the core to lose energy. If this energy loss
occurs too rapidly, the explosion cannot take place. In the remainder of
this section, we investigate this question further. We will first
consider the regime $\Lambda \gg T_{\rm SN}$, corresponding to a
compositeness scale well above the supernova temperature, and discuss
the limit $\Lambda \lesssim T_{\rm SN}$ later.

We begin by determining the conditions under which any composite singlet 
neutrinos that are produced are able to escape freely from the 
supernova. The scattering processes $\nu N \rightarrow NN$ and $\nu N 
\rightarrow \nu N$ act to prevent the composite neutrinos from escaping. 
The requirement that singlet neutrinos are trapped in the supernova by 
the $\nu N \rightarrow NN$ process corresponds to the condition,
 \begin{equation}
n_\nu\langle \sigma_{\nu N \rightarrow NN}\rangle L_{\rm SN} \gg 1 \;,
\label{NNtrap}
 \end{equation} 
 where $L_{\rm SN} \sim $ 10 km is the size of the supernova and 
$n_\nu$, the number density of neutrinos inside the core, can be 
estimated as $T_{\rm SN}^3$. The corresponding condition in the case of 
the $\nu N \rightarrow \nu N$ process takes the form,
 \begin{equation}
n_\nu\langle \sigma_{\nu N \rightarrow \nu N}\rangle L_{\rm SN} \gg 1 \;,
\label{nuNtrap}
 \end{equation}
 The inverse of the mean free path for the process $\nu N \rightarrow 
NN$ at a temperature $T_{\rm SN}$ can be estimated as,
 \begin{equation} 
n_\nu \langle \sigma_{\nu N \rightarrow N N}\rangle \approx 
\frac{\kappa^2}{4 \pi} \left(\frac{\lambda v_{\rm EW}}{M_N}\right)^2 
\frac{M_N^2 T_{\rm SN}^3}{\Lambda^4} {\rm exp}\left(-M_N/T_{\rm SN}\right)~, 
 \label{Nescape1} 
 \end{equation}
 where the exponential suppression arises because we are in the region 
of parameter space where the mass splitting between the initial and 
final states is greater than the temperature in the supernova core. 
Setting $\lambda$ to its lower bound in Eq.~(\ref{lambdabound}), we find 
that this process prevents the composite singlet neutrinos from escaping 
provided that the compositeness scale $\Lambda \lesssim$ 800 MeV. 
Conversely, if we set $\lambda$ to its upper bound in 
Eq.~(\ref{lambdabound}), we find that the composite singlet neutrinos 
remain trapped in the supernova for all compositeness scales $\Lambda 
\lesssim 1.7$ GeV. Turning our attention to the process $\nu N 
\rightarrow \nu N$, the inverse of the mean free path can be estimated 
as,
 \be
n_\nu \langle \sigma_{\nu N \rightarrow \nu N}\rangle \approx
\frac{\kappa^2}{4 \pi} \left(\frac{\lambda v_{\rm EW}}{M_N}\right)^4
\frac{M_N^2 T_{\rm SN}^3}{\Lambda^4} ~.
\label{Nescape2}
 \ee
 For $\lambda$ at its lower bound, we find that this process prevents 
the composite singlet neutrinos from escaping for all $\Lambda \lesssim$ 
300 MeV. For $\lambda$ at its upper bound, the corresponding limit on 
$\Lambda$ is $10^9$ GeV.

It follows from this discussion that for all $\Lambda$ below 800 MeV, 
the composite singlet neutrinos will be trapped in the supernova by the 
$\nu N \rightarrow N N$ process. For larger values of $\Lambda$, the 
composite singlet neutrinos may still be prevented from escaping but 
only if the coupling $\lambda$ is large enough. For smaller couplings of 
$\lambda$ any composite singlet neutrinos that are produced will free 
stream out of the supernova.

The scattering process $\nu N \rightarrow NN$ has the effect of 
increasing the number density of composite singlet neutrinos in the 
supernova. Therefore, in the regime that Eq.~(\ref{NNtrap}) is 
satisfied, their number density will increase exponentially quickly 
provided that a few seed composite singlet neutrinos are initially 
present. This will have the effect of bringing the composite singlet 
neutrinos into thermal and chemical equilibrium inside the core of the 
supernova. However, because their equilibrium number density is 
suppressed as ${\rm exp}\left(-M_N/T_{\rm SN}\right)$, we expect that 
their impact on supernova dynamics is limited. If Eq.~(\ref{NNtrap}) is 
not satisfied, the $N$-number changing process is slow. However, for 
large enough $\lambda$, Eq.~(\ref{nuNtrap}) may be satisfied so that the 
composite singlet neutrinos are trapped by the $\nu N \rightarrow \nu N$ 
process. In this regime, the composite singlet neutrinos would reach 
kinetic but not chemical equilibrium. For smaller values of $\lambda$, 
neither Eq.~(\ref{NNtrap}) nor Eq.~(\ref{nuNtrap}) is satisfied, and any 
composite singlet neutrinos that are produced will free stream out of 
the supernova. However, as we now argue, in this regime their production 
rate is exponentially suppressed and the resulting energy loss is too 
small to affect the supernova dynamics.

There are two processes that seed the initial production of composite
singlet neutrinos inside the supernova, $\nu \nu \rightarrow NN$ and
$\nu \nu \rightarrow N \nu$. The first process requires the total
energy in the incoming neutrinos to be at least $2 M_N$ but the
amplitude is suppressed by only two powers of the $\nu-N$ mixing. For
the second process, the total energy in the initial state particles need
only be $M_N$, but the amplitude is now suppressed by an additional
power of $\nu-N$ mixing. The production of the seed $N$ states must
occur on the timescales shorter than the lifetime of a supernova
($t_{\mathrm{SN}}\sim 10$ sec). The number of seed composite singlet
neutrinos produced through the process $\nu \nu \rightarrow NN$ over the
lifetime of the supernova can be estimated as,
 \be
\frac{\kappa^2}{4 \pi} \left(\frac{\lambda v_{\rm EW}}{M_N}\right)^4
\frac{M_N^2}{\Lambda^4}
\,e^{-2 M_N/T_{\rm SN}}\left(T_{\rm SN}^2 L_{\rm SN}\right)^3 t_{\rm SN}~.
 \label{seedNN}
 \ee
 Setting $\lambda$ to its lower bound in Eq.~(\ref{lambdabound}), we 
find that at least a few composite singlet neutrinos are produced in the 
supernova through this process provided $\Lambda \lesssim$ 2.5 GeV.  If 
$\lambda$ is at its upper bound, the corresponding upper bound on the 
compositeness scale is $\Lambda \lesssim$ 3 GeV. Turning our attention 
to the process $\nu \nu \rightarrow N \nu$, the number of seed composite 
singlet neutrinos produced can be estimated as,
 \be
\frac{\kappa^2}{4 \pi} \left(\frac{\lambda v_{\rm EW}}{M_N}\right)^6
\frac{M_N^2}{\Lambda^4}\,
e^{-M_N/T_{\rm SN}}\left(T_{\rm SN}^2 L_{\rm SN}\right)^3 t_{\rm SN}~.
 \label{seednuN}
 \ee
 For $\lambda$ at its lower bound in Eq.~(\ref{lambdabound}), we find 
that at least a few seed composite singlet neutrinos will be produced 
through this process in the supernova core on the relevant timescale 
provided $\Lambda \ltap 4\,\mathrm{GeV}$. For $\lambda$ at its upper 
bound, the corresponding limit is $\Lambda \ltap 5\,\mathrm{GeV}$.

In summary, when $\lambda$ is at its lower bound in 
Eq.~(\ref{lambdabound}), some number of composite singlet neutrinos are 
produced inside the supernova core provided $\Lambda \lesssim 4\, {\rm 
GeV}$. For $\Lambda \lesssim$ 800 MeV, these particles are trapped 
inside the supernova, and come into thermal and chemical equilibrium. 
For $\Lambda$ in the range $800\, {\rm MeV} \lesssim \Lambda \lesssim 
4\, {\rm GeV}$, the composite singlet neutrinos are able to stream 
freely out of the supernova. However, from Eqs.~(\ref{seedNN}) 
and~(\ref{seednuN}), we find that the resulting energy loss is too small 
to affect the supernova dynamics. Turning to the limit when $\lambda$ is 
at its upper bound in Eq.~(\ref{lambdabound}), some number of composite 
singlet neutrinos are produced inside the supernova core provided 
$\Lambda \lesssim 5\, {\rm GeV}$. Once produced, these particles remain 
trapped inside the core.  For $\Lambda \lesssim 1.7\, {\rm GeV}$, they 
come into thermal and chemical equilibrium. For values of $\Lambda$ 
above 1.7 GeV they enter into thermal equilibrium but not chemical 
equilibrium. For intermediate values of $\lambda$ the dynamics is 
expected to lie between these two extremes. We conclude that for 
$\Lambda \gg T_{\rm SN}$, there is no significant contribution to 
supernova cooling from the hidden sector.

We now turn our attention to the case when the compositeness scale is 
less than or comparable to the supernova temperature, $\Lambda \lesssim 
T_{\rm SN}$. In this regime composite singlet neutrinos are easily 
produced, and once produced, will be trapped inside the supernova. Now 
however, since $T_{\rm SN} \gtrsim \Lambda$, once the trapped singlet 
neutrinos approach thermal equilibrium with the SM inside the supernova, 
the composite dynamics may undergo a phase transition back to the 
conformal phase. If the number of degrees of freedom in the hidden 
sector is sizable, this could potentially have a large impact on the 
supernova dynamics. However, given that at present our understanding of 
supernova explosions is limited, it is difficult to use this to place a 
robust constraint.

Our discussion up till now has not taken into account the effects of the 
chemical potential of the electron neutrinos in the supernova core. This 
is of order 200 MeV, higher then $T_{\rm SN}$. The primary effect of the 
chemical potential is to make the supernova even more opaque to 
composite singlet neutrinos, and so the conclusion that there is no 
significant contribution to supernova cooling from the hidden sector 
remains unaltered. Now however, some part of the chemical potential 
could be passed on to the trapped composite singlet neutrinos, 
increasing their net abundance inside the supernova core. For 
compositeness scales of order 200 MeV or lower, this effect may be large 
enough to significantly affect the supernova dynamics. However, it is 
once again not clear that this leads to a robust constraint. We defer a 
careful study of the effects of this class of models on supernovae for
future work.

\section{Conclusions\label{s.conclusion}}

We have proposed a framework in which the SM neutrinos mix with the 
composite fermions of a strongly coupled hidden sector through the 
neutrino portal. The light neutrinos are then partially composite 
particles. An explicit breaking of lepton number in the hidden sector 
allows the neutrinos to obtain small Majorana masses. From the low 
energy perspective this framework leads to an inverse seesaw model for 
neutrino masses in which the masses of the singlet neutrinos are set by 
the compositeness scale. However, if probed at energies above the 
compositeness scale, where the constituents of the singlet neutrinos are 
strongly coupled, the phenomenology of this class of models differs 
greatly from that of a conventional inverse seesaw. For example, the 
decay of SM particles to the hidden sector can now result in multiple 
singlet neutrinos rather than just one. In addition, the large 
self-interactions of the composite singlet neutrinos allow them to 
annihilate away efficiently in the early universe, so that the severe 
cosmological bounds on singlet neutrinos with masses below a GeV are 
naturally satisfied.

We have outlined a specific realization of this scenario in which, in 
the ultraviolet, the composite sector is a strongly coupled CFT. The 
small parameters necessary to obtain realistic neutrino masses in the 
framework of a low-scale seesaw naturally arise as a consequence of the 
scaling dimensions of operators in the CFT. Neutrino masses could be 
small either because the mixing with composite states is dynamically 
suppressed, or because lepton number is an approximate symmetry of the 
hidden sector at the compositeness scale, with a continuum of 
possibilities between these two extremes. The approximate lepton number 
symmetry could also arise as a consequence of dynamical effects. This 
gives a concrete but very broad range of possibilities for the 
compositeness scale and for phenomenology. In particular, the scale of 
neutrino compositeness can be as low as an MeV. Below an MeV, we find 
that the cosmological limits on $N_\mathrm{eff}$ disfavor this scenario.

We have shown that this class of models has important implications for a 
wide range of experiments, including colliders and beam dumps, searches 
for lepton flavor violation and neutrinoless double beta decay, and 
precision observations of the CMB spectral shape. SM particles can now 
decay to final states that include multiple composite singlet neutrinos. 
These particles can be long-lived, resulting in striking displaced 
vertex signals at colliders and beam dumps. At loop level, the composite 
sector can contribute to lepton flavor violation processes such as $\mu 
\rightarrow e \gamma$ and $\mu \rightarrow e$ conversion at rates that 
can be probed in upcoming experiments such as COMET and Mu2e. If the 
compositeness scale is of order the pion mass or below, the rate of 
neutrinoless double beta decay will be suppressed by form factors. For 
compositeness scales below 50 MeV, the late decays of relic composite 
singlet neutrinos can give rise to spectral distortions in the CMB that 
are large enough to be observed in next generation experiments.

\acknowledgments
 ZC and ZL are supported in part by the National Science Foundation 
under Grant Number PHY-1914731. ZC is also supported in part by the 
US-Israeli BSF Grant 2018236. ZL is also supported in part by the 
Maryland Center for Fundamental Physics. ZC would like to thank the 
Fermilab Theory Group for hospitality during the completion of this 
work. ZC's stay at Fermilab was supported by the Fermilab Intensity 
Frontier Fellowship and the Visiting Scholars Award \#17-S-02 from the 
Universities Research Association. PF, RH, and ZL would like to thank 
the Aspen Center for Physics which is supported by National Science 
Foundation grant PHY-1607611, where part of the study was performed. PF 
and RH would like to thank the Galileo Galilei Institute for hospitality 
where part of this work was completed, and PF would like to thank the 
Simons Foundation (SIMONS, 341344 AL) for support. PF and RH are 
supported by Fermi Research Alliance, LLC under Contract 
DE-AC02-07CH11359 with the U.S. Dept. of Energy.

\appendix

\section{Matrix Elements and Widths}



For the process $W\rightarrow \ell\, \mathcal{U}$, the matrix element is,
\beq
i\mathcal{M}=\frac {g} {\sqrt 2} \frac {\hat \lambda v_{\rm EW}} {M_{\rm UV}^{\Delta_{\mathrm N}-3/2}} \bar \ell \gamma^\mu \frac 1 2 (1-\gamma_5) \frac {\slashed{N}} {N^2} N \epsilon_\mu (W),
\eeq
where $g$ is the $SU(2)$ gauge coupling. $\bar \ell$ and $N$ are the spinors of the charged lepton and the unparticle state, and $\epsilon_\mu (W)$ is the polarization vector of the $W$ boson with $W$ being its 4-momenta.

In evaluation of the spin and polarization averaged matrix element squared, we use the identity,
\beq
\sum_{\mathrm{pols.}} \epsilon_\mu\epsilon_\nu^*=-g_{\mu\nu}+\frac {W_\mu W_\nu} {m_W^2},
\eeq
and we obtain,
\bea
|\overline{\mathcal{M}}|^2
&=&g^2\frac {|\hat\lambda v_{\rm EW}|^2} {M_{\rm UV}^{2\Delta_{\mathrm N}-3}m_N^2} (\ell\cdot N) \left(2+\frac {m_N^2} {m_W^2}\right)\\
&=&g^2\frac {|\hat\lambda v_{\rm EW}|^2} {2M_{\rm UV}^{2\Delta_{\mathrm N}-3}m_N^2} (m_W^2-m_N^2) \left(2+\frac {m_N^2} {m_W^2} \right)
\label{eq:m2fermion}
\eea
where $\ell$ and $N$ are the 4-momenta of the corresponding states, and we have ignored the mass of $\ell$.  In the $W$ rest frame $m_N^2=m_W^2-2m_W E_\ell$ and the matrix element squared becomes,
\be
|\overline{\mathcal{M}}|^2 = g^2\frac {|\hat\lambda v_{\rm EW}|^2} {2M_{\rm UV}^{2\Delta_{\mathrm N}-3}} \frac{E_\ell}{m_W}\frac{3m_W-2E_\ell}{m_W-2E_\ell}~.
\ee




The only kinematic quantity this matrix element depends on is the lepton energy $E_\ell$.  Using the rest frame relation $p_N^2=m_W^2-2m_W E_\ell$, the  integration over lepton and unparticle phase space can be carried out,
\bea
\Gamma=&&\int_0^1\frac {g^2 m_W} {192\pi^2} \frac {|\hat \lambda v_{\rm EW}|^2} {m_W^2} \left(\frac {m_W} {M} \right)^{2\Delta_{\mathrm N}-3} 
A_{\Delta_{\mathrm N}-1/2}\left(\frac{3 m_W-2E_\ell}{m_W-2E_\ell}\right)\left(\frac {2 E_\ell} {m_W}\right)^2 \nonumber\\
&&\times\left (1-\frac {2E_\ell} {m_W}-\frac {\mu_{IR}^2} {m_W^2}\right)^{\Delta_{\mathrm N}-5/2} \theta(1-\frac {2E_\ell}{m_W} -\frac {\mu_{IR}^2} {m_W^2} ) d\frac {2 E_\ell} {m_W}\nonumber \\
=&& \frac {g^2 m_W} {192\pi^2} \frac {|\hat \lambda v_{\rm EW}|^2} {m_W^2} \left(\frac {m_W} {M} \right)^{2\Delta_{\mathrm N}-3} A_{\Delta_{\mathrm N}-1/2}\times f(\Delta_{\mathrm N},\delta),
\eea
where, we have defined $\delta\equiv {\mu_{IR}^2}/{m_W^2}$.  The full expression for $f(\Delta_{\mathrm N},\delta)$ is cumbersome,
\bea
f(\Delta_{\mathrm N},\delta)
&=& 
-4(1-\delta)^{\Delta_{\mathrm N}-\frac{3}{2}}
\frac{4(\Delta_{\mathrm N}-1)(\Delta_{\mathrm N}+2)-2(2\delta-1)(\delta-1)-4\delta \Delta_{\mathrm N}+7}{(2\Delta_{\mathrm N}-3)(2\Delta_{\mathrm N}-1)(2\Delta_{\mathrm N}+1)}\nonumber\\
&& + 2\delta^{\Delta_{\mathrm N}-\frac{5}{2}} B_{\frac{1}{\delta}-1}\left(\Delta_{\mathrm N}-\frac{5}{2}\right)
\label{eq:fnew}
\eea
where $B_z(a)$ is 
\be
B_z(a)=\int_0^z dt\,\frac{t^{a}}{1+t}~.
\ee
Some interesting limits for the product of $f(\Delta_{\mathrm N},\delta)$ and the phase volume factor $A_{\Delta_{\mathrm N}-1/2}$ are
\bea
\lim_{\Delta_{\mathrm N}\rightarrow\frac{3}{2}}A_{\Delta_{\mathrm N}-\frac{1}{2}}f(\Delta_{\mathrm N},\delta)&=&\frac{2 \pi  \left(\delta ^3-3 \delta +2\right)}{\delta }~,\\
\lim_{\delta\rightarrow 0}A_{\Delta_{\mathrm N}-\frac{1}{2}}f(\Delta_{\mathrm N},\delta)&=& 
(16\pi^2)^{2-\Delta_{\mathrm N}}\frac{\Gamma(\frac{5}{2}-\Delta_{\mathrm N})}{\Gamma(\Delta_{\mathrm N}-\frac{1}{2})}\, \delta^{\Delta_{\mathrm N}-\frac{5}{2}}+\mathcal{O}(\delta^0)~.
 \eea
 When the infrared cutoff $\mu_{IR}$ in the unparticle treatment is 
identified with the mass of the composite singlet neutrinos $M_N$, our 
calculations agree with the standard results for elementary HNLs in the 
$\Delta_{\mathrm N}\rightarrow 3/2$ limit.

 For meson decays, neglecting the lepton mass, the partial width is given by
 \bea
d\Gamma(\mathfrak{m}\rightarrow \ell \mathcal{U})&=& \frac {G_F^2 |V_{qq^\prime}|^2 f_{\mathfrak{m}}^2|\hat \lambda v_{\rm EW}|^2} {M_{\rm UV}^{2\Delta_{\mathrm N}-3}} (2 E_\ell) d\Phi_{(\Delta_{\mathrm N}-1/2)+1} \\
&=& \frac {m_\mathfrak{m}G_F^2 |V_{qq^\prime}|^2 f_{\mathfrak{m}}^2|\hat \lambda v_{\rm EW}|^2} {32\pi^2} \left(\frac {m_\mathfrak{m}} {M_{\rm UV}}\right)^{2\Delta_{\mathrm N}-3} A_{\Delta_{\mathrm N}-1/2}\left(\frac {2E_e} {m_\mathfrak{m}}\right)^2  \nonumber \\
&& \times \left(1-\frac{2E_\ell} {m_\mathfrak{m}}-\frac{\mu_{IR}^2} {m_\mathfrak{m}^2}\right)^{\Delta_{\mathrm N}-5/2} \theta(1-\frac {2E_\ell} {m_\mathfrak{m}}-\frac{\mu_{IR}^2} {m_\mathfrak{m}^2})d \frac {2E_\ell} {m_\mathfrak{m}},
 \eea
 where $m_\mathfrak{m}$ is the mass of the charged meson.

After integration, the partial width of $\mathfrak{m}\rightarrow e \mathcal{U}$ is,
\beq
\Gamma(\mathfrak{m}\rightarrow e \mathcal{U})
=\frac {m_\mathfrak{m}G_F^2 |V_{qq^\prime}|^2 f_{\mathfrak{m}}^2|\hat \lambda v_{\rm EW}|^2} {32\pi^2} \left(\frac {m_\mathfrak{m}} {M_{\rm UV}}\right)^{2\Delta_{\mathrm N}-3} A_{\Delta_{\mathrm N}-1/2}\times g\left(\Delta_{\mathrm N},\frac {\mu_{IR}^2} {m_\mathfrak{m}^2}\right),
\eeq
with
\beq
g\left(\Delta_{\mathrm N},\delta\right)=\frac{16 (1-\delta )^{\Delta _N+\frac{1}{2}}}{(2\Delta_{\mathrm N}-3)(2\Delta_{\mathrm N}-1)(2\Delta_{\mathrm N}+1)}~.
\label{eq:gnew}
\eeq
In this case the two limits of interest are,
\bea
\lim_{\Delta_{\mathrm N}\rightarrow\frac{3}{2}}A_{\Delta_{\mathrm N}-\frac{1}{2}}g(\Delta_{\mathrm N},\delta)&=&2\pi(1-\delta)^2~,\\
\lim_{\delta\rightarrow 0}A_{\Delta_{\mathrm N}-\frac{1}{2}}g(\Delta_{\mathrm N},\delta)&=& \frac{(16\pi^2)^{2-\Delta_{\mathrm N}}}{\Gamma(\Delta_{\mathrm N}+\frac{3}{2})\Gamma(\Delta_{\mathrm N}-\frac{1}{2})}+\mathcal{O}(\delta)~.
\eea

\bibliographystyle{utphys}
\bibliography{compnu}

\end{document}